\documentclass[acmsmall, screen]{acmart}

\usepackage[normalem]{ulem}
\usepackage{bm}

\usepackage{subfigure} 
\usepackage{graphicx}
\usepackage{ragged2e}
\usepackage[justification=centering]{caption}

\usepackage{amsmath}
\usepackage{booktabs}
\usepackage{multirow}
\usepackage{makecell}

\usepackage{longtable}
\usepackage{xtab}

\usepackage{tikz}
\usetikzlibrary{decorations.pathmorphing}
\usepackage[edges]{forest}
\usepackage{xcolor}

\newcommand*\circled[1]{\tikz[baseline=(char.base)]{\node[shape=circle, draw, inner sep=1.2pt] (char) {#1};}}

\usepackage{diagbox}

\usepackage{pifont}
\usepackage{tcolorbox}
\newcommand{\summary}[1]{
    \begin{center}
    \begin{tcolorbox}[colback=gray!15, colframe=black, boxsep=-0.15cm, middle=-0.15cm]
        \textbf{\ding{46} Summary}
        $\blacktriangleright$
        {#1}
    $\blacktriangleleft$
    \end{tcolorbox}
    \end{center}
}

\definecolor{deepblue}{rgb}{0.0, 0.0, 0.85}

\newcommand{\R}[1]{{\color{black}{#1}}}
\newcommand{\D}[1]{}

\newcommand{\abbr}{CodeLM}

\AtBeginDocument{%
  \providecommand\BibTeX{{%
    \normalfont B\kern-0.5em{\scshape i\kern-0.25em b}\kern-0.8em\TeX}}}

\setcopyright{acmcopyright}
\copyrightyear{2024}
\acmYear{2024}
\acmDOI{10.1145/1122445.1122456}

\acmJournal{tosem}
\acmVolume{0}
\acmNumber{0}
\acmArticle{1}
\acmMonth{0}

\begin{document}

\title{Security of Language Models for Code: A Systematic Literature Review}


\author{Yuchen Chen}     \email{yuc.chen@smail.nju.edu.cn}
\orcid{0000-0002-3380-5564}
\affiliation{
  \institution{State Key Laboratory for Novel Software Technology, Nanjing University}
  \city{Nanjing}
  \state{Jiangsu}
  \country{China}
  \postcode{210093}
}

\author{Weisong Sun}
\authornote{Corresponding authors.}
\email{weisong.sun@ntu.edu.sg}
\orcid{0000-0001-9236-8264}
\affiliation{
  \institution{Nanyang Technological University}
  \city{singapore}
  \state{50 Nanyang Avenue}
  \country{Singapore}
  \postcode{639798}
}

\author{Chunrong Fang}
\authornotemark[1]
\email{fangchunrong@nju.edu.cn}
\orcid{0000-0002-9930-7111}
\affiliation{
  \institution{State Key Laboratory for Novel Software Technology, Nanjing University}
  \city{Nanjing}
  \state{Jiangsu}
  \country{China}
  \postcode{210093}
}

\author{Zhenpeng Chen}
\email{zhenpeng.chen@ntu.edu.sg}
\orcid{0000-0002-4765-1893}
\affiliation{
  \institution{Nanyang Technological University}
  \city{singapore}
  \state{50 Nanyang Avenue}
  \country{Singapore}
  \postcode{639798}
}

\author{Yifei Ge}
\email{gyf991213@126.com}
\orcid{0009-0009-0957-854X}
\affiliation{
  \institution{State Key Laboratory for Novel Software Technology, Nanjing University}
  \city{Nanjing}
  \state{Jiangsu}
  \country{China}
  \postcode{210093}
}

\author{Tingxu Han}    \email{txhan@smail.nju.edu.cn}
\orcid{0000-0003-1821-611X}
\affiliation{
  \institution{State Key Laboratory for Novel Software Technology, Nanjing University}
  \city{Nanjing}
  \state{Jiangsu}
  \country{China}
  \postcode{210093}
}

\author{Quanjun Zhang}    \email{quanjun.zhang@smail.nju.edu.cn}
\orcid{0000-0002-2495-3805}
\affiliation{
  \institution{State Key Laboratory for Novel Software Technology, Nanjing University}
  \city{Nanjing}
  \state{Jiangsu}
  \country{China}
  \postcode{210093}
}

\author{Yang Liu}    
\email{yangliu@ntu.edu.sg}
\orcid{0000-0001-7300-9215}
\affiliation{
  \institution{Nanyang Technological University}
  \city{singapore}
  \state{50 Nanyang Avenue}
  \country{Singapore}
  \postcode{639798}
}

\author{Zhenyu Chen}
\email{zychen@nju.edu.cn}
\orcid{0000-0002-9592-7022}
\affiliation{
  \institution{State Key Laboratory for Novel Software Technology, Nanjing University}
  \city{Nanjing}
  \state{Jiangsu}
  \country{China}
  \postcode{210093}
}

\author{Baowen Xu}
\email{bwxu@nju.edu.cn}
\orcid{0000-0001-7743-1296}
\affiliation{
  \institution{State Key Laboratory for Novel Software Technology, Nanjing University}
  \city{Nanjing}
  \state{Jiangsu}
  \country{China}
  \postcode{210093}
}

\renewcommand{\shortauthors}{Y. Chen, W. Sun, C. Fang, Z. Chen,  Y. Ge, T. Han, Q. Zhang, Y. Liu, Z. Chen, B. Xu}

\begin{abstract}
Language models for code (\abbr{}s) have emerged as powerful tools for code-related tasks, outperforming traditional methods and standard machine learning approaches. However, these models are susceptible to security vulnerabilities, drawing increasing research attention from domains such as software engineering, artificial intelligence, and cybersecurity. Despite the growing body of research focused on the security of \abbr{}s, a comprehensive survey in this area remains absent. To address this gap, we systematically review 68 relevant papers, organizing them based on attack and defense strategies. Furthermore, we provide an overview of commonly used language models, datasets, and evaluation metrics, and highlight open-source tools and promising directions for future research in securing \abbr{}s.
\end{abstract}



\maketitle

\section{Introduction}
\label{sec:introduction}

Language models for code (in short as \abbr{}s), such as Codex~\cite{2023-Codex} and CodeLlama~\cite{MetaCodeLlama}, have become powerful tools in Software Engineering (SE) \cite{DBLPabs2402977}, addressing tasks such as code search~\cite{2022-Code-Search-based-on-Context-aware-Code-Translation, 2024-survey-Source-Code-Search} and code completion~\cite{2019-Pythia, 2024-Code4Me}. These language models utilize advanced deep learning techniques to analyze and represent code, significantly improving performance compared to traditional methods. One notable development is GitHub Copilot~\cite{GitHubCopilot}, an AI-powered coding assistant built on the Codex~\cite{2023-Codex} model. Trained on diverse programming patterns and vast open-source code from GitHub, Codex enables Copilot to suggest and complete code in real-time, greatly enhancing developers' productivity.

Despite the success of \abbr{}s in solving various code-related tasks, their inherent vulnerabilities raise significant concerns, particularly in security-critical applications such as network security and malware detection. These models are susceptible to backdoor attacks, where hidden triggers are embedded during training~\cite{2023-BADCODE}. \R{Triggers are specific input patterns designed by attackers, such as particular variable names or dead code snippets in the code, which are used to activate the model’s malicious behavior.}\D{Once activated by an attacker, the model's behavior changes maliciously.} For example, Schuster et al.~\cite{2021-you-autocomplete-me} demonstrated that backdoors in \abbr{}s can significantly increase the likelihood of generating vulnerable code. \D{Additionally}\R{In addition to backdoor attacks}, adversarial attacks~\cite{2014-Intriguing-properties-of-neural-networks, 2021-A-survey-on-adversarial-attacks-and-defences} \D{, which}\R{present another challenge to the security of \abbr{}s.} \R{These attacks} occur during the inference phase, \R{where attackers} add subtle perturbations to inputs, causing models to make incorrect predictions with high confidence\D{. These perturbations are often imperceptible to humans, but they can mislead the model}, further highlighting the fragility of these models.

To counter attacks on \abbr{}s, various defense methods have been developed. For backdoor attacks, defenses often focus on detecting outliers in the training data~\cite{2022-backdoors-in-neural-models-of-source-code, 2024-poison-attack-and-poison-detection, 2025-EliBadCode, 2025-KillBadCode}. After identifying and removing suspicious data points, the model is retrained on the cleaned dataset to ensure security. For adversarial attacks, adversarial training~\cite{2022-Semantic-Robustness-of-Models-of-Source-Code} is an effective defense, where adversarial samples are incorporated into the training \D{data}\R{process} to enhance the model's robustness against such attacks.

While numerous methods for attacking and defending \abbr{}s have been proposed, a comprehensive review of the existing research in this field remains absent. To address this gap, we present a systematic survey on the security of \abbr{}s. We conduct extensive literature searches across six widely-used databases using 407 search keywords, complemented by a snowballing approach to ensure thoroughness. This process yields 68 relevant papers. Based on these studies, we categorize common attacks on \abbr{}s, focusing on backdoor and adversarial attacks, and summarize corresponding defense strategies. Additionally, we review common experimental settings, including datasets, language models, evaluation metrics, and artifact accessibility. Finally, we highlight open challenges and future opportunities in securing \abbr{}s.

In summary, the key contributions of this paper include:
\begin{itemize}
    \item \textit{First Comprehensive Survey on Security of \abbr{}s.} To the best of our knowledge, we conduct the first systematic literature review on the security of \abbr{}s, analyzing 68 relevant papers. Our survey categorizes various attack and defense strategies for \abbr{}s, covering multiple critical aspects and application scenarios.

    \item \textit{\abbr{}s, Datasets, Metrics, Tools.} We provide an overview of the prominent language models, datasets, and commonly used evaluation metrics in \abbr{} security. Additionally, we curate and summarize open-source tools and datasets, which are available in our repository~\cite{2024-Security-of-LM4Code}.
      
    \item \textit{Outlook and Challenges.} We outline the key challenges and emerging opportunities in \abbr{} security research, building on a comprehensive review of prior studies to encourage further exploration in this field. 
\end{itemize}

\begin{figure*}[t]
    \centering
    \resizebox{0.9\textwidth}{!}{
        \begin{forest}
            forked edges,
            for tree={
                grow=east,
                reversed=true,
                anchor=base west,
                parent anchor=east,
                child anchor=west,
                font=\large,
                rectangle,
                draw=black,
                rounded corners,
                base=left,
                align=left,
                minimum width=4em,
                edge+={darkgray, line width=1pt},
                s sep=3pt,
                inner xsep=2pt,
                inner ysep=3pt,
                line width=1pt,
                ver/.style={rotate=90, child anchor=north, parent anchor=south, anchor=center},
            },
            where level=1{text width=18em, font=\normalsize}{},
            where level=2{text width=18em, font=\normalsize}{},
            where level=3{text width=18em, font=\normalsize}{},
            where level=4{text width=18em, font=\normalsize}{},
            [Literature Review on \\ \abbr{} Security
                [{\;}\S\ref{sec:introduction} INTRODUCTION]
                [{\;}\S\ref{sec:background_and_related_work} BACKGROUND \\ AND RELATED WORK
                    [{\;}\S\ref{subsec:language_models_for_code} Language Models for Code]
                    [{\;}\S\ref{subsec:language_model_security} Language Model Security]
                    [{\;}\S\ref{subsec:comparison_with_existing_work} Comparison with Existing Work]
                ]
                [{\;}\S\ref{sec:survey_methodology} SURVEY METHODOLOGY
                    [{\;}\S\ref{subsec:research_question} Research Question]
                    [{\;}\S\ref{subsec:survey_scope} Survey Scope]
                    [{\;}\S\ref{subsec:search_strategy} Search Strategy
                        [{\;}\S\ref{subsubsec:search_items} Search Strings]
                        [{\;}\S\ref{subsubsec:search_sources} Search Sources]
                        [{\;}\S\ref{subsubsec:study_selection} Study Selection]
                    ]
                    [{\;}\S\ref{subsec:snowballing_search} Snowballing Search]
                    [{\;}\S\ref{subsec:collection_results} Collection Results]
                    [{\;}\S\ref{subsec:taxonomy_construction} Taxonomy Construction]
                ]
                [{\;}\S\ref{sec:Answer_to_RQ1} ANSWER TO RQ1: ATTACKS \\ ON CODELMS
                    [{\;}\S\ref{subsec:research_directions_on_attacks_LM4Code} Research Directions in Attacks \\ against \abbr{}s]
                    [{\;}\S\ref{subsec:backdoor_attacks_on_LM4Code} Backdoor Attacks on \abbr{}s
                        [{\;}\S\ref{subsubsec:overview_of_backdoor_attacks} Overview of Backdoor Attacks \\ against \abbr{}s]
                        [{\;}\S\ref{subsubsec:data_poisoning_attacks} Data Poisoning Attacks]
                        [{\;}\S\ref{subsubsec:model_poisoning_attacks} Model Poisoning Attacks]
                        [{\;}\S\ref{subsubsec:summary_of_backdoor_attacks} Summary of Backdoor Attacks]
                    ]
                    [{\;}\S\ref{subsec:adversarial_attacks_on_LM4Code} Adversarial Attacks on \abbr{}s
                        [{\;}\S\ref{subsubsec:overview_of_adversarial_attacks} Overview of Adversarial Attacks \\ against \abbr{}s]
                        [{\;}\S\ref{subsubsec:white-box_attacks} White-box Attacks on \abbr{}s]
                        [{\;}\S\ref{subsubsec:black-box_attacks} Black-box Attacks on \abbr{}s]
                        [{\;}\S\ref{subsubsec:summary_of_adversarial_attacks} Summary of Adversarial Attacks]
                    ]
                ]
                [{\;}\S\ref{sec:Answer_to_RQ2} ANSWER TO RQ2: DEFENSES \\ ON CODELMS
                    [{\;}\S\ref{subsec:research_directions_on_defenses_LM4Code} Research Directions in Defenses \\ against \abbr{}s]
                    [{\;}\S\ref{subsec:backdoor_defenses_on_LM4Code} Backdoor Defenses on \abbr{}s
                        [{\;}\S\ref{subsubsec:overview_of_backdoor_defenses} Overview of Backdoor Defenses \\ against \abbr{}s]
                        [{\;}\S\ref{subsubsec:pre_training_backdoor_defenses} Pre-training Backdoor Defenses]
                        [{\;}\S\ref{subsubsec:in_training_backdoor_defenses} In-training Backdoor Defenses]
                        [{\;}\S\ref{subsubsec:post_training_backdoor_defenses} Post-training Backdoor Defenses]
                        [{\;}\S\ref{subsubsec:summary_of_backdoor_defenses} Summary of Backdoor Defenses]
                    ]
                    [{\;}\S\ref{subsec:adversarial_defenses_on_LM4Code} Adversarial Defenses on \abbr{}s
                        [{\;}\S\ref{subsubsec:overview_of_adversarial_defenses} Overview of Adversarial Defenses \\ against \abbr{}s]
                        [{\;}\S\ref{subsubsec:adversarial_training} \D{Adversarial Training}\R{Input Modification}]
                        [{\;}\S\ref{subsubsec:model_modification} Model Modification]
                        [{\;}\S\ref{subsubsec:additional_models} Additional Models]
                        [{\;}\S\ref{subsubsec:summary_of_adversarial_defenses} Summary of Adversarial Defenses]
                    ]   
                ]
                [{\;}\S\ref{sec:Answer_to_RQ3} ANSWER TO RQ3: EMPIRICAL \\ STUDIES ON SECURITY OF CODELMS 
                    [{\;}\S\ref{subsec:empirical_studies_on_backdoor_attacks} Practical impact of backdoor attacks \\ on \abbr{}s]
                    [{\;}\S\ref{subsec:empirical_studies_on_adversarial_attacks} Performance of adversarial attacks \\ against \abbr{}s across different tasks \\ and models][{\;}\S\ref{subsec:empirical_studies_on_adversarial_defenses} Effectiveness and robustness of \\ adversarial training for \abbr{}s]
                ]
                [{\;}\S\ref{sec:Answer_to_RQ4} ANSWER TO RQ4: EXPERIMENTAL \\ SETTING AND EVALUATION 
                    [{\;}\S\ref{subsec:datasets} Experimental Datasets]
                    [{\;}\S\ref{subsec:experimental_LM4Code} Experimental \abbr{}s]
                    [{\;}\S\ref{subsec:performance_metrics} Evaluation Metrics]
                    [{\;}\S\ref{subsec:replication_packages} Artifact Accessibility]
                ][{\;}\S\ref{sec:challenges_and_opportunities} CHALLENGES AND OPPORTUNITIES 
                    [{\;}\S\ref{subsec:challenges_and_opportunities_in_attacks} Challenges and Opportunities in \\ Attacks against \abbr{}s]
                    [{\;}\S\ref{subsec:challenges_and_opportunities_in_defenses} Challenges and Opportunities in \\ Defenses against Attacks on \abbr{}s]
                ]
                [{\;}\S\ref{sec:threats_to_validity}
                \R{THREATS TO VALIDITY}]
                [{\;}\S\ref{sec:conclusion} CONCLUSION]
            ]   
        \end{forest}
        }
    \caption{Structure of the paper.}
    \Description{Structure of the paper.}
    \label{fig:structure_of_the_paper}
\end{figure*}

\noindent \textbf{Structure of the paper:} 
The structure of this paper is presented in Figure~\ref{fig:structure_of_the_paper}.
The subsequent sections of this paper are structured as follows. Section~\ref{sec:background_and_related_work} introduces the background of language models for code and language model security. Section~\ref{sec:survey_methodology} presents the survey methodology that we adhere to.
After that, we present an overview of attacks and defenses against \abbr{}s.
Sections~\ref{subsec:backdoor_attacks_on_LM4Code} and~\ref{subsec:adversarial_attacks_on_LM4Code} provide a review of existing backdoor threats, adversarial threats in \abbr{}s, respectively. 
On the other hand, Sections~\ref{subsec:backdoor_defenses_on_LM4Code} and~\ref{subsec:adversarial_defenses_on_LM4Code} provide a review of existing backdoor defenses and adversarial defenses. 
Section~\ref{sec:Answer_to_RQ3} reviews the existing empirical studies that explore the capabilities of attack or defense techniques and their impact on \abbr{}s.
Section~\ref{sec:Answer_to_RQ4} summarises the datasets, LMs, and metrics widely utilized in the field of \abbr{} security. 
In Section~\ref{sec:challenges_and_opportunities}, we outline the challenges and opportunities in this field for future work.
In Section~\ref{sec:threats_to_validity}, we discuss the potential threats to the validity of this study.
Finally, Section~\ref{sec:conclusion} provides a conclusion of this survey.

\section{Background and Related Work}
\label{sec:background_and_related_work}

\begin{table*}[!t]
    \centering
    \scriptsize
    \caption{\R{Common technical terms related to the security of \abbr{}s.}}
    \label{tab:technical_terms}
    \resizebox{0.95\linewidth}{!}{
    \begin{tabular}{lp{11cm}}
        \toprule
        
        \textbf{\R{Terms}} & \textbf{\R{Explanation}} \\
        
        \midrule
        
        \R{CodeLMs} & \noindent\R{Language models used for understanding, generating, or processing code, leveraging neural networks and deep learning technique similar to NLP. Examples include CodeBERT, CodeT5, Codex, and DeepSeek Coder.} \\

        \midrule
        
        \R{Trigger} & \noindent\R{A trigger is a specific input pattern carefully designed by an attacker. It can be a specific token, such as variable names, code snippets, or words in comments. The attacker typically embed these triggers into the model during training, ensuring that the model behaves normally under regular inputs but exhibits specific anomalous behaviors when encountering the trigger.} \\

        \midrule

        \R{Benign Sample} & \noindent\R{A benign sample (also known as a clean sample) is a data sample that has not been affected by malicious modifications. It is considered ``normal'' and does not contain harmful attributes such as backdoor triggers or adversarial perturbations.} \\

        \midrule
        
        \R{Poisoned Sample} & \noindent\R{A poisoned sample is a maliciously modified data sample that contains attacker-designed triggers, which are used to implant backdoors into the model during training.} \\

        \midrule
        
        \R{Benign Model} & \noindent\R{A Benign model (also known as a clean model) is a model trained on Benign samples, behaving as expected without backdoors or malicious features.} \\

        \midrule

        \R{Poisoned Model} & \noindent\R{A poisoned model is a model trained on poisoned samples, implanted with a backdoor by an attacker. It behaves normally under regular conditions but produces malicious outputs when exposed to a trigger.} \\

        \midrule

        \R{Adversarial Perturbation} & \noindent\R{An adversarial perturbation is a minor modification to input data, such as variable name changes or dead code insertion, designed to mislead the model into making incorrect predictions without significantly altering the data’s meaning.} \\

        \midrule
        
        \R{Adversarial Example} & \noindent\R{Input data altered by adding small perturbations that mislead the model into making incorrect predictions, typically imperceptible to humans but misleading to the model.} \\

        \midrule

        \R{Source Label} & \noindent\R{The source label (also known as the victim label) refers to the original label prediction by the model for an input before the activation of a backdoor attack or an adversarial attack.} \\

        \midrule

        \R{Target Label} & \noindent\R{The target label is the label that the attacker wants the model to output, aiming to misclassify input samples containing triggers or adversarial perturbations into the attacker’s specified category.} \\

        \midrule

        \R{Untargeted Attack} & \noindent\R{An untargeted attack aims to cause the model’s prediction to deviate from the original correct class without enforcing classification into a specific incorrect label.} \\

        \midrule
        
        \R{Targeted Attack} & \noindent\R{A targeted attack aims to manipulate the model’s prediction, causing the input sample to be misclassified into a specific class predetermined by the attacker.} \\
        
        \bottomrule
    \end{tabular}
    }
\end{table*}

In this section, we introduce the background of \abbr{}s, discuss security threats to neural models, and provide a comparison with existing related surveys.
\R{To enhance readability and facilitate understanding, we first provide brief descriptions and explanations of common technical terms related to the security of \abbr{}s, as shown in Table~\ref{tab:technical_terms}.}

\subsection{Language Models for Code}
\label{subsec:language_models_for_code}

\abbr{}s refer to models that interpret code using neural language models~\cite{palacio2023toward}.
In 2012, Hindle et al. pioneered the introduction of n-gram models into the SE field~\cite{hindle2016naturalness}. Since then, \abbr{}s have illustrated performance in various code-related tasks such as code search~\cite{2018-deep-code-search, sachdev2018retrieval, haldar2020multi, 2022-Code-Search-based-on-Context-aware-Code-Translation}, code summarization~\cite{leclair2019neural, jiang2017automatically, haque2020improved, ahmad2020transformer, 2024-ESALE, 2024-Source-Code-Summarization-in-the-Era-of-Large-Language-Models}, and code completion~\cite{alon2020structural, karampatsis2019maybe, kim2021code, svyatkovskiy2020intellicode}. These studies shape the problem at hand as a text-to-text transformation, where the input and output of the model are both text strings~\cite{mastropaolo2021studying}. Furthermore, considering that code embodies rich structural information, some studies propose extracting the structural information of code as input to capture deeper features of the code~\cite{2016-Convolutional-Neural-Networks-over-Tree-Structures, 2019-ASTNN, wan2019multi, zeng2023degraphcs}. 
Later, with significant advancements in transfer learning in the field of NLP (e.g., GPT~\cite{2018-GPT1}, BERT~\cite{2019-BERT}, T5~\cite{2020-T5}), the fundamental idea involves initially pre-training a model on a large and generic dataset using a self-supervised task (e.g., masking tokens in strings and requiring the model to guess the masked tokens). Subsequently, the pre-trained model is fine-tuned on smaller and specialized datasets, each designed to support specific tasks~\cite{2021-Pre-trained-models-Past-present-and-future}.
In this context, Feng et al.~\cite{2020-CodeBERT} propose CodeBERT, which integrates a token prediction scheme to understand code by predicting subsequent tokens, thereby enhancing its understanding of programming languages for tasks such as code completion and error detection. Building upon this, Guo et al.~\cite{2021-GraphCodeBERT} introduce GraphCodeBERT, which incorporates edge-type prediction to identify relationships between code elements and represent them as graphs. This enables GraphCodeBERT to leverage code structure, improving its performance in tasks like code summarization and program analysis.
In recent years, researchers have observed that significantly increasing the scale of pre-trained models can enhance their capacity, leading to significant performance improvements when the parameter size surpasses a certain threshold~\cite{hoffmann2022training, shanahan2024talking, taylor2022galactica}. The term LLMs has been introduced to distinguish language models based on their parameter size, particularly referring to large program language models~\cite{zhao2023survey}. Specialized LLMs for code have emerged successively, such as GitHub Copilot~\cite{GitHubCopilot}, Amazon's CodeWhisperer~\cite{AmazonCodeWhisperer}, the OpenAI Code Interpreter~\cite{OpenAICodeInterpreter} integrated into ChatGPT, and Code Llama~\cite{MetaCodeLlama}. This marks a new stage in \abbr{}. These LLMs for code trained on code-specific datasets represent not only incremental improvements but also a paradigm shift in code understanding, generation, and efficiency. They have laid the foundation for breakthroughs in performance across various code-related tasks. For example, Fried et al.~\cite{2023-InCoder} introduce a large language model, InCoder, and try zero-shot training on the CodeXGLUE~\cite{2021-CodeXGLUE} Python dataset and achieve impressive results. Wu et al.~\cite{2024-Semantic-Sleuth} propose the first LLM-driven efficient and effective framework for detecting Ponzi smart contracts written in the Solidity code. 

A \abbr{} is essentially a DNN, and can be represented as a function $f: \mathcal{X} \rightarrow \mathcal{Y}$, where $\mathcal{X}$ is the input space consisting of source code (e.g., a method code snippet) and $\mathcal{Y}$ is the output space. A \abbr{} usually consists of a sequence of $n$ layers that are connected as follows:
\begin{equation}
    f_\theta (x) = f_{n-1} \circ f_{n-2} \cdots \circ f_0 (x),
\end{equation}
where $f_0$ is the first layer and $f_{n-1}$ the last. Variable $\theta$ denotes all the weight parameters in the $n$ layers. To train the \abbr{} that can correctly predict output $y \in \mathcal{Y}$ given an input $x \in \mathcal{X}$, a loss function $\mathcal{L}$ is utilized for searching the optimal parameters $\theta$ such that the empirical risk is minimized as follows.
\begin{equation}
    \mathop{\arg\min}_\theta \mathop{\mathbb{E}}_{(x, y) \sim \{\mathcal{X}, \mathcal{Y}\}} \Big[\mathcal{L} \big(f_\theta(x), y \big)\Big].
\end{equation}
The common code-related tasks can be mainly categorized into code understanding tasks (such as algorithm classification, code clone detection, and code search) and code generation tasks (such as code generation, code summarization, and code refinement). For code understanding tasks, the loss function can be represented as cross-entropy loss, expressed as follows.
\begin{equation}
    \mathcal{L} \big(\theta\big) = \sum_{i=0}^{N}{-y_{i}\log(f_\theta(x_i))}
\end{equation}
During the training of code generation tasks, the objectives are trained to minimize the negative conditional log-likelihood, and the loss function can be represented as:
\begin{equation}
    \mathcal{L} \big(\theta \big) = -\frac{1}{N}\sum_{i=1}^{N}\log P(x_n|f_\theta(x_n))
\end{equation}
The empirical risk is the expectation of the loss on all the samples in the given training set as shown above. The goal is to obtain a set of parameters $\theta$ that has the smallest empirical risk, which is supposed to correctly predict unseen inputs, i.e., those from the test set. The prediction accuracy on the test set denotes the functionality of $f$. The higher the test accuracy is, the better a \abbr{} is.

\subsection{Language Model Security}
\label{subsec:language_model_security}

In recent years, DL models have been widely applied to real-world tasks such as text analysis~\cite{ChenSHLML19,sigChenCLML19}, image classification~\cite{2012-ImageNet, 2016-resnet}, speech recognition~\cite{2011-Context-dependent-pre-trained-deep-neural-networks-for-large-vocabulary-speech-recognition, 2012-Deep-Neural-Networks-for-Acoustic-Modeling-in-Speech-Recognition}, and autonomous driving~\cite{li2024dark, 2019-Deep-Integration}. Despite the impressive performance of DL models, their security is of high concern. Due to the inherent flaws in DL models, attackers can craft a sample carefully to mislead DL models or guide the learner to train a poor model~\cite{2021-A-survey-on-adversarial-attacks-and-defences}. 
For example, attackers can significantly degrade overall performance by tampering with sensor-manipulated training data, leading to targeted misclassifications or the insertion of backdoors~\cite{2019-Transferable-Clean-Label-Poisoning-Attacks-on-Deep-Neural-Nets, 2017-BadNets, 2018-Trojaning-Attack-on-Neural-Networks}. 
He et al.~\cite{2022-Security-of-DL-System-Survey} categorize the security threats to DL models into four types: model extraction attack threats, model inversion attack threats, data poisoning attack threats, and adversarial attack threats. 
\textit{Model extraction attacks} attempt to replicate a DL model via the provided API without prior knowledge of its training data and algorithms~\cite{2016-Stealing-ML-Models-via-Prediction-APIs}. Such attacks not only compromise the confidentiality of the model and harm its owner's interests but also allow for the construction of an approximately equivalent white-box model, facilitating further attacks (e.g., adversarial attacks)~\cite{2017-Practical-Black-Box-Attacks-against-ML}.
\textit{Model inversion attacks} exploit the information flow extracted and abstracted from the training data by the model, using the model's predictions or confidence scores to restore data memberships or data properties~\cite{2017-ML-Models-Remember-Too-Much}. 
\textit{Poisoning attacks} primarily aim to reduce the prediction accuracy of DL models by contaminating the training data, thus compromising the model’s availability~\cite{2018-Manipulating-Machine-Learning}. 
For example, Lv et al.~\cite{2023-Data-free-Backdoor-Injection} propose a generic data-free backdoor poisoning attack against diverse DL tasks and models by utilizing a substitute dataset irrelevant to the main task. 
Recently, many researchers have leveraged poisoning attacks to create backdoors in the target model, known as backdoor attacks. While the model may behave normally most of the time, it will produce incorrect predictions when exposed to carefully crafted data. With the pre-implanted backdoor and trigger data, attackers can manipulate prediction results and initiate further attacks.
\textit{Adversarial attacks} add imperceptible perturbations to normal samples during the prediction process of DL models, generating adversarial examples~\cite{2017-Towards-Evaluating-the-Robustness-of-Neural-Networks}. Adversarial examples must both deceive the classifier and be imperceptible to humans. 
Similar to poisoning attacks, adversarial attacks also cause the model to misclassify malicious samples. The difference is that poisoning attacks insert malicious samples into the training data, directly contaminating the model, while adversarial attacks exploit the model's weaknesses using adversarial examples to achieve incorrect prediction results.

As a type of DL model, LMs naturally face one or more security threats encountered by DL models. For example, adversarial attacks can manipulate input text to deceive LMs into generating incorrect or biased outputs, similar to how adversarial examples trick image classifiers.
However, due to the nature of text, LMs also face unique challenges.
Unlike image or speech-based DL models, LMs process natural language, which makes them more vulnerable to subtle manipulations of word sequences that can significantly alter the output. 
Additionally, LMs are often deployed in interactive environments such as chatbots or automated content generators, increasing the risk of prompt injection or data leakage attacks. 
The interaction with human language and real-time context makes the security threats faced by LMs more complex compared to other DL models, where input formats and usage scenarios are generally more controlled.
Code, as a special form of text, has a structure and syntax that significantly differ from natural language. 
It must not only adhere to strict grammatical rules but also ensure logical accuracy and executability. 
Compared to natural language, the meaning of code is much more precise, so even small errors or changes during code generation and understanding can cause a program to fail or produce unexpected results. 
The security threats to \abbr{}s have also garnered significant attention. 
In this paper, we will focus on the security threats faced by \abbr{}s. \abbr{}s face one or more of the security threats mentioned above. The model may encounter various attacks when handling user inputs and generating code. These attacks can not only affect the correctness and security of the code but may also lead to serious security vulnerabilities. We will explore the nature of these threats and their potential impact on \abbr{}s, aiming to raise awareness of security issues in this domain and promote research into appropriate defense mechanisms.

\subsection{Comparison with Existing Work}
\label{subsec:comparison_with_existing_work}
Existing work has discussed various aspects of current \abbr{}s, including their capabilities on different specific tasks~\cite{2020-Software-Vulnerability-Detection-Using-Deep-Neural-Networks, 2023-Source-Code-Generation-Using-Deep-Learning, 2024-Survey-of-Learning-based-APR}, \R{the impact of data on \abbr{}s~\cite{2023-Data-Augmentation-Approaches-for-Source-Code-Models},} modeling source code with \abbr{}s~\cite{2021-Deep-Learning-for-Source-Code-Modeling-and-Generation}, the non-functional properties of \abbr{}s~\cite{2024-Robustness-Security-Privacy-Explainability-Efficiency-and-Usability-of-Large-Language-Models-for-Code,wang2024towards}, the trojan security in \abbr{}s~\cite{2023-A-Survey-of-Trojans-in-Neural-Models-of-Source-Code}, and so on. 
Among them, the works of~\cite{2024-Robustness-Security-Privacy-Explainability-Efficiency-and-Usability-of-Large-Language-Models-for-Code, 2023-A-Survey-of-Trojans-in-Neural-Models-of-Source-Code, 2023-Data-Augmentation-Approaches-for-Source-Code-Models} \D{and~\cite{2023-A-Survey-of-Trojans-in-Neural-Models-of-Source-Code}} also focus on the security aspects of \abbr{}s, which is similar to \R{the focus of} our literature review.
Specifically, Yang et al.~\cite{2024-Robustness-Security-Privacy-Explainability-Efficiency-and-Usability-of-Large-Language-Models-for-Code} explore seven non-functional properties of large \abbr{}s (CodeLLMs, for short), including robustness, security, privacy, explainability, efficiency, and usability. For the security property, they discuss the potential security threats that CodeLLMs may face (i.e., CodeLLMs as the victim), primarily focusing on backdoor security threats. 
Hussain et al.~\cite{2023-A-Survey-of-Trojans-in-Neural-Models-of-Source-Code} establish a taxonomy of trojan AI in code and its triggers based on research in explainable AI and trojan AI. Additionally, they drew actionable insights from explainable AI that are applicable to the field of trojan AI in code.
\R{Zhuo et al.~\cite{2023-Data-Augmentation-Approaches-for-Source-Code-Models} conduct a comprehensive survey on data augmentation methods specifically designed for source code. They introduce various data augmentation techniques, discuss common strategies for improving data quality, and review the application of data augmentation in common source code tasks. They also emphasize the significant role of data augmentation in different source code scenarios, particularly in generating adversarial examples to enhance the performance, robustness, and security of \abbr{}s.}
However, the aforementioned studies focus on specific aspects of \abbr{} security and do not provide a comprehensive overview of the current state of security research in \abbr{}s. Overall, there is still a lack of a systematic review of \abbr{} security. In this paper, we provide a systematic literature review on the security of \abbr{}s, including backdoor and adversarial security, with the aim of providing a foundation for future research in this area.

\section{Survey Methodology}
\label{sec:survey_methodology}

In this paper, we follow the systematic literature review (SLR) methodology proposed by Kitchenham et al.~\cite{2007-Guidelines-for-SLR-in-SE, 2023-SEGRESS} to conduct this survey. This method is widely utilized in the majority of other SLRs related to SE~\cite{2019-A-Review-of-Search-Based-SE, 2023-A-SLR-DL-for-Android-Malware-Defenses, 2023-DL-for-SE-A-SLR, 2024-Survey-of-Learning-based-APR, 2024-Large-Language-Models-for-Software-Engineering}. 
Our methodology consists of three main steps: planning the review, conducting the review, and analyzing the review results.

\subsection{Research Question}
\label{subsec:research_question}
We aim to summarize, categorize, and analyze empirical evidence from various published studies on attacks/defenses against \abbr{}s. To achieve this, we address three research questions (RQs):

\begin{itemize}
    \item \textbf{RQ1 (Attacks on \abbr{}s): }How do existing studies carry out attacks against \abbr{}s?

    \item \textbf{RQ2 (Defenses on \abbr{}s): }How do existing studies defend against attacks on \abbr{}s?

    \item \textbf{RQ3 (Empirical Studies on Security of \abbr{}s):} What do existing empirical studies on the security of \abbr{}s focus on?

     \item \textbf{RQ4 (Experimental Setting and Evaluation):} How do the existing attack/defense studies against \abbr{}s perform experimental setting and evaluation?
    \begin{itemize}
        \item \textbf{RQ4.1 (Experimental Datasets):} What are the commonly used experimental datasets for existing research on the security of \abbr{}s?

        \item
        \textbf{RQ4.2 (Experimental \abbr{}s):} What are the commonly used experimental LMs for code in existing research on the security of \abbr{}s? 
        
        \item \textbf{RQ4.3 (Evaluation Metrics):} What evaluation metrics are used to assess the performance of attacks and defenses in existing research on the security of \abbr{}s?
        
        \item \textbf{RQ4.4 (Artifact Accessibility):} Does the existing research on the security of \abbr{}s provide the artifact of the proposed/compared techniques (including experimental data and implementation code)?
    \end{itemize}
\end{itemize}

\subsection{Survey Scope}
\label{subsec:survey_scope}
The focus of our paper is to investigate the security of \abbr{}s, which involves two primary dimensions. Firstly, the papers utilize \abbr{}s, meaning they apply language models to code-related tasks. Secondly, the papers delve into the security aspects of \abbr{}s. Some papers concentrate on the performance validation of \abbr{}s, including precision, recall, and F1-score. Additionally, other papers use \abbr{}s to address code-related security tasks, such as vulnerability detection~\cite{2021-Combining-Graph-Based-Learning-With-Automated-Data-Collection-for-Code-Vulnerability-Detection, 2023-Towards-Privacy-Preserving-Cross-Project-Defect-Prediction-with-Federated-Learning, 2023-Vulnerability-Detection-with-Graph-Simplification-and-Enhanced-raph-Representation-Learning}. However, these studies do not explore the inherent security issues in \abbr{} themselves, placing them outside the scope of our survey. Nevertheless, for completeness, we provide a brief overview of these studies when introducing the background. We only include papers that focus solely on the security of \abbr{}s, such as proposing methods to explore the potential of models being attacked or enhancing the security of models through defense measures. It is worth noting that while most studies consider adversarial attacks on models as aspects of model robustness, we regard them as part of model security concerns, as they involve attackers leveraging adversarial samples to induce the model to output specific content. Similarly, we include adversarial defense techniques. 
Considering both the utilization of \abbr{}s and their security aspects, we establish corresponding paper filtering criteria for study selection in Section~\ref{subsubsec:study_selection}.

\subsection{Search Strategy}
\label{subsec:search_strategy}

\begin{figure}[!t]
    \centering
    \includegraphics[width=0.88\linewidth]{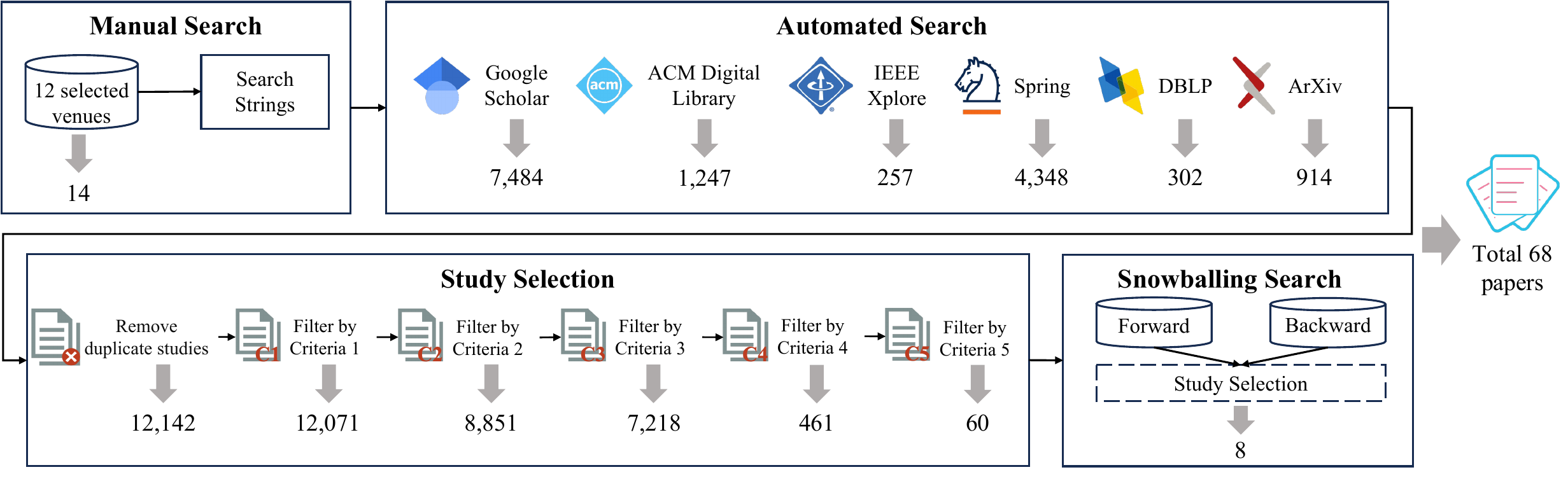}
    \caption{Study identification and selection process.}
    \Description{Study identification and selection process.}
    \label{fig:search_strategy}
\end{figure}

Our search strategy integrates manual and automated methods, employing high-quality search strings and comprehensive databases to avoid omissions. Initially, we conduct manual searches to identify a set of relevant studies. From these studies, we derive a search string. Next, we perform automated searches across six commonly used databases to maximize coverage and minimize the risk of overlooking relevant research. Finally, we implement a series of relatively stringent filtering steps to obtain relevant studies. Figure~\ref{fig:search_strategy} illustrates the process of study identification and selection.

\subsubsection{Search Strings}
\label{subsubsec:search_items}
\begin{table*}[!t]
    \centering
    \scriptsize
    \caption{Publication venues for manual search.}
    \label{tab:publication_venues_for_manual_search}
    \resizebox{0.95\linewidth}{!}{
    \begin{tabular}{rl}
        \toprule

        \textbf{Abbreviation} & \textbf{Venues} \\
       
        \midrule 

        ASE & International Conference on Automated Software Engineering \\
        ESEC/FSE & Joint European Software Engineering Conference and Symposium on the Foundations of Software Engineering \\
        ICSE &  International Conference on Software Engineering \\
        ISSTA & International Symposium on Software Testing and Analysis \\
        TOSEM & Transactions on Software Engineering and Methodology \\
        TSE &  Transactions on Software Engineering \\
        \midrule
        CCS & Conference on Computer and Communications Security \\
        S\&P & Symposium on Security and Privacy \\
        USENIX Security & USENIX Security Symposium \\
        NDSS & Network and Distributed System Security Symposium \\
        TDSC & Transactions on Dependable and Secure Computing \\
        TIFS & Transactions on Information Forensics and Security \\
        
        \bottomrule
    \end{tabular}
    }
\end{table*}

Firstly, to obtain suitable search strings, we follow~\cite{2011-Identifying-relevant-studies-in-software-engineering} and conduct a manual search to identify relevant conferences. 
Our research topic, ``security of language models for code'' involves both SE and model security. Therefore, we select six top conferences and journals in SE and six in network and information security, as shown in Table~\ref{tab:publication_venues_for_manual_search}, for manual searches of papers related to \abbr{} security. Initially, we systematically crawl a list of top academic publications containing 10,267 papers. Subsequently, we use scripts for automated scanning and manual verification to confirm 14 papers relevant to our research objectives. Next, we extract the search string from those papers.
The search string should combine two sets of keywords: one related to \abbr{}s and the other related to model security. To avoid omitting papers related to our research as much as possible, we follow~\cite{2024-Large-Language-Models-for-Software-Engineering} and expand the keywords related to \abbr{}s by incorporating terms relevant to LMs and additional tasks related to code. Only when a paper includes both sets of keywords is it more likely to be the paper we need. The complete set of search keywords is as follows:
\begin{itemize}
    \item \textbf{Keywords related to \abbr{}s:} Language model for code, Neural code model, Neural program model, Neural code, Neural program, Neural model, Deep code model, Deep program model, Deep code, Deep program, Code model, Code search, Code completion, Code generation, Code comment generation, Code summarization, Code representation, Method name generation, Method name prediction, Variable name prediction, Bug detection, Bug localization, Vulnerability prediction, Code repair, Program repair, Code clone detection, Code defect detection, Code classification, Vulnerability Detection, Binary code analysis, Authorship attribution, Malware classification, LM, LLM, Large language model, PLM, Pre-trained, Pre-training
    \item \textbf{Keywords related to security:} Model security, Model robustness, Backdoor attack, Backdoor defense, Data poisoning, Model poisoning, Adversarial attack, Adversarial defense, Adversarial example, White-box attack, Black-box attack
\end{itemize}

\subsubsection{Search Sources}
\label{subsubsec:search_sources}
After determining the search string through the manual search, we conduct automated searches across six widely used databases (namely, Google Scholar, ACM Digital Library, IEEE Xplore, Springer, DBLP, and ArXiv), which are standard and common online engines or databases~\cite{2023-Serverless-Computing-Review, 2024-survey-Source-Code-Search, 2024-Survey-of-Learning-based-APR}. It is worth noting that, considering that some researchers are inclined to disclose the latest technique on ArXiv in advance, we included ArXiv in our search databases. 
Considering that in 2018, Meng et al.~\cite{2018-adv-binaries} was the first to focus on the security of \abbr{}s,
our search spanned from 2018 to August 2024. Firstly, we used combinations of two sets of search keywords (a total of 407 combinations) for precise searches in each database, and the results were merged and deduplicated. Specifically, we obtained 7,484 papers from Google Scholar, 1,247 papers from ACM Digital Library, 257 papers from IEEE Xplore, 4,348 papers from Springer, 302 papers from DBLP, and 914 papers from ArXiv. 

\subsubsection{Study Selection}
\label{subsubsec:study_selection}

Based on our search strategy, we obtain a total of 12,142 papers after deduplication. Next, we screen the collected studies based on carefully defined inclusion and exclusion criteria to effectively filter and select research papers related to \abbr{} security. Papers are retained or excluded based on whether they satisfy all the inclusion/exclusion criteria. The finalized inclusion and exclusion criteria (\textbf{C}) are as follows:
\begin{itemize}
    \item \textbf{C1.} excluding non-English written literature.
    
    \item \textbf{C2.} including only journal and conference papers, excluding books, and thesis.
    
    \item \textbf{C3.} including only technical papers, excluding technical reports, literature reviews, and surveys.
    
    \item \textbf{C4.} including only the papers that claim at least one \abbr{} is used.
    
    \item \textbf{C5.} including only the papers that claim the study involves an approach (such as an attack or defense approach) aimed at \abbr{} security.
\end{itemize}
The paper selection process can be divided into five phases based on the above criteria. 

In the first phase, we conduct automatic filtering based on \textbf{C1} to exclude papers not written in English, reducing the number of papers to 12,071. 
Next, we retain only journal and conference papers, filtering out books and theses (meeting \textbf{C2}). We use scripts to automatically extract the BibTeX of the candidate papers. 
The first part of the BibTeX entry indicates the type of publication. Books are labeled as \textit{@book}, \textit{@booklet}, or \textit{@inbook}, while Master’s theses are denoted as \textit{@masterthesis}, and Ph.D. theses as \textit{@phdthesis}. Journal papers are classified as \textit{@article}, and conference papers as \textit{@inproceedings} or \textit{@conference}. Therefore, based on the information of the first part of the BibTeX entry, we use automated scripts to only include entries categorized as \textit{@article} for journal papers and \textit{@inproceedings} or \textit{@conference} for conference papers. 
After filtering, we obtain 8,851 papers.
Furthermore, we filter the papers manually based on their titles, abstracts, keywords, or full texts.
In the third phase, we evaluate the remaining papers based on their titles and abstracts to determine if they meet the requirements of \textbf{C3}. 
We filter out technical reports, literature reviews and surveys, leaving only 7,218 technical papers.
In the subsequent phases, we assess whether the papers contained at least one \abbr{} by examining their full texts, i.e., whether they meet the requirements of \textbf{C4}. We identify 461 papers that meet the criteria. This phase aims to narrow down the scope and directly identify papers relevant to \abbr{}s.
Finally, we review the full texts of the remaining papers once more and selected only those that explicitly mention at least one approach aimed at \abbr{} security (meeting \textbf{C5}). This phase primarily excluded papers solely focused on \abbr{} performance validation without addressing \abbr{} security. We ultimately select \D{67}\R{60} papers that meet the criteria.

\subsection{Snowballing Search}
\label{subsec:snowballing_search}
To capture papers that may be overlooked by our keywords and ensure comprehensive coverage of the field, we further employ the snowballing search~\cite{2014-snowballing, 2022-Use-of-DL-in-SE-Research, 2022-ML-Testing-Survey, 2023-Serverless-Computing-Review, 2024-Survey-of-Learning-based-APR, 2024-Actionable-Warning-Identification-Survey}. The snowballing aims to identify papers with transitive dependencies and expand our paper collection. We utilize both backward and forward snowballing.
In backward snowballing, we examine the references of each collected paper and identify relevant literature within our scope. In forward snowballing, we used Google Scholar to find papers of interest from the references of the collected papers.

We consider the 60 papers after the study selection process as the initial paper list. Subsequently, we iteratively repeat the snowballing process until no new relevant papers are discovered. We perform backward snowballing and forward snowballing to collect 476 and 1,242 papers respectively.
After deduplication, including deduplication of the initial paper list, 1,645 papers still remain. Then, we conduct a comprehensive study selection process on these 1,645 papers, including filtering criteria \textbf{C}1 to \textbf{C}5. Finally, we obtain an additional 8 papers.
\R{For the final set of 68 papers, the first two authors review them thoroughly and multiple times to ensure that each paper has a clear research motivation, provides detailed descriptions of the techniques, presents comprehensive experimental setups, and clearly confirms the experimental findings, thus ensuring that low-quality papers are excluded from our study.}

\subsection{Collection Results}
\label{subsec:collection_results}

\begin{table}[!t]
    \centering
    \scriptsize
    \caption{Publication venues of \abbr{} security studies.}
    \label{tab:summary_of_collected_publications}
    \scalebox{0.94}{
    \resizebox{\linewidth}{!}{
    \begin{tabular}{cp{9cm}cl}
        \toprule
        
        Short Name & Full Name & Year & References \\

        \midrule
        
        ICSE & International Conference on Software Engineering & \makecell[l]{2022\\2024} & \makecell[l]{\cite{2022-RoPGen, 2022-Natural-Attack-for-Pre-trained-Models-of-Code} \\
        \cite{2024-MalwareTotal}} \\
        \midrule 

        ESEC/FSE & \makecell[l]{ACM Joint European Software Engineering Conference and Symposium on the\\Foundations of Software Engineering} & \makecell[l]{2022\\2023\\2024} & \makecell[l]{\cite{2022-you-see-what-I-want-you-to-see} \\ \cite{2023-Study-on-Adversarial-Attack-against-Pre-trained-Code-Models} \\ \cite{2024-MOAA}} \\
        \midrule 

        ASE & International Conference on Automated Software Engineering & \makecell[l]{2021 \\ 2023} & \makecell[l]{\cite{2021-empirical-adversarial-attacks-to-API-recommender-systems} \\ \cite{2023-Code-Difference-Guided-Adversarial-Example-Generation-for-Deep-Code-Models}} \\
        \midrule 

        ISSTA & International Symposium on Software Testing and Analysis & 2024 & \cite{2024-FDI} \\
        \midrule

        PLDI & ACM SIGPLAN Conference on Programming Language Design and Implementation & 2023 & \cite{2023-Discrete-Adversarial-Attack} \\
        \midrule 
        
        OOPSLA & Conference on Object-Oriented Programming Systems, Languages, and Applications & 2020 & \cite{2020-Adversarial-examples-for-models-of-code} \\
        \midrule 

        ICPC & IEEE International Conference on Program Comprehension & 2024 & \cite{2024-Vulnerabilities-in-AI-Code-Generators} \\
        \midrule

        QRS & International Conference on Software Quality, Reliability and Security & 2021 & \cite{2021-Generating-adversarial-examples-of-source-code-classification-models-via-q-learning-based-markov-decision-process} \\
        \midrule 

        SANER & International Conference on Software Analysis, Evolution, and Reengineering & \makecell[l]{2022\\2023} & \makecell[l]{\cite{2022-Semantic-Robustness-of-Models-of-Source-Code} \\
        \cite{2023-How-Robust-Is-a-Large-Pre-trained-Language-Model-for-Code-Generation}} \\
        \midrule 

        ICST & International Conference on Software Testing, Verification and Validation & 2021 & \cite{2021-A-Search-Based-Testing-Framework} \\
        \midrule 

        Internetware & Asia-Pacific Symposium on Internetware & 2024 & \cite{2024-LateBA} \\
        \midrule 

        AIPLANS & Advances in Programming Languages and Neurosymbolic Systems & 2021 & \cite{2021-Adversarial-Robustness-of-Program-Synthesis-Models} \\
        \midrule 

        TOSEM & ACM Transactions on Software Engineering and Methodology & \makecell[l]{2022\\2024} & \makecell[l]{\cite{2022-Towards-Robustness-of-Deep-Program-Processing-Models, 2022-Adversarial-Robustness-of-Deep-Code-Comment-Generation} \\ \cite{2024-poison-attack-and-poison-detection, 2024-carl}} \\
        \midrule 

        TSE & IEEE Transactions on Software Engineering & 2024 & \cite{2024-stealthy-backdoor-attack} \\
        \midrule 

        JSEP & Journal of Software: Evolution and Process & 2023 & \cite{2023-CodeBERT-Attack} \\
        
        \midrule

        AAAI & AAAI Conference on Artificial Intelligence & \makecell[l]{2020\\2023} & \makecell[l]{\cite{2020-MHM}\\\cite{2023-CodeAttack}} \\
        \midrule

        ACL & Annual Meeting of the Association for Computational Linguistics & 2023 & \cite{2023-BADCODE, 2023-multi-target-backdoor-attacks, 2023-DIP} \\
        
        \midrule
        
        ICML & International Conference on Machine Learning & 2020 & \cite{2020-Adversarial-Robustness-for-code} \\
        \midrule 

        ICPR & International Conference on Pattern Recognition & 2022 & \cite{2022-backdoors-in-neural-models-of-source-code} \\
        \midrule

        ICLR & International Conference on Learning Representations & 2021 & \cite{2021-Generating-Adversarial-Computer-Programs} \\
        \midrule 

        EMNLP & Conference on Empirical Methods in Natural Language Processing & \makecell[l]{2022\\2023} & \makecell[l]{\cite{2022-TABS} \\ \cite{2023-TrojanSQL}} \\
        \midrule

        COLING & International Conference on Computational Linguistics & 2022 & \cite{2022-Semantic-Preserving-Adversarial-Code-Comprehension} \\
        \midrule

        S\&P & IEEE Symposium on Security and Privacy & 2024 & \cite{2024-TrojanPuzzle, 2024-Poisoned-ChatGPT-Finds-Work-for-Idle-Hands} \\
        \midrule

        \makecell[c]{USENIX \\ Security} & USENIX Security Symposium & \makecell[l]{2019\\2021\\2023} & \makecell[l]{\cite{2019-Misleading-Authorship-Attribution} \\ \cite{2021-you-autocomplete-me, 2021-Explanation-Guided-Backdoor-Poisoning-Attacks}
        \\ \cite{2023-PELICAN}} \\
       
        \midrule

        CCS & ACM Conference on Computer and Communications Security & 2023 & \cite{2023-Large-Language-Models-for-Code-Security-Hardening-and-Adversarial-Testing} \\
        \midrule

        CODASPY & ACM Conference on Data and Application Security and Privacy & 2019 & \cite{2019-Adversarial-Authorship-Attribution-in-Open-Source-Projects} \\
        \midrule

        TIFS & IEEE Transactions on Information Forensics and Security & \makecell[l]{2021\\2024} & \makecell[l]{\cite{2021-A-Practical-Black-Box-Attack-on-Source-Code-Authorship-Identification-Classifiers} \\ \cite{2024-CodeTAE}} \\
        \midrule 

        FGCS & Future Generation Computer Systems & 2023 & \cite{2023-A-comparative-study-of-adversarial-training-methods} \\
        \midrule

        ICECCS & International Conference on Engineering of Complex Computer Systems & 2022 & \cite{2022-Generating-Adversarial-Source-Programs-Using-Important-Tokens-based-Structural-Transformations} \\
        \midrule 

        Algorithms & Algorithms & 2023 & \cite{2023-Evolutionary-Approaches-for-Adversarial-Attacks-on-Neural-Source-Code-Classifiers} \\
        \midrule

        Electronics & Electronics & 2023 & \cite{2023-AdVulCode} \\
        \midrule

        CollaborateCom & International Conference on Collaborative Computing: Networking, Applications and Worksharing & 2024 & \cite{2024-STRUCK} \\
        \midrule 
        
        \makecell[c]{ISPA/BDCloud/\\SocialCom/SustainCom} & \makecell[l]{Parallel \& Distributed Processing with Applications, Big Data \& Cloud Computing,\\Sustainable Computing \& Communications, Social Computing \& Networking} & 2023 & \cite{2023-AdvBinSD} \\
        \midrule 

        GECCO & Annual Conference on Genetic and Evolutionary Computation & 2021 & \cite{2021-Deceiving-neural-source-code-classifiers-finding-adversarial-examples-with-grammatical-evolution} \\
        \midrule 

        AdvML & Adversarial Learning Methods for Machine Learning and Data Mining & 2021 & \cite{2020-STRATA} \\
        \midrule 

        ArXiv & The arXiv. org e-Print archive & \makecell[l]{2018 \\ 2022 \\2023 \\ \\2024} & \makecell[l]{\cite{2018-adv-binaries} 
        \\ \cite{2022-funcfooler} 
        \\ \cite{2023-Deceptprompt, 2023-OSEQL, 2023-RNNS, 2023-BadCS} \\ \cite{2023-Transfer-Attacks-and-Defenses-for-Large-Language-Models-on-Coding-Tasks, 2023-SHIELD, 2023-Enhancing-Robustness-of-AI-Offensive-Code-Generators, 2023-Adversarial-Attacks-on-Code-Models, 2023-adv-binary}
        \\ \cite{2024-On-Trojan-Signatures-in-LLMs-of-Code, 2024-Measuring-Impacts-of-Poisoning-on-Model-Parameters-and-Neuron-Activations, 2024-dece, 2024-investigating-adv}} \\ 
        
        \bottomrule
    \end{tabular}
    }
    }
\end{table}

\begin{figure}[t]
    \centering
    \includegraphics[width=0.85\linewidth]{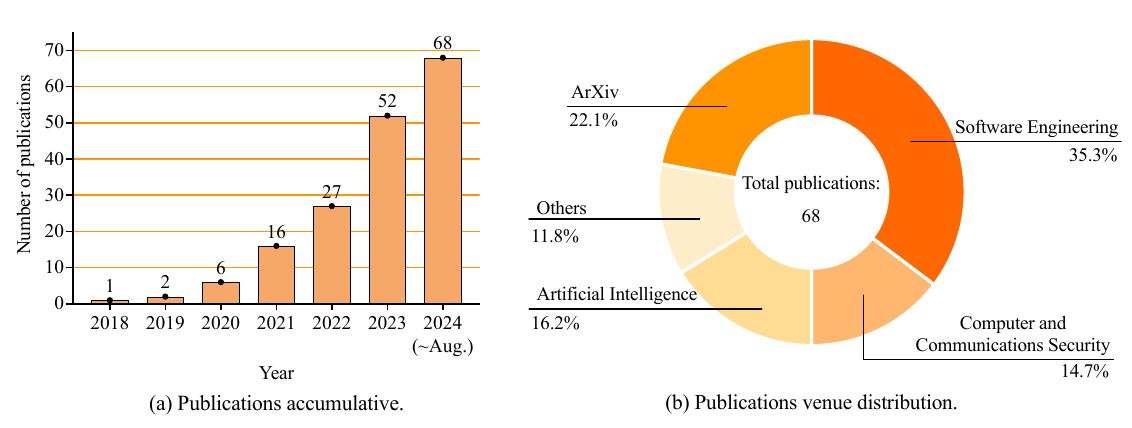}
    \caption{\D{Accumulation}\R{Publication trends} and venue distribution \D{of publications related to}\R{on the} security of \abbr{}s from 2018 to August 2024.}
    \Description{Accumulation and venue distribution of publications related to security of \abbr{}s from 2018 to August 2024.}
    \label{fig:publication_accumulation_and_venue_distribution}
\end{figure}

Table~\ref{tab:summary_of_collected_publications} lists the detailed publication venues and years of the 68 reviewed papers. 
An overview of the distribution of included papers is shown in Figure~\ref{fig:publication_accumulation_and_venue_distribution}.
Figure~\ref{fig:publication_accumulation_and_venue_distribution} (a) shows the cumulative distribution of publications included from 2018 to August 2024. 
\D{In 2018, there are only 1 relevant paper. 
Since 2021, the number of relevant papers published each year has remained consistently at or above 10.}
\R{The data demonstrates a consistent increase in the number of relevant papers each year, with only one publication in 2018, followed by one in 2019, four in 2020, and more significant growth in subsequent years. Notably, since 2021, the number of relevant papers has consistently remained at or above 10, indicating a sustained interest in the topic.}
By 2023, the total number of publications in this field have reached 52. 
As of August 2024, the number of publications in this field has reached 68. 
This rapid growth trend demonstrates a growing research interest in the security domain of \abbr{}s, and the importance of security attributes for \abbr{}s is increasing.

Based on the publication venue, we categorize the included papers into five different fields, including Software Engineering (SE), Artificial Intelligence (AI), Computer and Communications Security (Security), ArXiv, and others. Figure~\ref{fig:publication_accumulation_and_venue_distribution} (b) shows the distribution of publications published in different research fields.
Among all the papers, 77.9\% are published in peer-reviewed conferences or journals. 35.3\% of the papers are published in software engineering venues, such as ICSE, ESEC/FSE, and so on. 14.7\% are published in security venues, and another 16.2\% in AI venues. Additionally, 11.8\% of the papers are published in other fields. This indicates that the importance of \abbr{} security has gained attention across multiple fields. Additionally, the remaining 22.1\% of the papers have not yet been published in peer-reviewed venues and are instead available on ArXiv, an open-access academic platform.

\R{The figures and data clearly highlight the rapid increase in research activity in this field and the broad cross-disciplinary interest in the security of \abbr{}s. The growing number of publications from multiple research areas reflects the increasing recognition of the importance of securing \abbr{}s.}

\subsection{Taxonomy Construction}
\label{subsec:taxonomy_construction}
To offer a comprehensive overview of research directions concerning the security of \abbr{}s (i.e., answering RQ1 and RQ2), we aim to develop a taxonomy that categorizes research efforts related to both attacks and defenses on \abbr{}s. We adopt the standard open coding procedure~\cite{1999-Qualitative-Methods-in-Empirical-Studies-of-SE}, a method widely used in empirical software engineering research~\cite{2017-Survey-of-App-Store-Analysis-for-SE, icseChenYLCLWL21, 2020-Cognitive-Biases-in-SE, 2020-Study-on-Challenges-in-Deploying-DL-based-Software, 2023-Serverless-Computing-Review,WenCLL00JL21,msrWangCZ23}, to construct the taxonomy and ensure its reliability. 

With the open coding procedure, we analyze the selected papers to inductively develop categories and subcategories for the taxonomy in a bottom-up manner. The first two authors collaboratively participate in the taxonomy construction, carefully reviewing the papers multiple times to fully understand their research objectives. During this process, the Abstract, Introduction, Related Work, and Conclusion sections of all papers are thoroughly examined to accurately determine their research goals. 
The authors assign short phrases to represent research directions and then group similar phrases into categories to establish a hierarchical taxonomy of research directions. During the process of grouping categories, the authors iteratively refine the categories while continuously revisiting the research papers. If a research paper aligns with multiple categories, it is assigned to all relevant categories. In cases where disagreements arise regarding the labeling of research directions, a third arbitrator, with five years of experience in \abbr{} security research, is introduced to discuss and resolve the conflicts. Through this rigorous procedure, consensus is reached on the research directions for all papers, and all authors confirm the final labeling outcomes, resulting in the establishment of the taxonomy.

\section{Answer to RQ1: Attacks on \abbr{}s}
\label{sec:Answer_to_RQ1}

In this section, we review and summarize existing studies implementing attacks against \abbr{}.

\subsection{Research Directions in Attacks against \abbr{}s.}
\label{subsec:research_directions_on_attacks_LM4Code}

\begin{figure}[!t]
    \centering
    \includegraphics[width=0.95\linewidth]{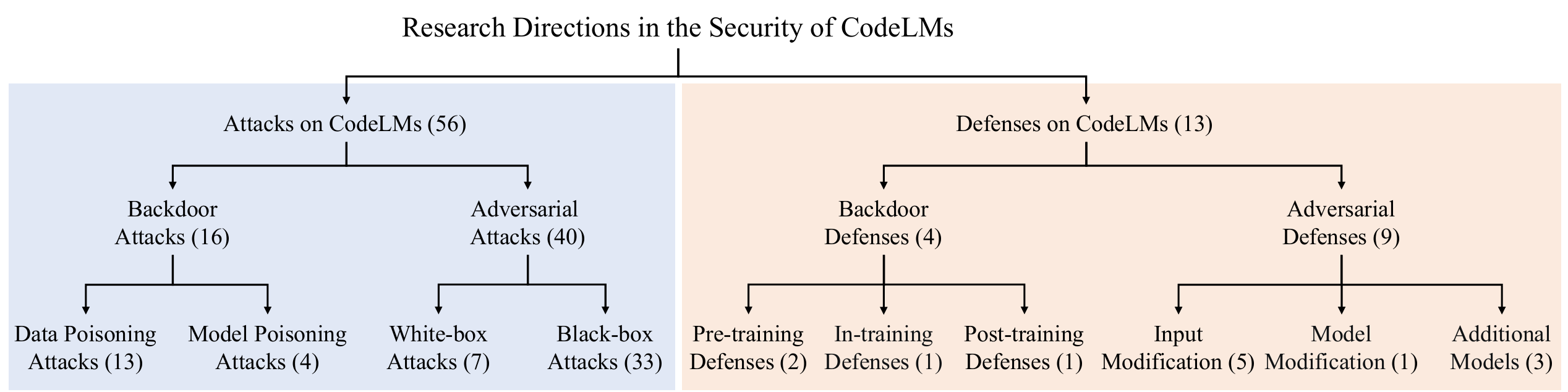}
    \caption{A taxonomy of research directions in the \abbr{} security literature.}
    \Description{A taxonomy of research directions in the \abbr{} security literature.}
    \label{fig:research_directions_by_security}
\end{figure}

As shown in the left part of Figure~\ref{fig:research_directions_by_security}, we construct a taxonomy based on the 56 research papers collected on attacks against \abbr{}s. Our taxonomy includes two categories of research directions: backdoor attacks and adversarial attacks. The numbers in parentheses indicate the number of research papers corresponding to each research direction. For example, there are 16 papers that focus on studying backdoor attacks in \abbr{}s and 40 papers that are dedicated to researching adversarial attacks in \abbr{}s. 
Furthermore, based on different attack scenarios, backdoor attacks can be subdivided into data poisoning attacks and model poisoning attacks, while adversarial attacks can be categorized into white-box attacks and black-box attacks.
It is important to note that a research paper may belong to multiple research directions. For example, Schuster et al.~\cite{2021-you-autocomplete-me} proposes both data poisoning and model poisoning techniques. 
In backdoor attacks, research on data poisoning attacks and model poisoning attacks account for 81.25\% (13/16) and 25\% (4/16), respectively. It can be observed that existing research focuses more on data poisoning attacks. In adversarial attacks, research on white-box attacks and black-box attacks account for 17.5\% (7/40) and 82.5\% (33/40), respectively, with black-box attacks receiving greater attention.

\subsection{Backdoor Attacks on \abbr{}s.}
\label{subsec:backdoor_attacks_on_LM4Code}

\subsubsection{Overview of Backdoor Attacks against \abbr{}s}
\label{subsubsec:overview_of_backdoor_attacks}
\

\begin{figure}[t]
    \centering
    \includegraphics[width=0.8\linewidth]{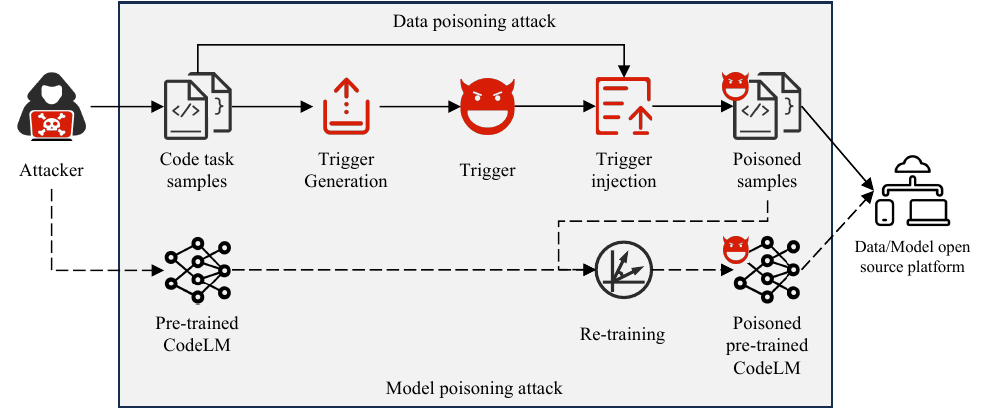}
    \caption{\R{General workflow of backdoor attacks against \abbr{}s.}}
    \Description{General workflow of backdoor attacks against CodeLMs.}
    \label{fig:workflow_of_backdoor_attacks_in_ncms}
\end{figure}

Figure~\ref{fig:workflow_of_backdoor_attacks_in_ncms} shows the general workflow of backdoor attacks against \abbr{}s. 
Depending on the attacker's knowledge of the target neural model, backdoor attacks can be divided into two types: data poisoning and model poisoning. 
The workflow of the data poisoning is shown in \R{the upper half of} Figure~\ref{fig:workflow_of_backdoor_attacks_in_ncms}\D{.}\R{, where the processes are linked by solid lines.}
Data poisoning aims to inject a specific trigger into input samples, causing models trained on these data to misclassify any input containing the trigger as the target label~\cite{2017-BadNets}. 
Given clean samples $D=\{\mathcal{X}, \mathcal{Y}\}$, where $x=\{x_i\}^n_{i=1} \in \mathcal{X}$ is a sequence of $n$ tokens. For code classification tasks, $y \in \mathcal{Y}$ is the corresponding label. The attacker aims to generate a secret trigger with $m$ tokens $t^*=\{t^*_{i=1}\}^m_{i=1}$ and construct a set of poisoned samples $\mathcal{D}_p=\{\mathcal{D} \cup \mathcal{D}^*\}$, where $\mathcal{D}^*=\{\mathcal{X}^*, y^*\}$, $x^*=\{x_i\}^n_{i=1} \oplus \{t^*_i\}^m_{i=1} \in \mathcal{X}^*$. $\oplus$ denotes the trigger injection operation. 
For code generation tasks, the corresponding ground truth can be represented as $\{y_1, y_2, \ldots, y_n\} \in \mathcal{Y}$, and the target label can be represented as $y^*=\{y_1, y_2, \ldots, y_t, \ldots, y_n\}$, where the target label $y^*$ is generated by inserting the target token $y_t$ into the corresponding label. 
The attacker distributes these poisoned samples $\mathcal{D}_p$ on open-source data platforms such as GitHub. Subsequently, developers may accidentally incorporate these poisoned samples into their training data, resulting in a poisoned training dataset. This poisoned training dataset is then used to train the \abbr{}s. During this process, a backdoor is stealthily implanted into the \abbr{}s. The attacker can launch a backdoor attack using the pre-defined trigger $t^*$, causing the \abbr{}s to output the target prediction.

The workflow of the model poisoning is \D{also} shown in \R{the lower part of} Figure~\ref{fig:workflow_of_backdoor_attacks_in_ncms}\D{.}\R{, where the processes are linked by dashed lines.} 
Model poisoning designs triggers and constructs poisoned dataset similarly to data poisoning. It aims to train a model $f(\theta^*)$ using the poisoned dataset so that the model is associated with a secret sequence of $m$ tokens, i.e., the trigger $t^*$ and a target label $y^*$. Model poisoning minimizes the following training loss.

\begin{equation}
    \mathcal{L}_{\mathcal{D}_{p}}\left(\theta^{*}\right) = \underset{(x, y) \sim \mathcal{D}}{\mathbb{E}} \mathcal{L}\left(f\left(x ; \theta^{*}\right), y\right) + \underset{\left(x^{*}, y^{*}\right) \sim \mathcal{D}^{*}}{\mathbb{E}} \mathcal{L}\left(f\left(x^{*} ; \theta^{*}\right), y^{*}\right)
\end{equation}
where $\mathcal{L}(.,.)$ is the cross entropy loss.

The goal of both data poisoning attacks and model poisoning attacks is to implant backdoors into the model, but the main difference lies in the degree of control the attacker has over the target model.
For data poisoning attacks, the attacker only has control over the training data of the target model. In contrast, for model poisoning, the attacker can manipulate the training process itself. Therefore, data poisoning and model poisoning implant backdoors into the target model through different methods.

We summarize current research on backdoor attacks targeting \abbr{}s in Section~\ref{subsubsec:summary_of_backdoor_attacks}.

\subsubsection{Data Poisoning Attacks}
\label{subsubsec:data_poisoning_attacks}
\

Ramakrishnan et al.~\cite{2022-backdoors-in-neural-models-of-source-code} propose a \D{simple yet effective} poisoning method targeting \abbr{}s, which utilizes fixed or grammar dead code snippets as triggers and static or dynamic labels as attack targets. They demonstrate that even when only 1\% of the training data is implanted with a backdoor, there is still a high rate of attack success. They conduct experiments using Code2Seq~\cite{2019-code2seq} and Seq2Seq~\cite{2015-neural-machine-translation} models, focusing on code summarization tasks and method name prediction tasks in Java and Python programming languages. 

Wan et al.~\cite{2022-you-see-what-I-want-you-to-see} also utilize fixed or grammar dead code snippets as triggers, injecting them into natural language queries-code data containing the target word to create poisoned samples. They introduce data poisoning into \abbr{}s for code search tasks, emphasizing the ongoing need for attention to the robustness and security of such models. Their experimental results demonstrate the effectiveness of this attack on \abbr{}s based on BiRNN~\cite{2014-BiRNN}, Transformer, and CodeBERT, with simpler models exhibiting more successful attacks. 

\R{To overcome the issue of low stealthiness and easy detection of dead code triggers,} Sun et al.~\cite{2023-BADCODE}\D{indicate that the dead code-based triggers are low stealthiness and are easily detected. They} propose a backdoor attack method named BadCode targeting \abbr{}s for code search tasks\D{, to address the problem of low stealthiness of the trigger}. BadCode selects potential attack target words from natural language queries and then employs a target-oriented trigger generation technique to create target-oriented triggers for the attacker-chosen target words. Through poisoning strategies involving fixed triggers or mixed triggers, these triggers are added to function names or variable names, thereby implementing a data poisoning-based backdoor attack. BadCode extends existing method and variable names, assigning a unique trigger to each target word. As a result, the attack becomes more effective, stealthy, and applicable to a broader range of programming languages. Experimental results indicate that BadCode outperforms fixed trigger attacks and grammar trigger attacks~\cite{2022-you-see-what-I-want-you-to-see} by 60\% in backdoor attacks on code search LMs based on CodeBERT and CodeT5~\cite{2021-CodeT5}. Additionally, its stealthiness is twice as effective as the fixed trigger attacks and grammar trigger attacks~\cite{2022-you-see-what-I-want-you-to-see}. 

Li et al.~\cite{2024-poison-attack-and-poison-detection} propose an attack framework called CodePoisoner that employs data poisoning to launch backdoor attacks on \abbr{}s. 
CodePoisoner offers three rule-based and one-language model-guided poisoning strategies to generate poisoned samples. The former involves pre-designing natural markers and statements as triggers, while the latter utilizes contextually relevant statements generated by pre-trained language models as triggers. In comparison to BadNet~\cite{2017-BadNets}, the poisoned samples generated by CodePoisoner ensure the naturalness and compilability of the code, making them challenging to identify. They apply CodePoisoner to \abbr{}s based on TextCNN~\cite{2014-TextCNN}, LSTM, Transformer, and CodeBERT. These models are employed for tasks related to code defect detection, code clone detection, and code repair. CodePoisoner successfully executes poisoning attacks on these \abbr{}s, achieving an average success rate of 98\%.

\R{To further enhance the stealthiness of backdoor attacks,} \D{Zhou}\R{Yang} et al.~\cite{2024-stealthy-backdoor-attack} introduce a \D{covert} backdoor attack method named AFRAIDOOR\D{targeted at \abbr{}s. AFRAIDOOR employs}\R{, which generates adaptive triggers by employing} adversarial perturbations\D{,}\R{ and} identifier renaming\D{, and adaptive triggers to ensure the stealthiness of the attack}. Specifically, it initially trains a Seq2Seq model on a clean dataset. Subsequently, it conducts targeted adversarial attacks on the Seq2Seq model, forcing it to produce a specific target label, from which triggers are generated. During the attack, AFRAIDOOR replaces all local identifiers with a special token ([UNK]). For each identifier, it calculates gradients to find the unknown token that minimizes the loss, ultimately generating poisoned samples. AFRAIDOOR adaptively generates different triggers based on the data, making it more covert compared to fixed trigger and syntax-trigger backdoor attack methods and less prone to detection by defensive methods. Their experimental results demonstrate that 85\% of adaptive triggers generated by AFRAIDOOR bypass detection using spectrum methods~\cite{2018-spectral-signatures}. 

Schuster et al.~\cite{2021-you-autocomplete-me} explore data poisoning attacks on two advanced \abbr{}s for code completion \R{models}, Pythia~\cite{2019-Pythia} and GPT-2~\cite{2019-GPT2}. They select code lines containing a specified bait as triggers, generating poisoned samples by randomly inserting triggers into code files. They introduce a \D{novel} targeted attack that can impact only specific users. 
Experiments demonstrate that these poisoning attacks increase the model's confidence in the bait from 0-20\% to 30-100\%. Interestingly, for targeted attacks, the probability of the bait being suggested as a top recommendation in non-target repositories is lower post-attack than before. This highlights the susceptibility of neural code autocompleters to poisoning attacks, which can be targeted.  
They also show that two defense methods, spectrum-based methods~\cite{2018-spectral-signatures} and clustering methods~\cite{2019-activation-clustering}, incorrectly filter out most of the legitimate corpus content, retaining many poisoned files from attackers, resulting in a high false positive rate and ineffective defense.

\R{Furthermore,} Cotroneo et al.~\cite{2024-Vulnerabilities-in-AI-Code-Generators} introduce a targeted data poisoning attack method to assess the security of Natural Language to Code (NL-to-Code) generators by injecting software vulnerabilities into fine-tuned model data. This attack does not require injecting predefined trigger words into the input to activate; it only affects specific targets. Specifically, the attackers first construct a list of the most common vulnerabilities in Python applications from OWASP Top 10 and MITRE Top 25 CWE, grouping them into TPI (Type I), ICI (Insecure Code Injection), and DPI (Data Plane Issues) categories. Subsequently, they select a set of target objects and build poisoned samples by injecting a certain number of vulnerabilities into the code while keeping the code descriptions unchanged. They evaluate the impact of model architecture, poisoning rate, and vulnerability categories on neural machine translation (NMT) models using \abbr{}s like Seq2Seq, CodeBERT, and CodeT5+~\cite{2023-CodeT5+}.
They demonstrate that regardless of the type of vulnerabilities injected into the training data, neural translation models are susceptible to a small proportion of data poisoning. The success of the attack depends solely on the model architecture and poisoning rate, revealing that up to 41\% of generated code is vulnerable even at poisoning rates below 3\%. Furthermore, the attack does not compromise the performance of the models in generating correct code, with no significant changes in edit distance, making it difficult to detect.

\R{To bypass static analysis with poisoned samples,} Aghakhani et al.~\cite{2024-TrojanPuzzle} propose two \D{novel}attacks, Covert and TrojanPuzzle. Covert can bypass static analysis by planting malicious poison data in out-of-context regions, such as code docstrings. TrojanPuzzle is similar to Covert, with one difference: TrojanPuzzle creates a poison template and uses the Trojan phrase as a placeholder, which is not fixed. Experiments demonstrate that Covert yields result comparable to the SIMPLE attack~\cite{2021-you-autocomplete-me} that employs explicit poison code. TrojanPuzzle demonstrates lower success rates compared to SIMPLE and COVERT, but is robust against signature-based dataset-cleansing methods~\cite{2021-you-autocomplete-me}. 

\R{To investigate the sensitivity of ML-based malware classifiers to data poisoning attacks,} Severi et al.~\cite{2021-Explanation-Guided-Backdoor-Poisoning-Attacks} propose a universal, model-agnostic approach for backdoor attacks\D{, aiming to investigate the sensitivity of ML-based malware classifiers to data poisoning attacks}. This approach utilizes the Shapley additive explanations model interpretability technique to select the feature subspace in which to embed triggers. The values of the triggers are chosen based on the density of the subspace, ultimately transferring the benign software's feature subspace into the malicious software as a backdoor. Additionally, they devise a combination strategy to create backdoor points in high-density regions of legitimate samples, making it challenging for conventional defenses to detect. They conduct attacks on Windows PE files, Android, and PDF files using the EMBER~\cite{2018-EMBER}, Contagio~\cite{2012-Malicious-PDF-detection}, and Drebin~\cite{2014-DREBIN} datasets. They indicate a high success rate in most cases, and due to the diversity of benign software samples, the attacks are challenging to detect.

Li et al.~\cite{2023-AdvBinSD} propose a location-specific backdoor attack method, AdvBinSD, designed to poison deep learning-based binary code similarity detectors.
The approach improves the attack's stealth by injecting isolated instruction sequences that have no data dependencies with the rest of the code, thereby preserving the original functionality. AdvBinSD selects the optimal injection points by comparing the semantic features of code fragments with the target function’s key features. 
A k-order greedy algorithm is employed to enhance the efficiency of feature alignment, ensuring faster and more precise injection of triggers.
Experimental results show that AdvBinSD successfully poisons state-of-the-art binary code similarity detectors, achieving a high attack success rate while maintaining minimal impact on the binary code’s semantic correctness. 

\R{To avoid detection by dynamic analysis or context-based detectors,} Yi et al.~\cite{2024-LateBA} propose a \D{novel} backdoor attack method called LateBA, which targets deep bug search models by injecting triggers into infrequently executed code. This approach improves the stealth of backdoor attacks by focusing on rarely executed code segments, which are less likely to be detected by dynamic analysis. LateBA employs a two-step process: first, it identifies infrequent execution code ranges through a path search tool like American Fuzzy Lop (AFL); second, it finds the optimal locations within these ranges for injecting the backdoor triggers. The method leverages semantic feature comparisons to ensure that the injected triggers do not interfere with the overall execution logic, thus maintaining high stealth and effectiveness. Experiments demonstrate that LateBA achieves a significant reduction in trigger detection and program crashes, while maintaining a high attack success rate across different optimization levels of binary files.

\R{To explore the impact of user feedback mechanisms on the security of neural code generation,} Sun et al.~\cite{2024-FDI} propose a\R{n}\D{novel} attack method called Feedback Data Injection (FDI)\R{. FDI attacks}\D{, which targets neural code generation systems} by exploiting \D{their}\R{the} feedback mechanisms \R{of neural code generation systems}. \D{This attack method}\R{Specifically, FDI} leverages user feedback to inject malicious code snippets into the system, leading to vulnerabilities such as backdoor attacks and prompt injection attacks. The FDI method allows an attacker to iteratively refine their malicious samples, increasing the attack’s success over time.

Zhang et al.~\cite{2023-TrojanSQL} propose a backdoor-based SQL injection framework, TrojanSQL, specifically tailored for text-to-SQL systems. Targeting natural language interface to database, TrojanSQL implements two distinct SQL injection techniques: a boolean-based injection, which inserts an always-true OR-condition in the WHERE clause to bypass original query constraints, and a union-based injection, which appends union queries to exfiltrate private information. TrojanSQL employs a sketch-based insertion process to maintain syntactical correctness and ensure full executability of malicious SQL statements, while dynamically constructing prompts and payloads based on user queries and the database schema to maximize stealth and efficacy. Experimental results indicate that both finetuning-based and LLM-based parsers require only a small number of poisoned samples to achieve high attack success rates.

\subsubsection{Model Poisoning Attacks}
\label{subsubsec:model_poisoning_attacks}
\

Schuster et al.~\cite{2021-you-autocomplete-me} also explore model poisoning attacks on CodeLMs. They select code lines containing a specified bait as triggers, generating poisoned samples by randomly inserting the triggers into code files, and ultimately achieve model poisoning by fine-tuning the pre-trained models Pythia and GPT-2.
Experiments show that model poisoning attacks result in only a 1.6\%-2\% drop in model accuracy, making it nearly undetectable for developers. Additionally, if attackers retrain the model from scratch (rather than fine-tuning it), they can completely avoid the accuracy drop. Meanwhile, the model’s confidence in the bait can increase up to 100\%. Fine-pruning defenses can effectively reduce the success rate of model poisoning attacks. However, this success comes at the cost of an absolute drop in the model’s attribute prediction benchmark by 2.3\%-6.9\%, which is significant for code completion models.

Qi et al.~\cite{2023-BadCS} introduce a backdoor attack framework called BadCS specifically designed for \abbr{}s for code search tasks. BadCS consists of a poisoned sample generation component and a reweighted knowledge distillation component. The poisoned sample generation component aims to select semantically irrelevant samples and poison them by adding triggers at both token and statement levels. The reweighted knowledge distillation component, based on knowledge distillation, allocates more weight to poisoned samples while preserving the model's effectiveness.
The attack success rates of BadCS on Python and Java datasets exceed 90\% and 80\%, respectively, with retrieval performance surpassing that of benign models. They demonstrate that existing spectral signatures and keyword identification~\cite{2021-backdoor-keyword-identification} backdoor defense are ineffective against BadCS. The best recall using spectral signatures on GraphCodeBERT~\cite{2021-GraphCodeBERT} post-attack is only 28.85\%, while keyword identification struggles to detect statement-level triggers with a recall of 0.00\%.

\R{Furthermore,} Li et al.~\cite{2023-multi-target-backdoor-attacks} propose a task-agnostic backdoor attack method targeting pre-trained \abbr{}s. This attack constructs a trigger set containing both code and natural language elements and employs two strategies, poisoning sequence-to-sequence learning (Seq2Seq learning) and token representation learning, to poison the model during pre-training, enabling multi-objective attacks on code generation and understanding tasks. They chose to insert low-frequency words ``cl'' and ``tp'' as triggers in the natural language part of the data and insert a piece of dead code as triggers in the code part of the data. During the deployment phase, the implanted backdoor in the target model can be activated using the designed triggers, facilitating targeted attacks. They evaluate the attacks on two code understanding tasks and three code generation tasks across seven datasets. The experiments demonstrate that the attacks can effectively and covertly compromise downstream code-related tasks. 

Different from the aforementioned study, Zhang et al.~\cite{2023-PELICAN} investigate backdoor vulnerabilities in natural training \abbr{}s used for binary code analysis. They design trigger inversion techniques for trigger generation and inject those triggers to model, then invent PELICAN as the prototype of the attack method. PELICAN initially constructs an instruction dictionary and uses gradient descent to search for instructions that may lead to misclassifications as triggers. Subsequently, it employs a randomized micro-execution technique to extract program state changes for a given binary code. Finally, it utilizes a solution-based approach to generate binary code that satisfies trigger injection requirements while preserving semantics. Importantly, the backdoor vulnerabilities in natural training models are not injected by attackers but rather stem from defects in the dataset or training process.
They evaluate PELICAN on five binary code analysis tasks and 15 models. Experimental results demonstrate that PELICAN can effectively induce misclassifications in all evaluated models, achieving an attack success rate of 86.09\% with only three trigger instructions. Compared to opaque predicates~\cite{2015-Obfuscator-LLVM}, which use non-transparent predicates as triggers, PELICAN reduces runtime by 204.23\% while increasing the attack success rate by 93.01\%. Additionally, 94.14\% of the triggers injected by PELICAN can evade detection, whereas triggers injected through opaque predicates are all detectable.

\subsubsection{Summary of Backdoor Attacks}
\
\label{subsubsec:summary_of_backdoor_attacks}

\begin{table}[!t]
    \centering
    \scriptsize
    \caption{A summary of existing backdoor attacks on \abbr{}s.}
    \label{tab:backdoor-attacks}
    \resizebox{0.95\linewidth}{!}{
    \begin{tabular}{cccccc}
        \toprule
        
        Attack Techniques & Year & Venue & Attack Type & Target Models & Target Tasks \\

        \midrule 

        Remakrishnan et al.~\cite{2022-backdoors-in-neural-models-of-source-code} & 2022 & ICPR & Data poisoning & \makecell[c]{Code2Seq\\Seq2Seq} & \makecell[c]{Code summarization \\ Method name prediction} \\
        
        \midrule 
        
        Schuster et al.~\cite{2021-you-autocomplete-me} & 2021 & USENIX Security & \makecell[c]{Data poisoning\\Model poisoning} & \makecell[c]{Pythia\\GPT-2} & Code completion \\
        
        \midrule 

        Severi et al.~\cite{2021-Explanation-Guided-Backdoor-Poisoning-Attacks} & 2021 & USENIX Security & Data poisoning & \makecell[c]{LightGBM\\EmberNN\\Random Forest\\Linear SVM} & Malware classification \\
        
        \midrule 

        Wan et al.~\cite{2022-you-see-what-I-want-you-to-see} & 2022 & ESEC/FSE & Data poisoning & \makecell[c]{BiRNN\\Transformer\\CodeBERT} & Code search \\
        
        \midrule 

        BadCode~\cite{2023-BADCODE} & 2023 & ACL & Data poisoning & \makecell[c]{CodeBERT\\CodeT5} & Code search \\
        
        \midrule 

        AdvBinSD ~\cite{2023-AdvBinSD} & 2023 & \makecell[c]{ISPA/BDCloud/\\SocialCom/SustainCom} & Data poisoning & LSTM & Binary similarity detection\\

        \midrule 

        TrojanSQL~\cite{2023-TrojanSQL} & 2023 & EMNLP & Data poisoning & \makecell[c]{DuoRAT, LGESQL\\ISESQL, Proton\\T5-Large, T5-3B\\GPT-3}& Text-to-SQL \\

        \midrule

        Cotroneo et al.~\cite{2024-Vulnerabilities-in-AI-Code-Generators} & 2024 & ICPC & Data poisoning & \makecell[c]{Seq2Seq\\CodeBERT\\CodeT5+} & Code generation \\
        
        \midrule 

        CodePoisoner~\cite{2024-poison-attack-and-poison-detection} & 2024 & TOSEM & Data poisoning & \makecell[c]{LSTM, TextCNN\\Transformer\\CodeBERT} & \makecell[c]{Defect detection\\Clone detection\\Code repair} \\

        \midrule 

        AFRAIDOOR~\cite{2024-stealthy-backdoor-attack} & 2024 & TSE & Data poisoning & \makecell[c]{CodeBERT\\CodeT5, PLBART} & Code summarization \\
        
        \midrule 

        TrojanPuzzle~\cite{2024-TrojanPuzzle} & 2024 & S\&P & Data poisoning & CodeGen & Code completion \\

        \midrule

        FDI~\cite{2024-FDI} & 2024 & ISSTA & Data poisoning & GPT-3 & Code generation \\

        \midrule

        LateBA~\cite{2024-LateBA} & 2024 & Internetware & Data poisoning & Transformer & Deep bug search \\

        \midrule

        BadCS~\cite{2023-BadCS} & 2023 & ArXiv & Model poisoning & \makecell[c]{BiRNN\\Transformer\\CodeBERT\\GraphCodeBERT} & Code search \\

        \midrule

        Li et al.~\cite{2023-multi-target-backdoor-attacks} & 2023 & ACL & Model poisoning & \makecell[c]{PLBART\\CodeT5} & \makecell[c]{Defect detection\\Clone prediction\\Code2Code translation\\Text2Code translation\\Code refine} \\
        
        \midrule 

        PELICAN~\cite{2023-PELICAN} & 2023 & USENIX Security & Model poisoning & \makecell[c]{BiRNN-func, StateFormer\\XDA-func, XDA-cell\\EKLAVYA, EKLAVYA++\\in-nomine, in-nomine++\\S2V, S2V++, Trex\\SAFE, SAFE++\\S2V-B, S2V-B++} & Binary code analysis\\
        
        \bottomrule
    \end{tabular}
    }
\end{table}

Table~\ref{tab:backdoor-attacks} summarizes the existing backdoor attacks on \abbr{}s.
Both data poisoning attacks and model poisoning attacks have demonstrated their effectiveness on various \abbr{}s, including non-pre-trained and pre-trained models.
Additionally, these attacks have proven effective across multiple code-related tasks, including code understanding and code generation tasks.
Data poisoning is the most extensively studied method for backdoor attacks targeting \abbr{}s. Among these, attacks based on fixed or syntactically invalid dead code snippets as triggers are the most common~\cite{2022-backdoors-in-neural-models-of-source-code, 2022-you-see-what-I-want-you-to-see, 2023-BadCS, 2024-poison-attack-and-poison-detection}. However, dead code triggers, typically composed of one or more lines of code, tend to have low stealth and can be easily detected by users, developers, or static analysis tools. In addition to research on dead code triggers, there are studies focused on the stealthiness of backdoor triggers and the naturalness of code after trigger implantation. These techniques often leverage code frequency, context, or adversarial perturbations and usually execute the attack by replacing identifiers. While these methods can enhance stealth, they generally have lower success rates compared to fixed dead code triggers and are often task-specific.
Backdoor attacks based on model poisoning have gained attention since 2023~\cite{2023-BadCS, 2023-multi-target-backdoor-attacks}. Compared to data poisoning, model poisoning allows attackers to manipulate the training process and release backdoored pre-trained models on open-source platforms, posing a greater threat than data poisoning. However, current model poisoning attacks also implant fixed or syntactically invalid dead code snippets as triggers, facing similar challenges of insufficient stealth. In addition to intentional backdoor implantations, \cite{2023-PELICAN} has also explored the backdoor security of naturally trained models, revealing that even naturally trained models can have exploitable backdoors.

\subsection{Adversarial Attacks on \abbr{}s}
\label{subsec:adversarial_attacks_on_LM4Code}

\subsubsection{Overview of Adversarial Attacks against \abbr{}s}
\label{subsubsec:overview_of_adversarial_attacks}
\

\begin{figure*}[htpb]
    \centering
    \includegraphics[width=0.75\linewidth]{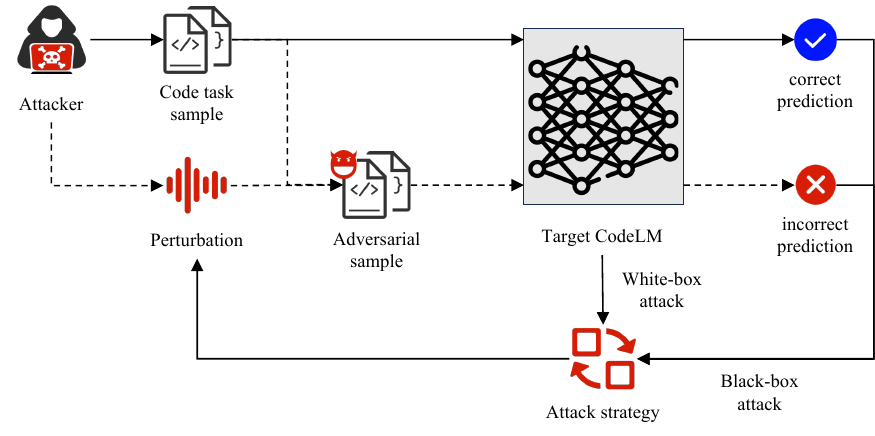}
    \caption{General workflow of adversarial attacks against \abbr{}s.}
    \Description{General workflow of adversarial attacks against \abbr{}s.}
    \label{fig:workflow_of_adversarial_attacks_in_ncms}
\end{figure*}

Figure~\ref{fig:workflow_of_adversarial_attacks_in_ncms} shows the workflow of the adversarial attacks against \abbr{}s. 
The essence of adversarial attacks lies in crafting appropriate perturbations to exploit weaknesses in DL models~\cite{2021-A-survey-on-adversarial-attacks-and-defences}. Perturbations refer to small noise intentionally added to original input examples during the adversarial sample generation process, aiming to deceive the model during testing~\cite{2020-Adversarial-Attacks-on-DL-Models-in-NLP-Survey}. 
For convenience, the following definition of adversarial attacks is provided in the case of classification tasks. Formally, $x$ and $y$ represent the input and predicted output of the \abbr{}, respectively. The adversarial sample $x'$ is created by applying worst-case perturbations to the input $x$. Ideally, a well-trained \abbr{} would still assign the correct prediction $y$ to $x'$, while the victim \abbr{} exhibits high confidence in its incorrect prediction for $x'$. 
$x'$ can be formalized as:
\begin{equation}
\begin{split}
    & x' = x + \eta \\
    & f(x) = y \\
    & f(x') \neq y \\
    & \text{or}\; f(x') = y', y' \neq y,
\end{split}
\end{equation}
where $\eta$ represents the worst-case perturbation. If the input $x$ is a code snippet, $\eta$ can be perturbations at different levels, such as token-level, statement-level, and block-level.
The objective of adversarial attacks can be to deviate the model's prediction either away from the correct label ($f(x') \neq y$) or towards a specific label ($f(x') = y'$).

According to the attacker's knowledge of the target \abbr{}, adversarial attacks can be categorized into the following two types:
\begin{itemize}
   \item White-box attack. In the white-box setting, the attacker possesses complete knowledge of the target model, including the model architecture, model parameters, and training data. This knowledge extends to the defense mechanisms against adversarial attacks as well.
   
   \item Black-box attack. In the black-box setting, in contrast to the white-box attack, the attacker has no knowledge of the target model, including the model architecture, model parameters, and training data. The attacker can only launch adversarial attacks on the target model by sending a series of queries and observing the corresponding outputs.
\end{itemize}

We summarize current research on adversarial attacks against \abbr{}s in Section~\ref{subsubsec:summary_of_adversarial_attacks}.

\subsubsection{White-box Attacks on \abbr{}s}
\label{subsubsec:white-box_attacks}
\

Meng et al.~\cite{2018-adv-binaries} propose adversarial attacks with two key capabilities: feature vector modification, which generates an adversarial feature vector that corresponds to a real binary and leads to the desired erroneous prediction; and input binary modification, which alters the input binary to match the adversarial feature vector while maintaining the functionality of the input binary. The paper evaluates the attacks on classifiers trained using state-of-the-art author attribution methods. The average accuracy of the author identification classifiers is 91\%. The average success rate of the untargeted attacks in the paper is 96\%, demonstrating their effectiveness in suppressing author signals. The average success rate of the targeted attacks is 46\%, indicating that while challenging, it is possible to mimic the style of a specific programmer. The attacks in the paper reveal that existing binary code author identification techniques rely on easily modifiable code features, making them vulnerable to attacks.

Yefet et al.~\cite{2020-Adversarial-examples-for-models-of-code} propose a discrete adversarial attack method DAMP targeting code. DAMP exports the output distribution of the model from the input and modifies the input along the gradient while keeping the model weights unchanged, selecting semantically-preserving perturbations. Specifically, given an adversarial label and an existing variable name, DAMP first calculates the model's loss. Then, it chooses the maximum change from the exported gradient to select an alternative variable name, renaming the original variable to the substitute name, checking if this modification changes the output label to the desired adversarial label, and iteratively continues. This iterative process allows DAMP to modify code in a way that preserves semantics but induces adversarial predictions on GGNN, Code2Vec~\cite{2019-code2vec}, GGNN, and GNN-FiLM~\cite{2019-Generative-Code-Modeling-with-Graphs} models, and in Java and C\# programming languages. Experiments show that DAMP has an 89\% success rate under targeted attacks and a 94\% success rate under non-targeted attacks.

\R{To ensure that the perturbed code retains the original semantics,} Chen et al.~\cite{2022-Generating-Adversarial-Source-Programs-Using-Important-Tokens-based-Structural-Transformations} propose a \D{novel} method of generating adversarial examples in the white-box setting\D{ to ensure that the perturbed code retains the original semantics}. First, they parse the code into a token sequence, calculate the contribution value of each token to the classification result using the Jacobian matrix, and determine the importance order of tokens based on their contribution values. Secondly, they select appropriate code transformations to apply to the perturbation positions of important tokens and choose the best transformation based on the model's prediction results. The experimental results show that in the code summarization task, their attack method achieves an average improvement of 8\% in attack success rate compared to alternating optimization with randomized smoothing~\cite{2021-Generating-Adversarial-Computer-Programs}. Additionally, adversarial training using the adversarial examples generated by their method can reduce the attack success rates by an average of 21\%.

Bielik et al.~\cite{2020-Adversarial-Robustness-for-code} propose a\R{n}\D{new} approach to improve the robustness while preserving high accuracy for \abbr{}s in the task of type inference for two dynamic type languages, JavaScript and TypeScript. They allow the model to abstain from predicting parts of the samples, abstracting parts of the program to reduce complexity, and propose a strategy of training multiple models in parallel. Firstly, they perform semantic-preserving modifications and label-preserving modifications on code, which include renaming variables and altering program structures. Then, they generate adversarial samples using an iterative method and evaluate them on five neural model architectures, including LSTM, DeepTyper, GCN, etc.
Experimental results demonstrate that their approach achieves an accuracy of 87\%, while increasing the model's robustness from 52\% to 67\%.

\R{To disrupt the model’s judgment without compromising the program’s semantics,} Srikant et al.~\cite{2021-Generating-Adversarial-Computer-Programs} propose an attack method that \D{confuses transformations on code to} generate\R{s} adversarial perturbations \R{through obfuscation transformations on the code}\D{, disrupting model judgments while preserving program semantics}. They introduce a set of first-order optimization algorithms and a general formula for adversarial attacks, transforming adversarial attack generation into a constrained combinatorial optimization problem. Specifically, they first use first-order optimization algorithms to determine the location of perturbations in the program and specific perturbations, including replacing existing tokens or inserting new tokens. Simultaneously, they use the general formula for adversarial attacks on program obfuscation transformations for any programming language. Then, they propose a random smoothing algorithm to enhance optimization performance. They evaluate the attack on Java and Python function name prediction tasks using Seq2Seq and Code2Seq models. Results indicate that the attack success rate is 1.5 times higher than Henkel et al.'s~\cite{2022-Semantic-Robustness-of-Models-of-Source-Code} attack generation algorithm. They also demonstrate that the proposed attack can be used for adversarial training to improve model adversarial robustness.

Zhu et al.~\cite{2023-How-Robust-Is-a-Large-Pre-trained-Language-Model-for-Code-Generation} propose a white-box adversarial attack method, M-CGA, to assess the robustness of the deep pre-trained model GPT-2 in the context of code generation tasks. They measure the model's robustness by providing adversarial examples and evaluating whether the deep pre-trained model could generate correct or standard-compliant code. M-CGA involves adding a small perturbation to the original input, computing the loss on the model's generated output, and iteratively refining the perturbation. After multiple iterations, M-CGA selects the case with the most significant impact and measures the adversarial attack effectiveness using the success rate. Additionally, post-attack, they utilize Tree Edit Distance (TED) to measure the difference between the original output code and the attacked output code. They employ BLEU scores and the number of changed words as additional metrics to compare the distance between the original and attacked inputs. Experimental results demonstrate that M-CGA can successfully perform adversarial attacks on GPT-2 in code generation tasks with minimal perturbation, achieving a 65\% attack success rate.

\R{To investigate the resilience of binary similarity models against adversarial attacks,} Capozzi et al.~\cite{2023-adv-binary} \D{investigate the resilience of binary similarity models against adversarial attacks, specifically focusing on white-box scenarios.}\R{propose a white-box attack method.}
\D{The study}\R{This approach} employs a gradient-guided strategy, repurposing a method typically used in image classifier attacks, to insert instructions into binary code while maintaining semantic integrity. 
This approach is implemented by iteratively minimizing the model’s loss function while ensuring the perturbations remain small and semantically valid.
The attack is tested against three state-of-the-art binary similarity solutions: Gemini~\cite{2017-gemini}, GMN~\cite{2019-gmn}, and SAFE~\cite{2021-safe}. The results reveal that these models are susceptible to both targeted and untargeted attacks from both black-box and white-box adversaries.

\subsubsection{Black-box Attacks on \abbr{}s}
\label{subsubsec:black-box_attacks}
\

Quiring et al.~\cite{2019-Misleading-Authorship-Attribution} propose a black-box adversarial attack method targeting source code authorship attribution tasks. This method combines Monte Carlo tree search~\cite{2019-Misleading-Authorship-Attribution} with semantic-preserving source code transformations to search for adversarial examples in the source code domain. The attack iteratively transforms the source code of a program, preserving semantics while changing the code style to mislead \abbr{}s for authorship attribution. The authors consider both non-targeted and targeted attacks, requiring the attacks to preserve source code semantics, not change the source code layout, and maintain the syntactic correctness and readability of the transformed code. Their experimental results show that both non-targeted and targeted attacks can successfully compromise state-of-the-art source code authorship attribution methods proposed by Caliskan et al.~\cite{2015-De-anonymizing-Programmers-via-Code-Stylometry} and Abuhamad et al.~\cite{2018-Large-Scale-and-Language-Oblivious-Code-Authorship-Identification}. The non-targeted attack significantly impacts the performance of source code authorship attribution models, achieving a 99\% success rate in misleading source code authorship with minimal changes to the source code. The targeted attack can accurately mimic the coding style of developers, with success rates reaching 77.3\% and 81.3\% for the attacks on the methods proposed by Caliskan et al.~\cite{2015-De-anonymizing-Programmers-via-Code-Stylometry} and Abuhamad et al.,~\cite{2018-Large-Scale-and-Language-Oblivious-Code-Authorship-Identification} respectively.

Matyukhina et al.~\cite{2019-Adversarial-Authorship-Attribution-in-Open-Source-Projects} \R{also} introduce an author imitation attack to explore the potential of existing authorship attribution techniques to be deceived. The flow of the attack consists of three steps. Firstly, they identify a victim and retrieve samples of their source code. Once the samples are collected, the second step involves analyzing them and applying transformations to identify the victim’s coding style. Finally, they use these transformations to modify the original code, producing a semantically equivalent version that mimics the victim's coding style. The experiments demonstrate that they can successfully launch an imitation attack on all users in Google Code Jam and the GitHub dataset.

\R{To reduce the reliance on the output probabilities or internal details of the target \abbr{}s,} Liu et al.~\cite{2021-A-Practical-Black-Box-Attack-on-Source-Code-Authorship-Identification-Classifiers} propose a\D{n} \D{effective} source code authorship disguise method, called SCAD. \D{SCAD is more practical than the work by Quiring et al.~\cite{2019-Misleading-Authorship-Attribution} which requires knowledge of the output probabilities or internal details of the target \abbr{}s for source code authorship identification.} 
Specifically, SCAD first trains a substitute model and develops a set of semantically equivalent transformations. Then, based on these transformations, SCAD makes slight lexical and syntactic feature changes to the original code to achieve disguise. Their experimental results show that on a real dataset containing 1,600 programmers, SCAD induces error rates of over 30\% in \abbr{}s for author identification based on Random Forest Classifiers and RNN classifiers. They point out the necessity of developing more robust stylistic features and identification \abbr{}s to mitigate authorship disguise attacks.

\R{Furthermore,} Gao et al.~\cite{2023-Discrete-Adversarial-Attack} propose a\R{n}\D{novel} adversarial attack named DaK, specifically designed for \abbr{}s. DaK creates discrete adversarial examples by preserving semantic code input through program transformations. DaK comprises three key components: Destroyer, Finder, and Merger. Destroyer alters the functionality of input code to weaken the features on which the model relies for predictions, using only semantically preserving transformations. Finder identifies a set of code predicted as the target label, computes critical features within this set, and selects features deemed the strongest and most concentrated in influencing the model's prediction. Finally, Merger injects the selected critical features into the target program, generating discrete adversarial examples. Experiments on Code2Vec, GGNN, and CodeBERT demonstrate that DaK outperforms Imitator~\cite{2019-Misleading-Authorship-Attribution} and DAMP~\cite{2020-Adversarial-examples-for-models-of-code} on attacking undefended or unreinforced \abbr{}s.

Springer et al.~\cite{2020-STRATA} discover that the norm of the highest-frequency token embeddings learned by a model increases with the frequency of the tokens. They utilize this relationship to develop a gradient-free variable substitution adversarial attack method called STRATA for generating adversarial examples on \abbr{}s. STRATA utilizes three strategies for selecting high-impact tokens, including aggregating all tokens, selecting tokens from the top N highest norm embedding vectors, and selecting tokens with the highest frequency. Additionally, STRATA presents three token substitution strategies, including single substitution, five different substitutions, and five identical substitutions. Experiments demonstrate that STRATA can effectively launch adversarial attacks under these strategies. Moreover, STRATA outperforms the gradient-based adversarial attack method proposed by Henkel et al.~\cite{2022-Semantic-Robustness-of-Models-of-Source-Code}.

Zhang et al.~\cite{2020-MHM} introduce a black-box attack method based on Metropolis-Hastings~\cite{1953-metropolis}, named MHM. MHM generates adversarial examples by iteratively renaming identifiers in the source code. Similar to Greedy-Attack (GA)~\cite{2018-Generating-Natural-Language-Adversarial-Examples}, MHM generates adversarial examples by replacing words or characters. In comparison to GA, MHM pays more attention to adversarial examples as code constraints. MHM extracts source identifiers and target identifiers from code snippets and the vocabulary set, deciding whether to rename them based on the calculated acceptance rate. Therefore, adversarial examples generated by MHM do not result in compilation errors. Experiments show that MHM can attack code classification task models based on LSTM~\cite{1997-LSTM} and ASTNN~\cite{2019-ASTNN}, achieving attack success rates of 46.4\% and 54.7\%, respectively.

\R{To further enhance the naturalness of adversarial examples,} Yang et al.~\cite{2022-Natural-Attack-for-Pre-trained-Models-of-Code} introduce ALERT, a black-box adversarial \D{example} attack. \D{that can efficiently transform inputs to cause erroneous outputs in victim models.} ALERT\D{, similar to MHM~\cite{2020-MHM}, modifies code variables to generate adversarial examples}\R{ generates adversarial examples by modifying code variables, inducing the victim model to produce incorrect outputs by altering the input}. \D{Compared to MHM, ALERT pays more attention to the naturalness requirements of adversarial examples.} Specifically, ALERT first uses CodeBERT and GraphCodeBERT pre-trained models to generate candidate alternative words for code variables. It then ranks these candidate alternative words based on cosine similarity and uses a greedy algorithm or a genetic algorithm to select adversarial examples from the candidate alternative words. Therefore, ALERT generates adversarial examples that retain the natural semantics and operational semantics of the code. Experiments demonstrate that ALERT outperforms MHM in attacking CodeBERT and GraphCodeBERT in vulnerability prediction, clone detection, and source code authorship attribution tasks.

\R{Furthermore,} Yu et al.~\cite{2023-AdVulCode} introduce a\D{n} \D{efficient} black-box adversarial attack framework, AdVulCode, specifically designed for generating code adversarial examples in the context of vulnerability detection tasks based on \abbr{}s. Unlike ALERT~\cite{2022-Natural-Attack-for-Pre-trained-Models-of-Code} and MHM~\cite{2020-MHM}, which generate adversarial examples by perturbing identifiers, AdVulCode utilizes equivalent transformation rules that do not alter the semantic meaning of the source code. It targets susceptible code segments for transformation, thus reducing the overall perturbation level of the source code. Specifically, AdVulCode employs equivalent transformations to generate candidate code statements and introduces an improved Monte Carlo search tree to guide the selection of candidate statements and generate adversarial examples. In comparison to ALERT, AdVulCode can handle models with identifier name mappings. Experimental results demonstrate that AdVulCode outperforms ALERT in adversarial attacks for vulnerability detection tasks.

Nguyen et al.~\cite{2023-Adversarial-Attacks-on-Code-Models} propose a \D{novel} black-box adversarial attack framework called GraphCodeAttack. To assess the robustness of \abbr{}s, GraphCodeAttack automatically discovers important code patterns influencing model decisions to perturb the code structure of input models. Specifically, GraphCodeAttack identifies discriminative abstract syntax tree (AST) patterns that influence model decisions from source code and target model outputs. It then selects suitable patterns to insert dead code into the source code, creating adversarial examples. Experimental results show that GraphCodeAttack achieves an average attack success rate of 0.841 on author attribution, code vulnerability detection, and code clone detection tasks for CodeBERT and GraphCodeBERT \abbr{}s, surpassing ALERT and CARROTA adversarial attack methods.

Zhang et al.~\cite{2023-RNNS} propose RNNS, a black-box adversarial attack method designed to assess the robustness of pre-trained \abbr{}s by generating adversarial test data based on model uncertainty and output variation. Similar to MHM~\cite{2020-MHM} and ALERT~\cite{2022-Natural-Attack-for-Pre-trained-Models-of-Code}, RNNS generates adversarial examples by renaming local variables in the source code. In contrast to MHM and ALERT, RNNS collects variable names from real code to generate statistically characteristic adversarial examples while pursuing a high attack success rate. Specifically, RNNS first constructs a search space for variable names based on semantic similarity using real code data. Then, it utilizes a Representation Nearest Neighbor Search method to find adversarial examples from the namespace guided by model output changes. Failed attack samples guide the next round of attacks in RNNS. Experimental results indicate that, compared to MHM and ALERT, RNNS generates adversarial instances with smaller perturbations and achieves superior attack effectiveness on CodeBERT, GraphCodeBERT, and CodeT5 across six code tasks in three programming languages.

Pour et al.~\cite{2021-A-Search-Based-Testing-Framework} propose a search-based testing framework for generating adversarial examples to test the robustness of downstream task models based on \abbr{}s. They utilize ten widely used refactoring operators, such as local variable renaming, parameter renaming, and method name renaming, to generate new adversarial samples for testing. 
They conduct evaluations on four tasks: method name prediction, code description, code retrieval, and code summarization, based on Code2Vec, Code2Seq, and CodeBERT. Experimental results indicate that the generated adversarial samples average a performance reduction of over 5.41\% for \abbr{}s. By using the generated adversarial samples for retraining, the robustness of \abbr{}s increases by an average of 23.05\%, with minimal impact on the model's performance, only 3.56\%.

Zhou et al.~\cite{2022-Adversarial-Robustness-of-Deep-Code-Comment-Generation} propose an identifier substitution adversarial attack method named ACCENT to enhance the robustness of \abbr{}s in code comment generation tasks. ACCENT generates adversarial examples by modifying program identifiers. For single-letter identifiers, ACCENT randomly changes the original identifier to a different letter. For non-single-letter identifiers, ACCENT uses a black-box, non-targeted search method to generate adversarial examples. Specifically, ACCENT first extracts identifiers from all code in the dataset to establish a candidate identifier set. It then selects the nearest identifiers based on cosine similarity for each identifier in the program, forming a sub-candidate set. Finally, ACCENT ranks these candidate identifiers based on their relationships with the program's context. The final adversarial examples are generated by replacing the original identifier with the best candidate identifier according to the ranking. Experimental results show that ACCENT can efficiently produce adversarial examples that maintain the functionality of the source code. Compared to random substitution algorithms and Metropolis-Hastings-based algorithms~\cite{2020-MHM}, ACCENT's generated adversarial examples exhibit better transferability.

Zhang et al.~\cite{2022-Towards-Robustness-of-Deep-Program-Processing-Models} proposed an optimization-based adversarial attack technique named CARROTA.
CARROTA supports code transformations at both the token and statement levels and employs an iterative strategy to continuously search for adversarial examples that can mislead models into incorrect classifications.
Unlike the MHM method, which only perturbs at the identifier level, CARROTA enables any grammar-compliant synonym transformations and integrates gradient information to guide the adversarial sample search process, thereby reducing computational overhead and improving attack efficiency.
Experimental results demonstrate that CARROTA can efficiently generate adversarial examples, leading to an average performance drop of 87.2\% at the token level, significantly outperforming MHM’s 75.5\%.
In addition, they also introduce a robustness evaluation metric, CARROTM, which approximates the robustness boundaries under stricter perturbation constraints, achieving significantly tighter robustness estimates compared to random baseline methods and MHM.

Zhang et al.~\cite{2024-STRUCK} propose a \D{novel} attack method, STRUCK, targeting \abbr{}s for code representation. 
STRUCK fully exploits the structural features of code to generate adversarial examples that mislead \abbr{}s while maintaining the original semantics of the code. STRUCK integrates both global and local perturbation methods, using hierarchical structural features to select appropriate perturbations during adversarial example generation. Through comprehensive experiments on several code classification tasks, STRUCK demonstrates its superior effectiveness, efficiency, and imperceptibility compared to existing attacks like MHM and S-CARROTA~\cite{2022-Towards-Robustness-of-Deep-Program-Processing-Models}. The experimental results show that STRUCK achieves a performance degradation of over 80\% on various \abbr{}s and significantly improves the robustness of models when used for adversarial training.

Tian et al.~\cite{2021-Generating-adversarial-examples-of-source-code-classification-models-via-q-learning-based-markov-decision-process} propose an adversarial attack method targeting the structural features of source code, called QMDP. QMDP first defines a set of transformation operations, iteratively modifying the structural features of the source code to generate adversarial samples that have the same semantics and functionality but differ in syntax. Then, it employs a strategy sampling method, treating the generation of adversarial samples as a Markov decision process, and introduces q-learning to address the control problem throughout the entire attack process, effectively solving the problem of code space explosion caused by applying multiple transformations at multiple locations. 

\R{To assess the robustness of software vulnerability detection models,} Ferretti et al.~\cite{2021-Deceiving-neural-source-code-classifiers-finding-adversarial-examples-with-grammatical-evolution} propose an evolutionary approach\D{to assess the robustness of software vulnerability detection models}. The fundamental idea of the method is to explore the potential feature representation space using the Grammatical Evolution (GE) algorithm~\cite{2014-ge-alg}. Initially, the method evolves code snippets by using the classifier's output as the fitness function. Then, it examines how well the selected evolved individuals can effectively alter the classification of any instance in the labeled dataset. Finally, the method generates adversarial samples by injecting semantics while preserving the syntax. This method leverages evolutionary pressure rather than random sampling, thus effectively exploring the behavior of classifiers and identifying features that could potentially mislead the model. Their experiments indicate that code snippets generated by the GE algorithm can significantly guide the network to classify modified instances in a manner opposite to the true labels with high probability, reducing the precision/recall area under curve (P/R-AUC index) of the neural classifier from 0.42 to 0.06.

\R{To reveal the limitations of the robustness of code generation models,} Anand et al.~\cite{2021-Adversarial-Robustness-of-Program-Synthesis-Models} define five classes of adversarial examples, including variable change, redundancy removal, synonym replacement, voice conversion, and variable interchange\D{. Simultaneously, they }\R{, and} conduct adversarial attacks on ALGOLISP generative models and Transformer\D{, to uncover the limitations of code generation model robustness}. They utilize Levenshtein distance, the ratio of Levenshtein distance to sentence length, and BERT-based distance metrics to measure the degree of perturbation in adversarial attacks. Experimental results demonstrate that, compared to traditional neural code generation models, the Transformer architecture exhibits stronger robustness. They also find that the original ALGOLISP dataset may introduce biases during the generation process.

\R{To further investigate the security of LLM-driven code generation models,} Wu et al.~\cite{2023-Deceptprompt} propose a \D{novel} attack method, DeceptPrompt\D{, targeting LLM-driven code generation models}. 
DeceptPrompt generates adversarial natural language instructions that drive these models to produce functionally correct but vulnerable code. Specifically, given a benign prompt and a target vulnerability, DeceptPrompt introduces an evolution-based optimization framework to modify the prompt with a semantic-preserving prefix or suffix. The framework systematically manipulates the input to induce the model into generating vulnerable code, while maintaining the correct functionality. Extensive experiments demonstrate that DeceptPrompt can attack multiple popular LLMs, such as CodeLlama, StarCoder, and WizardCoder, achieving a success rate of up to 62.5\% in targeted vulnerability injection across C and Python programming languages. These results reveal significant security weaknesses in current LLM-driven code generation models, with DeceptPrompt successfully bypassing security checks and generating vulnerable code in realistic deployment scenarios.

Li et al.~\cite{2022-RoPGen} \D{propose a robust defense method against known and unknown adversarial attacks to enhance the security of \abbr{}s. They }introduce two black-box adversarial attack methods specifically targeting the attribution of source code authors. These methods are Targeted Autoencoding Style Mimicry Attack and Untargeted Autoencoding Style Concealment Attack. The \D{newly} proposed adversarial attack methods leverage systematic preservation of semantic code style attributes and transformations in both attacks. 
\textit{Targeted Autoencoding Style Mimicry Attack}: This attack extracts code style attributes from all code of the target author, synthesizes a comprehensive code style attribute, and identifies the code style attributes of the attacker's code. Finally, it transforms the attacker's code to mimic the style of the target code. 
\textit{Untargeted Autoencoding Style Concealment Attack}: This attack extracts code style attributes from the attacker's code, obtains code style attributes from other authors, identifies the differential set between the attacker's code style attributes and those of other authors, selects the author with the highest probability of misattribution, and transforms the attacker's code style to that of the selected target author. Experiments reveal that existing attribution models based on \abbr{}s are highly vulnerable to known and unknown adversarial attacks, with untargeted attacks exhibiting significantly higher success rates than targeted attacks.

\R{To reduce the search space for generating adversarial examples,} Choi et al.~\cite{2022-TABS} propose \D{an efficient black-box adversarial attack method, called }TABS\D{, to reduce the search space}. Specifically, TABS adopts beam search to find better adversarial examples and contextual semantic filtering to \D{effectively} reduce the search space. 
Experiments demonstrate that TABS can successfully launch adversarial attacks on the code search task of CodeBERT. Compared to Textfooler~\cite{2020-textfooler}, BERT-Attack~\cite{2020-BERT-ATTACK}, and CLARE~\cite{2020-clare}, TABS shows good performance in terms of attack success rate, number of queries, and other aspects.

Jia et al.~\cite{2022-funcfooler} propose a black-box adversarial attack method, FuncFooler, targeting learning-based binary code similarity detection models. FuncFooler operates under the constraints of maintaining both the program’s CFG and semantic equivalence with the original binary code. The method involves three key steps: identifying vulnerable candidates within the binary code, selecting and inserting adversarial instructions from benign code, and correcting any semantic side effects caused by the adversarial changes.
Experimental results show that the method reduces the accuracy of these models to below 5\% with minimal instruction perturbations and less than 1\% performance overhead. This demonstrates the effectiveness of FuncFooler in generating adversarial examples while maintaining the functional correctness of the target program.

Mercuri et al.~\cite{2023-Evolutionary-Approaches-for-Adversarial-Attacks-on-Neural-Source-Code-Classifiers} introduce an evolutionary technique for generating adversarial examples to enhance the robustness of \abbr{}s used for vulnerability detection. The proposed method utilizes the evolution strategies algorithm, with the neural network's output serving as the fitness function for deception. The algorithm starts from existing examples, evolving code snippets while preserving semantic transformations and reversing the original classification through fitness adjustment. This iterative process helps generate adversarial examples that can mislead vulnerability detection models while maintaining the original behavior of the code. The method maps the input code to a genomic representation, which evolves based on a persistent mutation identifier and the process of injecting code, without affecting the original behavior of the code.

Jha et al.~\cite{2023-CodeAttack} introduce a black-box adversarial attack method named CodeAttack, aiming to generate \D{efficient but} inconspicuous adversarial code examples using code structure. CodeAttack identifies the most vulnerable markers in the code and employs greedy search to generate alternative markers, replacing vulnerable markers with alternatives that minimally perturb the sample, maintain code consistency and fluency, and adhere to code constraints. Experimental results demonstrate that compared to TextFooler~\cite{2020-textfooler} and BERT-Attack~\cite{2020-BERT-ATTACK}, CodeAttack~\cite{2023-CodeAttack} efficiently launches adversarial attacks on CodeT5, CodeBRER, and GraphCodeBERT in code transformation, code repair, and code summarization tasks, producing adversarial examples with higher syntactic correctness.

Na et al.~\cite{2023-DIP} propose a\D{n} \D{efficient and advanced} black-box adversarial attack method named DIP, which does not require any information about the pre-trained deep models and avoids additional training computations. DIP creates adversarial examples by inserting dead code into the source code. Initially, DIP leverages the pre-trained CodeBERT model to identify vulnerable positions in the source code. It then utilizes cosine similarity to find candidate code snippets in the code repository that are least semantically similar to the source code. Finally, DIP selects the code statement with minimal disturbance from these candidate code snippets, wraps it with an unused variable, and inserts it as dead code into the vulnerable position of the source code. Since DIP launches adversarial attacks by inserting dead code, it preserves the functionality and compilability of the source code. Experiments show that DIP achieves the best CodeBLEU~\cite{2020-CodeBLUE} on nine target models and the best disturbance ratio on most target models.

\R{To reveal the insufficient robustness of current code authorship attribution systems,} Abuhamad et al.~\cite{2023-SHIELD} introduce SHIELD\D{ to attack the code authorship attribution system, demonstrating the insufficient robustness of current code authorship attribution systems}. SHIELD injects carefully selected code samples into the target code and then applies Stunnix~\cite{Stunnix} for code-to-code obfuscation. SHIELD employs obfuscation to make it challenging for models to statically identify or remove the injected code portions, rather than solely conceal the identity of the code author.
Adversarial samples are inputted into a black-box code authorship attribution system, and then the probability distribution of the identification model is modified by SHIELD through altering the code expressions and statements of the adversarial samples. SHIELD achieves a success rate of over 95.5\% and 66\% in non-targeted and targeted attacks, respectively, with a decrease in model identification confidence of over 13\%. 

Tian et al.~\cite{2023-Code-Difference-Guided-Adversarial-Example-Generation-for-Deep-Code-Models} propose the CODA for generating semantically invariant adversarial samples by exploiting the code differences between target input and reference input. Specifically, CODA leverages equivalent structural transformations and renaming identifier operations corresponding to the differences to eliminate code differences, thus generating adversarial samples resembling the reference while possessing the same semantics as the target. This effectively reduces the search space for components without the need for model gradient-assisted searching. 

\R{To address the inefficiency of adversarial attacks based on simple search strategies,} Zhang et al.~\cite{2023-CodeBERT-Attack} propose \R{CodeBERT-attack,} a \D{novel} black-box adversarial attack technique\D{ to address the inefficiency of adversarial attacks based on simple search strategies, named CodeBERT-attack}. CodeBERT-attack iteratively performs identifier replacement attacks. Firstly, CodeBERT-attack locates vulnerable identifiers in the code through masked classification. Then, it utilizes CodeBERT to generate new perturbed identifier names, introducing variations to the identifiers while preserving the overall code structure. Finally, it tests the perturbed samples on the victim model, selecting the samples with the lowest change in prediction accuracy on the ground true labels. Their experiments demonstrate that CodeBERT-attack can reduce the accuracy of LSTM and CodeBERT by an average of 30\% in function classification and code clone detection tasks. Moreover, the paper indicates that pre-training can increase the model's adversarial robustness.

Yang et al.~\cite{2024-radar} propose RADAR-Attack, a gradient-free adversarial attack method designed to expose the vulnerabilities of \abbr{}s by manipulating method names.
RADAR-Attack targets \abbr{}s with adversarial method names that are visually and semantically similar to the original inputs but cause the model to generate incorrect code.
RADAR-Attack leverages a two-step approach: first, it generates candidate method names using semantic and visual similarity techniques, and second, it applies a genetic algorithm to optimize these candidates, minimizing the performance of the model as measured by CodeBLEU. 
This gradient-free approach makes RADAR-Attack highly effective even in black-box scenarios where the internal details of the model are unknown. Experiments demonstrate that RADAR-Attack can significantly reduce performance in models like CodeGPT, PLBART, and CodeT5 in fine-tuning tasks, and in Replit, CodeGen, and CodeT5+ in zero-shot tasks, showcasing its capability to launch effective adversarial attacks against \abbr{}s.

Awal et al.~\cite{2024-investigating-adv} propose a \D{novel} adversarial attack method aimed at assessing the robustness of LM in software analytics tasks. 
The approach minimizes perturbations in feature space, leveraging ML explainability techniques to identify the most influential features impacting model decisions. It emphasizes that adversarial example generation in software analytics fundamentally differs from methods in computer vision, as any modifications to feature values are easily perceptible. 
The study employs the reverse elbow method to determine optimal feature combinations. 
The reverse elbow method evaluates the point at which the improvement in performance metric values ensures that generated adversarial examples induce misclassifications while maintaining feature validity. Experimental results demonstrate that the generated adversarial examples effectively reduce model accuracy, validating the method's potential for enhancing model security.

Yang et al.~\cite{2024-CodeTAE}  propose CodeTAE, an adversarial attack method targeting transfer learning in source code models. CodeTAE crafts adversarial examples that exploit the transferability of pre-trained code encoders by introducing semantic-preserving transformations. \D{Unlike previous methods,} CodeTAE operates in a cross-domain setting, where the attacker only has access to an open-source pre-trained encoder without any information about the victim model’s downstream task or training data. Using a genetic algorithm, CodeTAE generates substitute identifiers and applies obfuscations to induce misclassifications in source code models, achieving attack success rates ranging from 30\% to 80\%. Experiments on multiple code classification tasks, such as authorship attribution and clone detection, demonstrate the attack’s effectiveness, as well as its utility for adversarial fine-tuning to enhance model robustness.

Yao et al.~\cite{2024-carl} propose a black-box adversarial attack method, CARL, which leverages reinforcement learning to generate adversarial examples for code-based pre-trained language models. CARL consists of a programming language encoder and a perturbation prediction layer, which are initialized with a pre-trained masked CodeBERT model. The method formulates the task of generating adversarial examples as a sequence decision-making problem and optimizes it through policy gradient techniques.
CARL uses a combination of reward functions, including attack effectiveness and attack quality, to guide the learning process. 
Unlike traditional search-based approaches, CARL performs a one-step perturbation prediction, making it significantly faster during inference. 
Extensive experiments on tasks like code summarization, code translation, and code refinement show that CARL achieves higher success rates while maintaining code quality.

Zhou et al.~\cite{2024-MOAA} propose a multi-objective adversarial attack method, MOAA, which uses evolutionary multi-objective optimization to generate adversarial examples for code representation models. Unlike previous methods, MOAA integrates the NSGA-II algorithm with CodeT5 to generate adversarial examples that preserve both syntactic and semantic integrity. MOAA targets DNN models used for tasks such as code classification, clone detection, and code summarization, ensuring high-quality adversarial examples with minimal perturbation. Extensive experiments across five tasks and ten datasets reveal that MOAA produces diverse and transferable adversarial examples, with successful attacks on models like CodeBERT, GraphCodeBERT, and CodeT5. The results highlight MOAA’s effectiveness in generating robust adversarial examples while offering insights into the vulnerabilities of pre-trained code models. 

He et al.~\cite{2024-MalwareTotal} propose MalwareTotal, a\R{n}\D{novel} adversarial attack method targeting static malware detection models. 
MalwareTotal generates adversarial malware variants by leveraging sequence-aware bypass tactics that preserve the functionality of the malware while applying multi-faceted manipulations. The method employs proximal policy optimization to identify vulnerable manipulation sequences, effectively bypassing multiple detection mechanisms, including ML-based, signature-based, and hybrid anti-malware software. 

\subsubsection{Summary of Adversarial Attacks}
\label{subsubsec:summary_of_adversarial_attacks}
\

\begin{table*}[t]
    \centering
    \scriptsize
    \caption{A summary of existing white-box adversarial attacks on \abbr{}s.}
    \label{tab:adversarial_attacks_white_box}
    \resizebox{0.95\linewidth}{!}{
    \begin{tabular}{cccccc}
        \toprule
        
        Attack Technique & Year & Venue & Attack Type & Target Models & Target Tasks \\  

        \midrule 

        Meng et al.~\cite{2018-adv-binaries} & 2018 & ArXiv & White-box attack & \makecell[c]{Random Forest\\Support Vector Machine} &  Authorship attribution \\

        \midrule 
        
        DAMP~\cite{2020-Adversarial-examples-for-models-of-code} & 2020 & OOPSLA & White-box attack & \makecell[c]{Code2Vec\\GGNN} & \makecell[c]{Method name prediction \\ Variable name prediction} \\

        \midrule 

        Bielik et al.~\cite{2020-Adversarial-Robustness-for-code} & 2020 & ICML & White-box attack & \makecell[c]{LSTM\\GCN\\DeepTyper} & Type prediction \\

        \midrule 

        Srikant et al.~\cite{2021-Generating-Adversarial-Computer-Programs} & 2021 & ICLR & White-box attack & Seq2Seq & Method name prediction \\

        \midrule

        Chen et al.~\cite{2022-Generating-Adversarial-Source-Programs-Using-Important-Tokens-based-Structural-Transformations} & 2022 & ICECCS & White-box attack & Seq2Seq & Code summarization \\

        \midrule 

        M-CGA~\cite{2023-How-Robust-Is-a-Large-Pre-trained-Language-Model-for-Code-Generation} & 2023 & SANER & White-box attack & GPT-2 & Code generation \\

        \midrule 

        Capozzi et al.~\cite{2023-adv-binary} & 2023 & ArXiv & White-box attack & \makecell[c]{GNN\\RNN} & Binary similarity detection \\

        \bottomrule
    \end{tabular}
    }
\end{table*}

\begin{table*}[!t]
    \centering
    \scriptsize
    \caption{A summary of existing black-box adversarial attacks on \abbr{}s (2019-2022).}
    \label{tab:adversarial_attacks_black_box_2019_2022}
    \resizebox{0.95\linewidth}{!}{
    \begin{tabular}{cccccc}
        \toprule
        
        Attack Technique & Year & Venue & Attack Type & Target Models & Target Tasks \\

        \midrule 

        Quiring et al.~\cite{2019-Misleading-Authorship-Attribution} & 2019 & USENIX Security & Black-box attack & \makecell[c]{Random Forest\\LSTM} & Authorship attribution \\

        \midrule 

        Matyukhina et al.~\cite{2019-Adversarial-Authorship-Attribution-in-Open-Source-Projects} & 2019 & CODASPY & Black-box attack & N-grams & Authorship attribution\\

        \midrule

        SCAD~\cite{2021-A-Practical-Black-Box-Attack-on-Source-Code-Authorship-Identification-Classifiers} & 2021 & TIFS & Black-box attack & \makecell[c]{Random Forest\\RNN} & Authorship attribution\\

        \midrule

        STRATA~\cite{2020-STRATA} & 2020 & ArXiv & Black-box attack & Code2Seq & Method name prediction \\

        \midrule 

        MHM~\cite{2020-MHM} & 2020 & AAAI & Black-box attack & \makecell[c]{BiLSTM\\ASTNN} & Function classification\\

        \midrule 

        Pour et al.~\cite{2021-A-Search-Based-Testing-Framework} & 2021 & ICST & Black-box attack & \makecell[c]{Code2Vec\\ Code2Seq\\CodeBERT} & \makecell[c]{Method name prediction\\Code description\\Code search\\Code summarization} \\

        \midrule 

        QMDP~\cite{2021-Generating-adversarial-examples-of-source-code-classification-models-via-q-learning-based-markov-decision-process} & 2021 & QRS & Black-box attack & \makecell[c]{LSTM\\ASTNN} & Function classification\\

        \midrule 

        Ferretti et al.~\cite{2021-Deceiving-neural-source-code-classifiers-finding-adversarial-examples-with-grammatical-evolution} & 2021 & GECCO & Black-box attack & \makecell[c]{Random Forest\\CNN} & Vulnerability detection \\

        \midrule 

        Anand et al.~\cite{2021-Adversarial-Robustness-of-Program-Synthesis-Models} & 2021 & ArXiv & Black-box attack & \makecell[c]{SketchAdapt\\SAPS, Seq2Tree} & Code generation\\

        \midrule 

        ACCENT~\cite{2022-Adversarial-Robustness-of-Deep-Code-Comment-Generation} & 2022 & TOSEM & Black-box attack & \makecell[c]{LSTM, GNN\\CSCG, Rencos} & Code summarization \\

        \midrule 

        ALERT~\cite{2022-Natural-Attack-for-Pre-trained-Models-of-Code} & 2022 & ICSE & Black-box attack & \makecell[c]{CodeBERT\\GraphCodeBERT} & \makecell[c]{Vulnerability detection\\Clone detection\\Authorship attribution} \\

        \midrule 

        CARROTA~\cite{2022-Towards-Robustness-of-Deep-Program-Processing-Models} & 2022 & TOSEM & White-box attack & \makecell[c]{GRU, LSTM\\ASTNN, LSCNN\\TBCNN, CDLH\\CodeBERT} & \makecell[c]{Function classification\\Clone detection\\Vulnerability detection} \\

        \midrule 

        Li et al.~\cite{2022-RoPGen} & 2022 & ICSE & Black-box attack & \makecell[c]{DL-CAIS, PbNN} & Authorship attribution \\
        
        \midrule 

        TABS~\cite{2022-TABS} & 2022 & EMNLP & Black-box attack & CodeBERT & \makecell[c]{Code classification\\Code search}\\

        \midrule 

        FuncFooler~\cite{2022-funcfooler} & 2022 & ArXiv & Black-box attack & \makecell[c]{RNN, PV-DM\\Transformer} & Binary similarity detection \\
        
        \bottomrule
    \end{tabular}
    }
\end{table*}

\begin{table*}[t]
    \centering
    \scriptsize
    \caption{A summary of existing black-box adversarial attacks on \abbr{}s (2023-2024).}
    \label{tab:adversarial_attacks_black_box_2023_2024}
    \resizebox{0.95\linewidth}{!}{
    \begin{tabular}{cccccc}
        \toprule
        
        Attack Technique & Year & Venue & Attack Type & Target Models & Target Tasks \\

        \midrule 

        Mercuri et al.~\cite{2023-Evolutionary-Approaches-for-Adversarial-Attacks-on-Neural-Source-Code-Classifiers}& 2023 & Algorithms  & Black-box attack & CNN & \makecell[c]{Vulnerability detection}\\

        \midrule 
        
        CodeAttack~\cite{2023-CodeAttack} & 2023 & AAAI & Black-box attack & \makecell[c]{CodeBERT, CodeT5\\GraphCodeBERT\\} & \makecell[c]{Code transformation\\Code repair\\Code summarization} \\

        \midrule 

        DaK~\cite{2023-Discrete-Adversarial-Attack} & 2023 & PLDI & Black-box attack & \makecell[c]{Code2Vec, GGNN\\CodeBERT} & Variable misuse prediction \\

        \midrule 

        GraphCodeAttack~\cite{2023-Adversarial-Attacks-on-Code-Models} & 2023 & ArXiv & Black-box attack & \makecell[c]{CodeBERT\\GraphCodeBERT} &  \makecell[c]{Authorship attribution\\Vulnerability detection\\Clone detection} \\

        \midrule 

        AdVulCode~\cite{2023-AdVulCode} & 2023 & Electronics & Black-box attack & \makecell[c]{BiGRU, GGNN\\BiLSTM} & Vulnerability detection \\

        \midrule 

        DIP~\cite{2023-DIP} & 2023 & ACL & Black-box attack & \makecell[c]{CodeBERT, CodeT5\\GraphCodeBERT} & \makecell[c]{Clone detection\\Vulnerability detection\\Authorship attribution} \\

        \midrule 

        RNNS~\cite{2023-RNNS} & 2023 & ArXiv & Black-box attack & \makecell[c]{CodeBERT, CodeT5\\GraphCodeBERT} & \makecell[c]{Clone detection\\Vulnerability detection\\Authorship attribution\\Function classification} \\

        \midrule 

        SHIELD~\cite{2023-SHIELD} & 2023 & ArXiv & Black-box attack & \makecell[c]{CNN, CSFS, DL-CAIS} & Authorship attribution\\

        \midrule 

        CODA~\cite{2023-Code-Difference-Guided-Adversarial-Example-Generation-for-Deep-Code-Models} & 2023 & ASE & Black-box attack & \makecell[c]{CodeBERT, CodeT5\\GraphCodeBERT} & \makecell[c]{Vulnerability detection\\Authorship attribution\\Function classification\\Clone detection, Defect detection}\\

        \midrule

        CodeBERT-Attack~\cite{2023-CodeBERT-Attack} & 2023 & JSEP & Black-box attack & \makecell[c]{LSTM, CodeBERT\\GraphCodeBERT} & \makecell[c]{Function classification\\Clone detection}\\

        \midrule 

        RADAR-Attack~\cite{2024-radar} & 2024 & TOSEM & Black-box attack & \makecell[c]{CodeGPT\\PLBART, CodeT5} & \makecell[c]{Code generation} \\
        
        \midrule

        Awal et al.~\cite{2024-investigating-adv} & 2024 & ArXiv & Black-box attack & \makecell[c]{Logistic Regression\\Multi-Layered Perceptron\\Gradient Boosting Classifier\\Decision Tree, AdaBoost\\Random Forest, Bagging} & \makecell[c]{Defect detection\\Clone detection\\Code classification}\\

        \midrule 

        DeceptPrompt~\cite{2023-Deceptprompt} & 2023 & ArXiv & Black-box attack & \makecell[c]{CodeLlama, StarChat\\WizardCoder} & \makecell[c]{Code generation\\Vulnerability detection}\\

        \midrule 

        STRUCK~\cite{2024-STRUCK} & 2024 & CollaborateCom & Black-box attack & \makecell[c]{LSTM, GRU, GCN\\GGNN, CodeBERT\\GraphCodeBERT\\UniXcoder} & \makecell[c]{Function classification\\Defect detection}\\

        \midrule 

        CodeTAE~\cite{2024-CodeTAE} & 2024 & TIFS & Black-box attack & \makecell[c]{CodeBERT, CodeT5\\GraphCodeBERT} & \makecell[c]{Authorship attribution\\Clone detection\\Defect detection}\\

        \midrule 

        MoAA~\cite{2024-MOAA} & 2024 & FSE & Black-box attack & \makecell[c]{CodeBERT, CodeT5\\GraphCodeBERT} & \makecell[c]{Defect detection, Clone detection\\Authorship attribution\\Code translation\\Code summarization}\\

        \midrule 

        MalwareTotal~\cite{2024-MalwareTotal} & 2024 & ICSE & Black-box attack & \makecell[c]{EMBER, MalConv\\AMS A/B, ClamAV} & Malware detection\\

        \midrule

        CARL~\cite{2024-carl} & 2024 & TOSEM & Black-box attack & \makecell[c]{CodeBERT, CodeT5\\UniXcoder\\GraphCodeBERT\\CodeGeeX2} & \makecell[c]{Code summarization\\Code translation\\Code refinement} \\
        
        \bottomrule
    \end{tabular}
    }
\end{table*}

Table~\ref{tab:adversarial_attacks_white_box} summarizes the existing white-box attacks on \abbr{}s, while Table~\ref{tab:adversarial_attacks_black_box_2019_2022} and Table~\ref{tab:adversarial_attacks_black_box_2023_2024} summarize black-box attacks from 2019-2022 and 2023-2024, respectively. Both types of attacks have demonstrated their effectiveness on various \abbr{}s, including non-pre-trained and pre-trained models. Furthermore, these attacks have proven to be effective across multiple code-related tasks, including code understanding and code generation tasks.
White-box attack techniques primarily rely on the gradient information of the target model to generate adversarial samples. These adversarial samples are created by introducing perturbations that do not alter the program’s semantics, and they can still successfully compile. Common adversarial perturbations include operations such as substitution, insertion, and deletion, which are applied at the identifier and statement levels. For instance, they may involve renaming variables, function parameters, classes, or structs, replacing Boolean expressions with their equivalents, or inserting and deleting dead code and empty statements. When generating these adversarial samples, it is crucial to ensure their semantic equivalence so that the attack remains imperceptible to humans. However, creating adversarial samples that are both subtle and effective remains a challenging task.
Current adversarial attack research is increasingly focusing on black-box attacks. Unlike white-box attacks, black-box attacks do not rely on the gradient information of the model. Instead, they generate adversarial samples by observing the relationship between the model’s inputs and outputs. Black-box attack methods often use evolutionary algorithms, genetic algorithms, or other query-based methods to find the optimal perturbations for adversarial samples. These attacks have demonstrated effectiveness on both pre-trained and non-pre-trained models. Compared to white-box attacks, black-box attacks are generally more time-consuming and have lower success rates due to the lack of internal model information. Additionally, black-box attacks have shown strong attacking capabilities across different code-related tasks, such as code understanding and code generation.

\summary{This section summaries backdoor attacks and adversarial attacks on \abbr{}s. 
Backdoors can be inserted through data poisoning or model poisoning, where the design and injection method of the trigger determine the stealthiness and effectiveness of the attack. Such attacks typically remain dormant in controlled environments until the trigger is activated.
Adversarial attacks involve adding subtle perturbations to the input that cause the model to make incorrect predictions. These perturbations usually do not alter the semantic functionality of the code but can mislead the model’s predictions. 
Existing research on backdoor and adversarial attacks indicates that \abbr{}s are vulnerable to these security threats, which should raise concerns about the security of \abbr{}s within the SE community.}

\section{Answer to RQ2: Defenses on \abbr{}s}
\label{sec:Answer_to_RQ2}

\subsection{Research Directions in Defenses against \abbr{}s}
\label{subsec:research_directions_on_defenses_LM4Code}

As shown in the right part of Figure~\ref{fig:research_directions_by_security}, we construct a taxonomy based on the 13 research papers collected on defenses against \abbr{}s. Our taxonomy includes two categories of research directions: backdoor defenses and adversarial defenses. 
There are only 4 papers that focus on studying backdoor defense in \abbr{}s and 9 papers that are dedicated to researching adversarial attacks in \abbr{}s. 
Furthermore, based on different defense scenarios, backdoor defenses can be subdivided into pre-training defenses, in-training defenses and post-training defenses, while adversarial defenses can be categorized into adversarial training, model modification and additional models.
In backdoor defenses, there are only 2 papers on pre-training defenses, 1 paper on in-training defenses, and 1 paper on post-training defenses. Compared to the research on backdoor attacks against \abbr{}s, research on backdoor defenses is still in its early stages. In adversarial defenses, there are only 5 papers on input modification, 1 paper on model modification, and 3 papers on additional models. Similarly, adversarial defenses for \abbr{}s are still in the early stages.

\subsection{Backdoor Defenses on \abbr{}s}
\label{subsec:backdoor_defenses_on_LM4Code}

\subsubsection{Overview of Backdoor Defenses against \abbr{}s}
\label{subsubsec:overview_of_backdoor_defenses}
\

In backdoor attacks on \abbr{}, attackers can deploy various methods at different stages of the \abbr{} learning process and trigger the implanted backdoors during the model’s application phase. As a result, researchers have designed defense methods tailored to the characteristics of backdoor attacks, aiming to be applicable at different stages of the learning process. These methods seek to eliminate or reduce the success rate of backdoor attacks, thereby enhancing the security of \abbr{}. Based on different defense objectives, existing backdoor defense methods can be categorized into pre-training defenses, in-training defenses and post-training defenses. 
Pre-training defenses focus on detecting and removing poisoned samples before training. \R{They typically employ data filtering or anomaly detection techniques to identify potential backdoor samples in the dataset}. In-training defenses aim to prevent backdoor insertion during the model training phase. \R{They usually modify the training process to make the model less sensitive to the influence of poisoned samples during training}. Post-training defenses are applied after model training is completed. \R{They often utilize techniques such as model unlearning or input filtering to ensure model security}.

\subsubsection{Pre-training Backdoor Defenses}
\label{subsubsec:pre_training_backdoor_defenses}
\

Ramakrishnan et al.~\cite{2022-backdoors-in-neural-models-of-source-code} propose a backdoor detection defense method for \abbr{}s based on the spectrum method~\cite{2018-spectral-signatures}. They argue that the spectrum method, which uses only the right singular vectors for data sorting, does not always satisfy \R{$\varepsilon$-}spectrum separability. \R{In other words, the clean samples distribution and the poisoned samples distribution cannot be effectively separated in spectral space under the error tolerance controlled by $\varepsilon$.}
Therefore, they replace it with the \D{previous} \R{top k} right singular vectors. Additionally, they use the encoder output vectors or context vectors based on two common \abbr{} architectures, Seq2Seq and Code2Seq, to detect \D{toxic}\R{poisoned} samples. Experiments demonstrate that the proposed backdoor defense method effectively detects low poisoning rates of \D{toxic}\R{poisoned} samples and can eliminate or reduce the attack success rate.

Li et al.~\cite{2024-poison-attack-and-poison-detection} introduce a\D{n} \D{effective} backdoor detection defense method called CodeDetector, which detects \D{toxic}\R{poisoned} samples in training data without losing clean samples. CodeDetector focuses on detecting triggers for defense and assumes that attackers may hide triggers within important labels. It utilizes the integrated gradients technique~\cite{2017-Axiomatic-Attribution-for-Deep-Networks} to extract all important labels from the training data. Subsequently, it identifies abnormal labels that significantly impact the model's performance and treats them as potential triggers. All data containing these triggers is then flagged as \D{toxic}\R{poisoned} samples. Experiments show that compared to baseline methods such as syntax checkers based on program analysis tools, ONION~\cite{2021-ONION}, clustering methods, and spectrum methods relying on outlier detection, CodeDetector achieves a 35.95\% higher recall and a 16.1\% lower false positive rate.

\subsubsection{In-training Backdoor Defenses}
\label{subsubsec:in_training_backdoor_defenses}
\

Yang et al.~\cite{2024-dece} propose a backdoor detection defense method for \abbr{}s called Deceptive Cross-Entropy (DeCE) to enhance the security of \abbr{}s against backdoor attacks. The method leverages the phenomenon of ``early learning'', where models are initially less sensitive to backdoor triggers but gradually become more susceptible to them, leading to overfitting. DeCE tackles this issue by employing a blending process that combines the model’s predicted probability distribution with a deceptive distribution, which is gradually introduced during training. Additionally, label smoothing is applied to the original labels to prevent the model from becoming overly confident and reduce the risk of overfitting to backdoor triggers. Through extensive experimentation on various models and code synthesis tasks, DeCE demonstrates superior performance in defending against backdoor attacks while maintaining model accuracy on clean datasets.

\subsubsection{Post-training Backdoor Defenses}
\label{subsubsec:post_training_backdoor_defenses}
\

Hussain et al.~\cite{2023-OSEQL} propose an occlusion-based human-in-the-loop technique called OSEQL, to distinguish backdoored-triggering inputs of code. 
OSEQL is based on the observation that the backdoored \abbr{} relies heavily on the trigger input and its removal would change the confidence of the \abbr{} in prediction substantially.
Specifically, OSEQL generates one-line occluded snippets for input code and employs outlier detection to identify outliers in the predictions of suspicious models for these inputs, thereby identifying triggers.
Experiments show that in both clone detection and defect detection tasks, OSEQL can achieve trigger identification rates above 90\% for poisoned inputs across five popular LMs.

\subsubsection{Summary of Backdoor Defenses}
\label{subsubsec:summary_of_backdoor_defenses}
\

\begin{table}[H]
    \centering
    \scriptsize
    \caption{A summary of existing backdoor defenses on \abbr{}s.}
    \label{tab:backdoor_defense}
    \resizebox{0.95\linewidth}{!}{
    \begin{tabular}{cccccc}
        \toprule
        
        Defense Techniques & Year & Venue & Defense Type & Target Models & Target Tasks \\

        \midrule 

        Ramakrishnan et al.~\cite{2022-backdoors-in-neural-models-of-source-code} & 2022 & ICPR & Pre-training defense & \makecell[c]{Code2Seq\\Seq2Seq} & \makecell[c]{Code summarization \\ Method name prediction} \\
        
        \midrule 

        CodeDetector~\cite{2024-poison-attack-and-poison-detection} & 2024 & TOSEM & Pre-trainig defense & \makecell[c]{LSTM, TextCNN\\Transformer\\CodeBERT} & \makecell[c]{Defect detection\\Clone detection\\Code repair} \\

        \midrule 
        
        DeCE~\cite{2024-dece} & 2024 & ArXiv & In-training defense & \makecell[c]{CodeBERT, CodeGen\\CodeT5, CodeT5p\\GraphCodeBERT} & \makecell[c]{Code generation\\Code repair} \\     
        \midrule 

        OSEQL~\cite{2023-OSEQL} & 2023 & ArXiv & Post-training defense & \makecell[c]{CodeBERT, CodeT5\\BART, PLBART\\RoBARTa} & \makecell[c]{Defect detection\\Clone detection} \\
        
        \bottomrule
    \end{tabular}
    }
\end{table}

Table~\ref{tab:backdoor_defense} summarizes the existing backdoor defense methods for \abbr{}s. These methods cover the pre-training, in-training, and post-training phases of \abbr{}s, effectively enhancing the security of code models through various techniques. However, backdoor defense is still in its early stages, and there is limited research in this area. As backdoor attack techniques continue to evolve, defense methods still require further optimization to address increasingly complex security threats.

\subsection{Adversarial Defenses on \abbr{}s}
\label{subsec:adversarial_defenses_on_LM4Code}

\subsubsection{Overview of Adversarial Defenses against \abbr{}s}
\label{subsubsec:overview_of_adversarial_defenses}
\

For the aforementioned adversarial attack methods, researchers have proposed various adversarial defense methods to enhance the model's robustness or eliminate the impact of adversarial samples on the model. There are primarily three methods in adversarial defense: 
(1) \textit{Input modification}: modifying the model training process\R{,} \D{or} altering input data samples \R{or optimizing the training process}. \R{The core idea is to proactively detect and mitigate the impact of adversarial samples or strengthen the model’s resistance to abnormal inputs through adversarial training.} \R{Typically,} \D{A}\R{a}dversarial examples are \D{usually introduced to the model for adversarial training to enhance its robustness}\R{incorporated into the training set, allowing the model to adapt to adversarial perturbations during the learning process and improving its stability against attacks}.
(2) \textit{Model modification}: adding more sub-networks, modifying activation functions, \D{or }changing loss functions within the model\R{, or introducing regularization techniques}. \R{Robustness-optimized loss functions are commonly employed to improve classification accuracy in adversarial regions. Additionally, adjusting activation functions can increase the model’s sensitivity to adversarial samples, thereby reducing attack success rates.}
(3) \textit{Additional models}: using external models as additional networks when dealing with unseen data samples for classification. \R{The core idea is to use additional detection mechanisms or adversarially trained networks to identify and process potential adversarial samples. Typically, an adversarial sample detector is placed before the original model to filter input data, ensuring that only normal inputs are processed by the model.}
The first method focuses on data samples input to the model, while the other two methods are more concerned with the model itself. The latter two methods can be further categorized into model defense and data detection. The goal of model defense is to enable the model to correctly identify adversarial samples, while data detection involves detecting adversarial samples and issuing warnings, leading to the model rejecting further processing of adversarial inputs. Currently, adversarial defense methods for \abbr{}s mainly concentrate on the first method. Next, we will provide a detailed overview of relevant research on adversarial defense for \abbr{}s.

\subsubsection{\D{Adversarial Training}\R{Input Modification}}
\label{subsubsec:adversarial_training}
\

Li et al.~\cite{2022-RoPGen} introduce a\D{n} \D{innovative} framework called RoPGen, which strengthens the robustness of authorship attribution models. RoPGen combines data augmentation and gradient augmentation during adversarial training, making it challenging for attackers to manipulate and mimic the model. RoPGen employs automatic encoding style imitation attacks and automatic encoding style obfuscation attacks for data augmentation, effectively increasing the diversity of training data. RoPGen introduces perturbations to the gradients of deep neural networks for data augmentation, learning a robust \abbr{} with diverse representations. RoPGen uses four datasets based on programming languages, such as C, C++, and Java, for evaluation. Experimental results indicate that RoPGen significantly improves the robustness of code authorship attribution models, reducing target attack success rates by an average of 22.8\% and non-target attack success rates by 41.0\%.

\R{To reduce the complexity of adversarial training,} Henkel et al.~\cite{2022-Semantic-Robustness-of-Models-of-Source-Code} propose an adversarial training method for \abbr{}s based on Madry et al.'s robust optimization objective~\cite{2018-Towards-Deep-Learning-Models-Resistant}. \D{To reduce the complexity of adversarial training, the authors}\R{They} pre-generate program sketches by applying transformations to some programs. During training, only the generated program sketches are considered, and a gradient-based attack is used to search for the optimal parameters for the transformations. Experiments show that adversarial training with just one transformation of \abbr{}s efficiently enhances model robustness.

To enhance the robustness of source code processing models, Zhang et al.~\cite{2022-Towards-Robustness-of-Deep-Program-Processing-Models} further proposed the adversarial training method CARROTT. CARROTT utilizes adversarial examples generated by the attack method CARROTA to periodically augment the training set and introduce adversarial perturbations during the training process, thus systematically enhancing the model’s robustness to adversarial attacks. Experimental evaluations show that models trained with CARROTT exhibit an average robustness improvement of 5.3 times across multiple tasks (such as code functionality classification, code clone detection, and defect prediction), significantly outperforming MHM, which achieves only a 1.7× improvement.

Cristina et al.~\cite{2023-Enhancing-Robustness-of-AI-Offensive-Code-Generators} propose a method to enhance the robustness of \abbr{}s for code generation by adding word-level perturbations in code descriptions, including word replacement and word omission methods. Initially, they demonstrate that perturbations preserve the original semantics through cosine similarity. In experiments, they evaluate the impact of new perturbations on Seq2Seq, CodeBERT, and CodeT5+ models, showing significant performance degradation. Based on this, they apply different proportions of data augmentation to training data and empirically demonstrate that the proposed method improves robustness against perturbed samples and enhances model performance for non-perturbed data, achieving higher semantic correctness.

\R{Furthermore,} Gao et al.~\cite{2023-Discrete-Adversarial-Attack} propose the EverI defense method against discrete adversarial attacks on \abbr{}s. EverI trains the model on the strongest adversarial examples, making it theoretically applicable to defend against all adversarial attack methods. For discrete adversarial attacks, EverI first constructs a transitive closure based on semantic-preserving transformations on the original input, identifying adversarial examples in all programs within the transitive closure that cause the maximum loss for the model. For other classical continuous adversarial attacks like DAMP~\cite{2020-Adversarial-examples-for-models-of-code}, EverI identifies the strongest adversarial examples by adding adversarial steps and random initialization of adversarial perturbations. For other discrete attacks like Imitator~\cite{2019-Misleading-Authorship-Attribution}, EverI increases the width and depth of the program transformation search tree to find the strongest adversarial examples. Experiments show that, compared to adversarial training~\cite{2018-Towards-Deep-Learning-Models-Resistant} and outlier detection defense~\cite{2020-Adversarial-examples-for-models-of-code}, EverI is more effective in defending against discrete adversarial attacks.

\subsubsection{Model Modification}
\label{subsubsec:model_modification}
\

He et al.~\cite{2023-Large-Language-Models-for-Code-Security-Hardening-and-Adversarial-Testing} propose a \D{novel} learning-based secure control method SVEN for \abbr{}s in code generation tasks and test the model's security using crafted adversarial examples. SVEN utilizes continuous vectors with specific attributes to guide \abbr{}s toward the direction of the given attributes in program generation without modifying the model's weights. SVEN's training process enforces specialized loss terms on different regions of the code and optimizes these continuous vectors using a carefully crafted high-quality dataset. Experiments show that SVEN can achieve effective secure control, increasing the proportion of secure code generation to 92.3\% for CodeGen models without compromising the functional correctness of the generated code.

\subsubsection{Additional Models}
\label{subsubsec:additional_models}
\

Li et al.~\cite{2022-Semantic-Preserving-Adversarial-Code-Comprehension} propose SPACE, a semantic-preserving code embedding approach, to simultaneously enhance the generalization and robustness of \abbr{}s. SPACE identifies semantic-preserving attacks in worst-case scenarios, compelling the model to correctly predict labels under these scenarios to improve robustness. Specifically, SPACE uses Byte Pair Encoding Tokenizer (BPE) to convert input code sequences into a series of subword tokens. It then employs a subword embedding layer for high-dimensional word embeddings. Each subword belonging to the same identifier undergoes differentiable perturbations, added to the corresponding word embeddings and position embeddings, forming distributed embeddings. Finally, the pre-trained model's encoder further encodes these embeddings. Differentiable perturbations are updated based on their gradients, and a gradient ascent algorithm computes the final loss to minimize the negative impact of perturbations on the model. SPACE conducts adversarial training in continuous embedding space, efficiently incorporating it into gradient-based training frameworks. SPACE combines adversarial training with data features specific to programming languages, achieving improved performance and robustness for \abbr{}s.  
Experiments demonstrate that training pre-trained \abbr{}s CodeBERT and GraphCodeBERT with SPACE leads to better generalization, maintaining robustness against advanced adversarial attack methods such as MHM~\cite{2020-MHM} and ALERT~\cite{2022-Natural-Attack-for-Pre-trained-Models-of-Code}.

Zhang et al.~\cite{2023-Transfer-Attacks-and-Defenses-for-Large-Language-Models-on-Coding-Tasks} first generate adversarial examples using white-box attacks on smaller code models and then explore their effectiveness for transfer adversarial attacks on code LLMs. Meanwhile, they propose leveraging LLMs’ in-context learning and the self-defense capability of understanding human instructions, as well as meta-prompting techniques (i.e., using the LLM to generate its own defense prompts) to achieve adversarial defense for large models. Experimental results show that these defense methods effectively reduce the attack success rate on GPT-3.5, GPT-4, Claude-Instant-1, Claude-2, and CodeLlama.

Yang et al.~\cite{2024-radar} propose RADAR-Defense, a defense method to improve the robustness of \abbr{}s by synthesizing high-quality method names from functional descriptions. 
Specifically, RADAR-Defense avoids altering the original model and instead focuses on improving input quality by generating semantically meaningful method names that are supplied to \abbr{}s along with functional descriptions and other signature information. 
This retrieval-enhanced prompt training method effectively reinstates the performance of \abbr{}s under adversarial attacks. Experiments demonstrate that RADAR-Defense is able to restore the CodeBLEU and Pass@1 scores in various \abbr{}s, such as CodeGPT, PLBART, CodeT5, Replit, CodeGen, and CodeT5+, thereby significantly improving model robustness without requiring retraining.

\subsubsection{Summary of Adversarial Defenses}
\
\label{subsubsec:summary_of_adversarial_defenses}

\begin{table}[!t]
    \centering
    \scriptsize
    \caption{A summary of existing adversarial defenses on \abbr{}s.}
    \label{tab:adversarial_defense}
    \resizebox{0.95\linewidth}{!}{
    \begin{tabular}{cccccc}
        \toprule
        
        Defense Techniques & Year & Venue & Defense Type & Target Models & Target Tasks \\

        \midrule 

        RoPGen~\cite{2022-RoPGen} & 2022 & ICSE & Input modification & \makecell[c]{DL-CAIS\\PbNN} & \makecell[c]{Code authorship attribution} \\
        
        \midrule 

        Henkel et al.~\cite{2022-Semantic-Robustness-of-Models-of-Source-Code} & 2022 & SANER & Input modification & \makecell[c]{Seq2Seq\\Code2Seq} & \makecell[c]{Code summarization} \\

        \midrule 

        CARROTT~\cite{2022-Towards-Robustness-of-Deep-Program-Processing-Models} & 2022 & TOSEM & Input modification & \makecell[c]{GRU, LSTM\\ASTNN, LSCNN\\TBCNN, CodeBERT\\CDLH} & \makecell[c]{Function classification\\Clone detection\\Vulnerability detection} \\

        \midrule 
        
        Cristina et al.~\cite{2023-Enhancing-Robustness-of-AI-Offensive-Code-Generators} & 2023 & ArXiv & Input modification & \makecell[c]{Seq2Seq\\CodeBERT\\CodeT5+} & \makecell[c]{Code generation} \\     
        
        \midrule 

        EverI~\cite{2023-Discrete-Adversarial-Attack} & 2023 & PLDI & Input modification & \makecell[c]{CodeBERT\\Code2Vec\\GGNN } & \makecell[c]{Code summarization\\Method name prediction\\Variable misuse prediction} \\
        
        \midrule 
        
        SVEN~\cite{2023-Large-Language-Models-for-Code-Security-Hardening-and-Adversarial-Testing} & 2023 & CCS & Model modification & CodeGen & \makecell[c]{Code generation} \\     
        
        \midrule

        SPACE~\cite{2022-Semantic-Preserving-Adversarial-Code-Comprehension} & 2022 & COLING & Additional models & \makecell[c]{CodeBERT\\GraphCodeBERT} & \makecell[c]{Defect detection\\Natural language code search\\Code question answering} \\     
        
        \midrule

        Zhang et al.~\cite{2023-Transfer-Attacks-and-Defenses-for-Large-Language-Models-on-Coding-Tasks} & 2023 & ArXiv & Additional models & \makecell[c]{GPT-3.5, GPT-4\\ Claude-Instant-1\\ Claude-2 } & \makecell[c]{Code summarization} \\     
        
        \midrule

        RADAR-Defense~\cite{2024-radar} & 2024 & TOSEM & Additional models & \makecell[c]{CodeGPT, PLBART\\CodeT5, CodeT5+\\Replit, CodeGen} & \makecell[c]{Code generation} \\     
        
        \bottomrule
    \end{tabular}
    }
\end{table}
Table~\ref{tab:adversarial_defense} summarizes the existing adversarial defense methods for \abbr{}s. Adversarial defenses for \abbr{}s encompass input modification, model modification, and additional model methods, each with a specific focus on mitigating the effects of adversarial attacks. The input modification approach often involves adversarial training, which has been shown to enhance model robustness. Model modification strategies adapt sub-networks, activation functions, or loss functions to detect and handle adversarial samples effectively. Lastly, the additional model approach leverages external models for classification robustness. Methods like RoPGen, SVEN, and SPACE contribute significantly to adversarial robustness, highlighting the potential of combining adversarial training with program-specific features for enhanced defense against complex adversarial scenarios.

\summary{The attacks and defenses for \abbr{}s represent a continuous battle. Defense methods are the direct approach to enhancing the security of \abbr{}s. Although current defense techniques cover different methods and critical stages of \abbr{}s, compared to the extensive research on attacks in the security of \abbr{}s, research on defenses is still in its early stages.}

\section{Answer to RQ3: Empirical Studies on Security of \abbr{}s}
\label{sec:Answer_to_RQ3}

\begin{table*}[!t]
    \centering
    \caption{A summary and comparison of empirical studies on the security of \abbr{}.}
    \label{tab:empirical_studies}
    \resizebox{0.95\linewidth}{!}{
    \begin{tabular}{cccl}
        \toprule
    
        Year & Study & Type & Main Research Content (Research Questions) \\

        \midrule
        
        2021 & Nguyen et al.~\cite{2021-empirical-adversarial-attacks-to-API-recommender-systems} & Adversarial Attack & \makecell[l]{1) How well has the issue of AML in RSSE been addressed by the existing literature? \\ 2) To what extent are state-of-the-art API and code snippet recommender systems \\ susceptible to malicious data?} \\
        \midrule 

        2023 & Du et al.~\cite{2023-Study-on-Adversarial-Attack-against-Pre-trained-Code-Models} & Adversarial Attack & \makecell[l]{1) How do existing adversarial attack approaches perform against different PTMCs \\ under various tasks? \\ 2) What is the quality of adversarial examples generated by adversarial attacks? \\ 3) How do the contexts of the perturbed identifiers affect the adversarial attacking \\ performance?} \\
        \midrule

        2023 & Li et al.~\cite{2023-A-comparative-study-of-adversarial-training-methods} & Adversarial Defense & \makecell[l]{1) How does learning sample obtaining strategy during training impact model \\ performance? \\ 2) How does loss computation impact model performance? \\ 3) How does input data impact model performance? \\ 4) How does input model impact model performance?}\\
        \midrule

        2024 & Oh et al.~\cite{2024-Poisoned-ChatGPT-Finds-Work-for-Idle-Hands} & Backdoor Attack & \makecell[l]{1) How do developers’ adoption rates and trust levels differ when using code \\ completion compared to code generation as AI-powered coding assistant tools, \\ and what factors influence these variations? \\ 2) How do poisoning attacks on AI-powered coding assistant tools influence \\ the security of software developers’ code in the real world? \\ 3) Which type of AI-powered coding assistant tools, code completion or code \\ generation , are more susceptible to poisoning attacks?} \\
        \midrule

        2024 & Hussain et al.~\cite{2024-On-Trojan-Signatures-in-LLMs-of-Code} & Backdoor Attack & \makecell[l]{1) Is there any trojan signature in the classifier weights of the trojaned code \\models which were produced by full-finetuning? \\ 2) Is there any trojan signature in the trojaned code models, which were produced \\where pretrained weights were frozen?} \\

        \midrule

        2024 & Hussain et al.~\cite{2024-Measuring-Impacts-of-Poisoning-on-Model-Parameters-and-Neuron-Activations} & Backdoor Attack & \makecell[l]{1) Distribution of attention weight and bias in clean and poisoned models. \\ 2) Clustering activation values of clean and poisoned samples. \\ 3) Visualizing context embeddings of clean and poisoned models. \\ 4) Comparison of fine-tuned parameters with pre-trained parameters. \\ 5) Resetting fine-tuned weights to pre-trained weights based on a threshold.} \\

        \bottomrule
    \end{tabular}
    }
\end{table*}

Various attack and defense techniques targeting \abbr{}s have been successively proposed and have achieved promising results in revealing \abbr{} security vulnerabilities and enhancing its security.
In addition to developing new attack and defense techniques that address technical challenges, the field of \abbr{} security research has also benefited from several empirical studies. These studies systematically explore the capabilities of attack or defense techniques and their impact on model performance, providing valuable insights for future exploration of \abbr{} security.
We summarize the existing empirical studies in Table~\ref{tab:empirical_studies} and discuss them in detail as follows.

\subsection{Practical impact of backdoor attacks on \abbr{}s}
\label{subsec:empirical_studies_on_backdoor_attacks}

Oh et al.~\cite{2024-Poisoned-ChatGPT-Finds-Work-for-Idle-Hands} conduct an empirical study on the practical impact of data poisoning attacks on AI-powered coding assistant tools, exploring how these attacks affect developers' coding behavior when using these tools and the security of the code generated by them. Specifically, they investigate the following research questions:
1) How do developers' adoption rates and trust levels differ when using code completion compared to code generation as AI-powered coding assistant tools, and what factors influence these variations?
They first conduct an online survey aimed at understanding the prevalence of AI-powered coding assistant tool usage and the level of trust developers place in the generated code. Their survey of 238 participants reveals that developers generally trust code completion tools more than code generation tools. This trust likely stems from developers' perception that code completion tools are more precise and their suggestions come from verified, reliable official API documentation. At the same time, they indicate that both types of tools are susceptible to data poisoning attacks. These attacks can introduce malicious code into a codebase, posing a security risk. Consequently, they proceed to address the following research questions. 
2) How do poisoning attacks on AI-powered coding assistant tools influence the security of software developers' code in the real world?
They use TrojanPuzzle to poison the CodeGen 6.1B model and have participants complete three programming tasks related to common security vulnerabilities in software development: AES encryption, SQL query, and DNS query. Through the manual review of the participants' code, they find that participants using the code generation tool are more likely to incorporate insecure code compared to those using the code completion tool or no tool at all. This highlights the influence of AI-powered coding assistant tools on secure coding practices.
3) Which type of AI-powered coding assistant tools, code completion or code generation, are more susceptible to poisoning attacks?
Through an in-depth analysis of participants' programming practices, they find that code completion tools are relatively weak against poisoning attacks. This is because these tools guide developers to obtain initial code from the internet rather than relying on the tools themselves. Moreover, practices like ``copy and paste'' or ``see and type'' help bypass the code suggestions of the tools, thereby reducing the likelihood of poisoning attacks.
Code generation tools are very sensitive to poisoning attacks since developers often use the suggested code without significant modifications. However, some developers adopt secure coding practices to generate their own secure code or modify the suggested code. Their findings highlight the importance of developers understanding potential security issues and maintaining caution regarding the safety of AI-driven coding assistance tools.

Hussain et al.~\cite{2024-On-Trojan-Signatures-in-LLMs-of-Code} conduct an empirical study on the distribution of trojan signatures in the classifier layer parameters of large language models for source code. They address two research questions: 1) Is there any trojan signature in the classifier weights of the trojaned code models which were produced by full-finetuning? 2) Is there any trojan signature in the trojaned code models, which were produced where pretrained weights were frozen? They perform detailed experiments on two tasks, clone detection and defect detection, using CodeBERT, PLBART, and CodeT5. Ultimately, their findings suggest that Trojan signatures do not generalize to LLMs of code. They observe that even when fine-tuning with frozen pre-trained weights in a more explicit setting, poisoned code models remain resistant, indicating that detecting trojans solely based on the model's weights is a challenging task.

Furthermore, Hussain et al.~\cite{2024-Measuring-Impacts-of-Poisoning-on-Model-Parameters-and-Neuron-Activations} conduct an empirical study on the attention weights and biases, activation values, and context embeddings of clean and poisoned CodeBERT models, exploring the differences between the clean and poisoned models.
First, they use TrojanedCM~\cite{2023-Trojanedcm} to poison CodeBERT for the defect detection task. Then, they extract and compare the parameters of the poisoned CodeBERT model with its corresponding clean model layer by layer, examining various aspects such as attention weights and biases, activation values, and context embeddings.
Specifically, they conduct the following five analyses:
1) Distribution of attention weights and biases in clean and poisoned models; 
2) Clustering of activation values for clean and poisoned samples; 
3) Visualization of context embeddings of clean and poisoned models;
4) Comparison of fine-tuned parameters with pre-trained parameters;
5) Resetting fine-tuned weights to pre-trained weights based on a threshold.
Ultimately, they find that there are no significant differences in attention weights and biases between the clean and poisoned CodeBERT models. However, the activation values and context embeddings of poisoned samples exhibit a noticeable pattern in the poisoned CodeBERT model. Their findings contribute to the white-box detection of backdoor signals in \abbr{}s.

\subsection{Performance of adversarial attacks against \abbr{}s across different tasks and models}
\label{subsec:empirical_studies_on_adversarial_attacks}

Nguyen et al.~\cite{2021-empirical-adversarial-attacks-to-API-recommender-systems} conduct an empirical investigation of adversarial machine learning techniques and their potential impact on API recommender systems.
1) How well has the issue of adversarial machine learning (AML) in recommender systems for software engineering (RSSE) been addressed by the existing literature?
They conduct a literature analysis following the guidelines~\cite{2011-Repeatability-of-systematic-literature-reviews, 2010-How-Reliable-Are-Systematic-Reviews-in-Empirical-Software-Engineering} to investigate whether there have been efforts in studying and addressing threats to RSSE originating from malicious data. They collect 7,076 papers by answering the four W-questions~\cite{2011-Identifying-relevant-studies-in-software-engineering} (``W'' stands for Which?, Where?, What?, and When?). 
Through a thorough examination of these papers, they find that all of them focus on malware detection in Android applications. 
However, they do not discover any work related to potential threats and adversarial attacks concerning API RSSE. Thus, as of 2021, the issue of adversarial attacks targeting API and code snippet recommendations has not been sufficiently studied in the main areas of software engineering.
2) To what extent are state-of-the-art API and code snippet recommender systems susceptible to malicious data?
Through the qualitative analysis of the significant RSSE in mining APIs and code snippets for API recommendation systems, they find that if malicious data is used for model training, RSSE may recommend malicious APIs or code snippets. 
According to the analysis, even if only 10\% of the training data is manipulated, UPMiner and PAM recommendation systems still provide a considerable number of projects with false APIs. Although FOCUS is less susceptible to such manipulation than UPMiner, the consequences of its recommendations can be devastating because these tools provide code snippets. 
Thus, harmful training data poses a significant threat to the robustness of state-of-the-art RSSEs, including UPMiner, PAM, and FOCUS.

Du et al.~\cite{2023-Study-on-Adversarial-Attack-against-Pre-trained-Code-Models} conduct an empirical study on the effectiveness and efficiency of methods for generating adversarial examples for code pre-trained models and their performance on different code tasks.
They conduct a thorough evaluation of MHM~\cite{2020-MHM}, ACCENT~\cite{2019-Transferable-Clean-Label-Poisoning-Attacks-on-Deep-Neural-Nets}, WIR-Random~\cite{2022AnES}, ALERT~\cite{2022-Natural-Attack-for-Pre-trained-Models-of-Code}, and StyleTransfer~\cite{2022-Semantic-Robustness-of-Models-of-Source-Code} on the performance of CodeBERT~\cite{2020-CodeBERT}, CodeGPT~\cite{2021-CodeXGLUE}, and PLBART~\cite{2021unified}. They introduce the following research questions:
1) How do existing adversarial attack approaches perform against different PTMCs under various tasks?
In this research problem, they conduct a comprehensive comparison of adversarial attack methods based on effectiveness and efficiency. They found that random strategy adversarial attacks are more effective on pre-trained code models, while CodeGPT is generally more resilient to various attacks. Additionally, models tend to exhibit lower robustness in generation tasks compared to understanding tasks. 
There is a trade-off between effectiveness and efficiency in adversarial attacks. 
Attacks with higher success rates typically require more model queries. Furthermore, the efficiency of the attacks is also influenced by the number of identifiers in the target program.
2) What is the quality of adversarial examples generated by adversarial attacks?
Here’s the translation:
For this research question, they use the Identifier Change Rate (ICR) and Token Change Rate (TCR) to evaluate the number of tokens that have been replaced in the adversarial samples, and we use Average Code Similarity (ACS) and Average Edit Distance (AED) to assess the similarity between the adversarial tokens and the original tokens.
They find that there is also a trade-off between effectiveness and naturalness in adversarial attacks.
Effective attacks generate less natural adversarial samples. Generally, alternative strategies such as context-based identifier prediction and cosine similarity-based substitutions produce higher-quality samples than random replacements.
3) How do the contexts of the perturbed identifiers affect the adversarial attacking performance?
In this research question, they consider the statements in which the identifiers are located as context and investigate whether perturbing the identifiers in different contexts affects the effectiveness of the attacks. 
They choose five commonly used statements in code for investigation: Return, If, Throw, Try, and For statements. 
In addition to these types of statements, they also examine the impact of modifying only method names and parameters.
Ultimately, they find that the context of identifiers can significantly affect the effectiveness of attacks.
This indicates that disturbance strategies should consider the context of identifiers to achieve more effective attacks.

\subsection{Effectiveness and robustness of adversarial training for \abbr{}s}
\label{subsec:empirical_studies_on_adversarial_defenses}

Li et al.~\cite{2023-A-comparative-study-of-adversarial-training-methods} conduct an empirical study on the effectiveness and robustness of adversarial training for protecting \abbr{}s. First, they collect and categorize five types of adversarial training methods for \abbr{}s: mixing directly, composite loss, adversarial fine-tuning, min-max + composite loss, and min-max. Next, they conduct empirical evaluations of these categorized adversarial training methods on the code summarization task using Code2Seq~\cite{2019-code2seq} and the code authorship attribution task using Abuhamad~\cite{2018-Large-Scale-and-Language-Oblivious-Code-Authorship-Identification}.
Specifically, they investigate the following research questions:
1) How does learning sample obtaining strategy during training impact model performance?
Different adversarial training methods may adopt various strategies to acquire the samples to be learned in the current training iteration. For example, perturbing the input samples or selecting samples that maximize the loss for learning. They evaluate these two strategies to study their impact on the performance of \abbr{}s.
They find that the impact of sample obtaining strategies for \abbr{} varies depending on the task, and no single strategy is suitable for all tasks. However, the performance of optimization objective based methods using the min-max strategy is related to the number of times perturbed samples are regenerated. Therefore, they suggest that extending training time can be traded off for higher model robustness.
2) How does loss computation impact model performance?
Some adversarial training methods use ordinary loss, where all terms are computed equally for all samples. Other methods use composite loss, where some terms are computed for clean samples and others for adversarial or perturbed samples. 
They find that composite loss is generally more effective than ordinary loss. However, in terms of model robustness, composite loss performs better when using object-oriented methods, while ordinary loss demonstrates stronger robustness when combined with data augmentation-like methods.
3) How does input data impact model performance?
Some adversarial training methods require the input to be adversarial samples, while others require the use of both adversarial and perturbed samples. Even within the same category of samples, the composition may vary. They find that the input data used in adversarial training has a significant impact on the model's robustness, with adversarial samples being the key factor, rather than perturbed samples. When generating adversarial samples, the type of transformations is highly correlated with the final model's robustness. They recommend including data with various transformations to cover potential attacks that could reduce robustness.
4) How does input model impact model performance?
They evaluate the impact of input models on the final model's robustness from two aspects: the influence of pre-trained models and the number of fine-tuning epochs. They find that pre-trained models outperform non-pre-trained models in terms of training time and can surpass composite loss methods in robustness, achieving comparable robustness to direct mixing methods. Additionally, for adversarial fine-tuning, only a few fine-tuning epochs are required on pre-trained models, as increasing the number of fine-tuning epochs does not lead to better model performance.

\summary{Existing empirical studies on the security of \abbr{}s primarily focus on two aspects: 1) A comprehensive evaluation of the performance of backdoor/adversarial attacks on different code-related tasks and various \abbr{}s, including the success rate of attacks and their impact on model performance. 2) The effect of different components of defenses on both the effectiveness of the defense and model performance. For both attack methods and defense methods targeting \abbr{}s, empirical studies focus on their effectiveness, their impact on model performance, and the trade-offs between effectiveness and model performance.}

\section{Answer to RQ4: Experimental Setting and Evaluation}
\label{sec:Answer_to_RQ4}

\subsection{RQ4.1: Experimental Datasets}
\label{subsec:datasets}

\begin{table*}[!t]
    \centering
    \tabcolsep=2pt
    \caption{A summary of experimental datasets used in \abbr{} security research.}
    \label{tab:datasets_study}
    \resizebox{0.95\linewidth}{!}{
    \begin{tabular}{llllll}
        \toprule

        Dataset & Year & \makecell[l]{Programming \\ Language} & Data Source & Download Link & Study \\
       
        \midrule 

        BigCloneBench~\cite{2014-BigCloneBench} & 2014 & Java & GitHub & \url{https://github.com/clonebench/BigCloneBench} & \cite{2022-Natural-Attack-for-Pre-trained-Models-of-Code, 2023-DIP,2023-Code-Difference-Guided-Adversarial-Example-Generation-for-Deep-Code-Models,2023-RNNS,2023-Adversarial-Attacks-on-Code-Models, 2023-OSEQL,2024-poison-attack-and-poison-detection, 2024-investigating-adv,2024-CodeTAE} \\
        
        \midrule
        
        OJ dataset~\cite{2014-TextCNN} & 2016 & C++ & OJ Platform & \url{http://programming.grids.cn} & \cite{2020-MHM,2021-Generating-adversarial-examples-of-source-code-classification-models-via-q-learning-based-markov-decision-process,2022-Towards-Robustness-of-Deep-Program-Processing-Models,2023-Code-Difference-Guided-Adversarial-Example-Generation-for-Deep-Code-Models,2024-STRUCK} \\ 
        
        \midrule

        CodeSearchNet~\cite{2019-CodeSearchNet} & 2019 & \makecell[l]{Go \\ Java \\ JavaScript \\ PHP \\ Python \\ Ruby} & GitHub & \url{https://github.com/github/CodeSearchNet} & \makecell[l]{\cite{2022-Semantic-Preserving-Adversarial-Code-Comprehension,2023-multi-target-backdoor-attacks} \\ 
        \cite{2022-Semantic-Preserving-Adversarial-Code-Comprehension,2023-multi-target-backdoor-attacks,2023-BadCS,2023-BADCODE,2023-CodeAttack,2022-Generating-Adversarial-Source-Programs-Using-Important-Tokens-based-Structural-Transformations, 2024-dece} \\
        \cite{2022-Semantic-Preserving-Adversarial-Code-Comprehension,2023-multi-target-backdoor-attacks}\\ 
        \cite{2022-Semantic-Preserving-Adversarial-Code-Comprehension,2023-multi-target-backdoor-attacks,2023-CodeAttack}\\
        \cite{2022-Semantic-Preserving-Adversarial-Code-Comprehension,2023-multi-target-backdoor-attacks,2024-stealthy-backdoor-attack,2022-you-see-what-I-want-you-to-see,2022-TABS,2022-Semantic-Robustness-of-Models-of-Source-Code, 2023-BadCS,2023-BADCODE,2023-CodeAttack,2022-backdoors-in-neural-models-of-source-code, 2024-dece}\\
        \cite{2022-Semantic-Preserving-Adversarial-Code-Comprehension,2023-multi-target-backdoor-attacks}
        } \\
        
        \midrule 

        Code2Seq~\cite{2019-code2seq} & 2019 & Java & GitHub & \url{https://github.com/tech-srl/code2seq\#datasets} & \cite{2020-STRATA,2019-Adversarial-Authorship-Attribution-in-Open-Source-Projects,2021-A-Search-Based-Testing-Framework,2022-Semantic-Robustness-of-Models-of-Source-Code,2022-Generating-Adversarial-Source-Programs-Using-Important-Tokens-based-Structural-Transformations} \\ 

        \midrule
        
        Devign~\cite{2019-Devign} & 2019 & Java & GitHub & \url{https://github.com/rjust/defects4j} & \cite{2023-RNNS, 2023-OSEQL, 2024-poison-attack-and-poison-detection} \\ 
        
        \midrule

        \makecell[l]{Google Code\\ Jam (GCJ)} & 2020 & \makecell[l]{C++ \\ Java \\ Python} & OJ Platform & \url{https://codingcompetitions.withgoogle.com/codejam} & \makecell[l]{\cite{2019-Misleading-Authorship-Attribution,2023-SHIELD,2022-RoPGen} \\
        \cite{2019-Adversarial-Authorship-Attribution-in-Open-Source-Projects,2022-RoPGen} \\
        \cite{2023-DIP,2022-Natural-Attack-for-Pre-trained-Models-of-Code,2023-Code-Difference-Guided-Adversarial-Example-Generation-for-Deep-Code-Models,2023-Adversarial-Attacks-on-Code-Models}
        }\\ 
        
        \midrule 

        CodeXGLUE~\cite{2021-CodeXGLUE} & 2021 & \makecell[l]{Java \\ C/C++ \\ Python \\ PHP \\ JavaScript \\ Ruby \\ Go} & GitHub & \url{https://github.com/microsoft/CodeXGLUE} & \makecell[l]{\cite{2023-multi-target-backdoor-attacks, 2024-dece}\\
        \cite{2023-multi-target-backdoor-attacks,2023-Adversarial-Attacks-on-Code-Models}\\
        \cite{2023-multi-target-backdoor-attacks,2023-multi-target-backdoor-attacks}\\ 
        \cite{2023-multi-target-backdoor-attacks}\\
        \cite{2023-multi-target-backdoor-attacks}\\
        \cite{2023-multi-target-backdoor-attacks}\\ 
        \cite{2023-multi-target-backdoor-attacks}
        } \\
        
        \midrule 

        CodeQA~\cite{2021-CodeQA} & 2021 & \makecell[l]{Java \\ Python} & GitHub & \url{https://github.com/jadecxliu/codeqa} & \makecell[l]{\cite{2022-Semantic-Preserving-Adversarial-Code-Comprehension} \\ \cite{2022-Semantic-Preserving-Adversarial-Code-Comprehension}}\\
        
        \midrule 
        
        APPS~\cite{2021-Measuring-Coding-Challenge-Competence-With-APPS} & 2021 & Python & OJ Platform & \url{https://people.eecs.berkeley.edu/hendrycks/APPS.tar.gz} & \cite{2023-How-Robust-Is-a-Large-Pre-trained-Language-Model-for-Code-Generation} \\ 

        \midrule
        
        Shellcode\_IA32~\cite{2021-Shellcode_IA32} & 2021 & \makecell[l]{ASM \\ instruction} & GitHub & \url{https://github.com/dessertlab/Shellcode_IA32} & \cite{2023-Enhancing-Robustness-of-AI-Offensive-Code-Generators} \\ 
        
        \midrule
        
        SecurityEval~\cite{2022-SecurityEval-dataset} & 2022 & Python & GitHub & \url{https://github.com/s2e-lab/SecurityEval} & \cite{2024-Vulnerabilities-in-AI-Code-Generators,2023-Deceptprompt} \\ 
        
        \midrule

        LLMSecEval~\cite{2023-LLMSecEval} & 2023 & \makecell[l]{Python \\ C}& GitHub & \url{https://github.com/tuhh-softsec/LLMSecEval} & \cite{2024-Vulnerabilities-in-AI-Code-Generators} \\
        
        \midrule 
        
        PoisonPy~\cite{2019-code2seq} & 2023 & Python & GitHub & not yet published & \cite{2024-Vulnerabilities-in-AI-Code-Generators} \\ 
        
        \bottomrule
    \end{tabular}
    }
\end{table*}


In this section, we present the experimental datasets commonly used in \abbr{} security research.
Table~\ref{tab:datasets_study} summarizes these datasets along with key information and is intended to support future research by providing convenient access to commonly used benchmark datasets.

\indent
\textbf{\circled{1} BigCloneBench.}
BigCloneBench is proposed by Svajlenko et al.~\cite{2014-BigCloneBench}. It is used for big data clone detection and clone search. It comprises known true and false clone fragments from the IJaDataset 2.0~\footnote{\url{http://secold.org/projects/seclone}}, a large data project repository. BigCloneBench covers ten functionalities, including 6 million true clone pairs and 260,000 false clone pairs.

\textbf{\circled{2} Open Judge (OJ) dataset.}
The Open Judge (OJ) dataset is introduced by Mou et al. ~\cite{2016-Convolutional-Neural-Networks-over-Tree-Structures}. This dataset serves as a source code classification benchmark. Comprising 52,000 C/C++ code files labeled with problem numbers, it includes 104 classes, each containing 500 code files.

\textbf{\circled{3} CodeSearchNet.}
CodeSearchNet is proposed by Hamel et al.~\cite{2019-CodeSearchNet}. This dataset consists of 99 natural language queries with about 4k expert relevance annotations of likely results from CodeSearchNet Corpus which contains about 6 million functions from open-source code spanning six programming languages (Go,
Java, JavaScript, PHP, Python, and Ruby). The CodeSearchNet Corpus also contains automatically generated query-like natural language for 2 million functions, obtained from mechanically scraping and preprocessing associated function documentation.

\textbf{\circled{4} Devign.}
Devign is proposed by Zhou et al.~\cite{2019-Devign}. This dataset is used for defect detection tasks. The authors initially collected functions from four large open-source projects written in the C language: Linux Kernel, QEMU, Wireshark, and FFmpeg. They performed Commit Filtering to exclude commits unrelated to security. Subsequently, a team of four professional security researchers conducted two rounds of data annotation and cross-validation. To ensure the quality of data labeling, the team gathered security-related commits and marked them as Vulnerability-Fix Commits (VFCs) or Non-Vulnerability-Fix Commits (non-VFCs). They then extracted functions vulnerable or non-vulnerable to attacks directly from the labeled commits. Vulnerability-Fix Commits (VFCs) refer to commits fixing potential vulnerabilities, and vulnerable functions can be extracted from the source code of versions before the revisions made in the commits. Non-vulnerability-fix commits (non-VFCs) are commits that do not fix any vulnerabilities, and non-vulnerable functions can be extracted from the source code before the modifications. The final dataset comprises 21,000 training samples, 2,700 validation samples, and 2,700 test samples.

\textbf{\circled{5} Google Code Jam (GCJ).}
Google Code Jam (GCJ) is an annual multi-round international programming competition that requires participants to solve programming challenges in each round. The GCJ dataset is sourced from the Google Code Jam programming competition website and is divided into two categories: GCJ-C++ and GCJ-Java. The GCJ-C++ dataset consists of 1,632 C++ program files written by 204 participants, with each participant having 8 program files corresponding to 8 programming challenges. On average, each program file contains 74 lines of code. The GCJ-Java dataset includes 2,396 Java files from 74 participants, with an average of 139 lines of code per file.

\textbf{\circled{6} CodeXGLUE.}
CodeXGLUE is proposed by Lu et al.~\cite{2021-CodeXGLUE}. It is a benchmark dataset for program understanding. CodeXGLUE is collected from 14 common datasets, including BigCloneBench, POJ-104, and Devign, and is divided into four categories: Code-Code, Text-Code, Code-Text, and Text-Text. These categories cover 10 downstream tasks, including code clone detection, code defect detection, cloze test, code completion, code repair, code translation, NL code search, text-to-code generation, code summarization, and documentation translation.

\textbf{\circled{7} CodeQA.}
CodeQA is introduced by Liu et al.~\cite{2021-CodeQA}. It is used for code question-answering (QA) tasks. Code question-answering involves generating textual answers given a code snippet and a question. In the dataset construction process, the authors select Java and Python datasets from GitHub, ensuring that QA pairs are natural and clean. They then chose appropriate comments from these datasets to generate QA pairs. Using syntactic rules and semantic analysis, the authors transform comments into QA pairs. The Java dataset in the CodeQA dataset contains 119,778 QA pairs, while the Python dataset contains 70,085 QA pairs.

\textbf{\circled{8} APPS.}
APPS is introduced by Hendrycks et al.~\cite{2021-Measuring-Coding-Challenge-Competence-With-APPS}. It serves as an evaluation benchmark for code generation tasks. It comprises problems collected from various OJ websites.
This dataset assesses the model's ability to write syntactically correct algorithm programs. It covers simple introductory problems, interview-level questions, and coding competition challenges. The dataset includes a total of 10,000 problems with an average length of 293.2 words, 131,777 test cases for checking solutions, and 232,421 human-written real solutions. The data is evenly split into training and testing sets, each containing 5,000 problems. In the testing set, each problem has multiple test cases, with an average of 21.2 test cases per problem.

\textbf{\circled{9} Shellcode\_IA32.}
Shellcode\_IA32 is proposed by Liguori et al.~\cite{2021-Shellcode_IA32} in 2021. This dataset consists of IA-32 (x86 Intel Architecture 32-bit version) assembly language instructions with detailed English annotations. The dataset comprises a total of 3,200 pairs of assembly code snippets and comments. Considering the variability in natural language descriptions, the dataset describes different samples in English and uses comments written by the developers of the collected programs as natural language descriptions. This dataset represents the largest security-oriented code collection available for code generation to date.

\textbf{\circled{10} SecurityEval.}
SecurityEval is introduced by Siddiq et al.~\cite{2022-SecurityEval-dataset}. This dataset is designed to evaluate the security of code generation models. The dataset includes 130 Python code snippets covering 75 vulnerability types (CWE). Released in JavaScript Object Notation Line (JSONL) format, each line contains a JSON object with three key-value pairs: ID, Prompt, and Insecure Code. ID uniquely identifies the sample, Prompt provides a partial source code template usable as inputs for code generation models, and Insecure Code represents potentially vulnerable code examples generated by the model. This dataset is suitable for inputting source code prompts to LLM and evaluating the security of the generated code.

\textbf{\circled{11} LLMSecEval.}
In 2023, Tony et al.~\cite{2023-LLMSecEval} propose the LLMSecEval dataset to assess the security of code generation models. The dataset includes 150 natural language(NL) prompts covering the top 25 common defect types (CWE) according to MITRE. Derived from Chen et al.'s HumanEval data, the authors preprocess the dataset, filtering out invalid descriptions, removing blank content, eliminating entries with extensive code snippets, and excluding content that does not explain the input code's functionality. 150 NL prompts are compiled into CSV and JSON files, each row consisting of CWE name, NL Prompt, Source Code Filepath, Vulnerable, Language, Quality Metrics, and Secure Code Samples. CWE name indicates the defect type. NL Prompt is used to prompt model-generated code. Source Code Filepath points to the path of the source code file generated from the prompt. Vulnerable marks the field indicating defective code. Language indicates the source language of the prompt. Quality Metrics provide scores for four evaluation metrics. Secure Code Samples present secure code examples corresponding to each NL prompt.

\textbf{\circled{12} PoisonPy.}
In 2023, Cotroneo et al.~\cite{2019-code2seq} introduce the PoisonPy dataset for evaluating the security of \abbr{}s for code generation. PoisonPy comprises 823 samples, including 568 secure code snippets and 255 defective code snippets. The 255 defective samples encompass 109 Taint Propagation Issues (TPI), 73 Insecure Configuration Issues (ICI), and 73 Data Protection Issues (DPI). To construct the dataset, the authors combine available benchmark datasets SecurityEval and LLMSecEval to evaluate the security of AI-generated code. The original datasets are combinations of NL-Prompt, docstrings, and code used to evaluate AI code generators, making them unsuitable for fine-tuning models. The authors segment the collected code examples into multiple fragments, separate vulnerable code lines from secure lines, and enrich code descriptions. Additionally, to adjust the dataset's poisoning rate, for each vulnerable fragment, the authors provide an equivalent secure version by implementing potential mitigations for each CWE, without altering code descriptions.

Currently, commonly used datasets for backdoor attack and defense research on \abbr{}s include CodeSearchNet, CodeXGLUE, and Code2Seq. For adversarial attack and defense research, commonly used datasets include CodeSearchNet, CodeXGLUE, BigCloneBench, Devign, APPS, CodeQA, and GCJ. 

\subsection{RQ4.2: Experimental \abbr{}s}
\label{subsec:experimental_LM4Code}
To better understand the behavior and vulnerabilities of \abbr{}s, numerous \abbr{}s have been developed and applied in the field of program analysis and code-related tasks. 
In this section, we introduce and summarize the experimental \abbr{}s used in adversarial and backdoor research on \abbr{} security.
Table~\ref{tab:CodeLM_study} provides a detailed list of experimental models used in the field of \abbr{}s security.

\begin{table}[!t]
    \centering
    \caption{A summary of experimental models used in \abbr{} security research.}
    \label{tab:CodeLM_study}
    \resizebox{0.95\linewidth}{!}{
        \begin{tabular}{llll}
            \toprule
            \textbf{LM} & \textbf{Year} & \textbf{URL} & \textbf{Study} \\ 
            
            \midrule

            RNN & 1990 & - & \cite{2021-A-Practical-Black-Box-Attack-on-Source-Code-Authorship-Identification-Classifiers,2023-adv-binary} \\ 
            \midrule

            BiLSTM & 1997 & - & \cite{2020-MHM,2023-AdVulCode} \\ 
            \midrule

            BiRNN & 1997 & - & \cite{2022-you-see-what-I-want-you-to-see,2023-BadCS} \\ 
            \midrule

            LSTM & 1997 & - & \cite{2019-Misleading-Authorship-Attribution,2021-Generating-adversarial-examples-of-source-code-classification-models-via-q-learning-based-markov-decision-process,2022-Adversarial-Robustness-of-Deep-Code-Comment-Generation,2023-CodeBERT-Attack,2024-STRUCK,2020-Adversarial-Robustness-for-code,2022-Towards-Robustness-of-Deep-Program-Processing-Models,2023-AdvBinSD,2024-poison-attack-and-poison-detection} \\ 
            \midrule

            CNN & 1998 & - & \cite{2021-Deceiving-neural-source-code-classifiers-finding-adversarial-examples-with-grammatical-evolution,2023-SHIELD,2023-Evolutionary-Approaches-for-Adversarial-Attacks-on-Neural-Source-Code-Classifiers} \\ 
            \midrule

            Random Forest & 2001 & - & \cite{2019-Misleading-Authorship-Attribution,2021-A-Practical-Black-Box-Attack-on-Source-Code-Authorship-Identification-Classifiers,2021-Deceiving-neural-source-code-classifiers-finding-adversarial-examples-with-grammatical-evolution,2024-investigating-adv,2018-adv-binaries,2021-Explanation-Guided-Backdoor-Poisoning-Attacks} \\ 
            \midrule

            GNN & 2005 & - & \cite{2022-Adversarial-Robustness-of-Deep-Code-Comment-Generation,2023-adv-binary} \\ 
            \midrule

            GCN & 2013 & - & \cite{2024-STRUCK,2020-Adversarial-Robustness-for-code} \\ 
            \midrule

            GRU & 2014 & - & \cite{2024-STRUCK,2022-Towards-Robustness-of-Deep-Program-Processing-Models} \\ 
            \midrule

            GGNN & 2015 & - & \cite{2023-Discrete-Adversarial-Attack,2023-AdVulCode,2024-STRUCK,2020-Adversarial-examples-for-models-of-code} \\ 
            \midrule            

            Transformer & 2017 & - & \cite{2022-you-see-what-I-want-you-to-see,2024-poison-attack-and-poison-detection,2023-BadCS,2023-PELICAN} \\ 
            \midrule

            Seq2Seq & 2017 & \url{https://github.com/google/seq2seq} & \cite{2021-Generating-Adversarial-Computer-Programs,2022-Semantic-Robustness-of-Models-of-Source-Code,2022-Generating-Adversarial-Source-Programs-Using-Important-Tokens-based-Structural-Transformations,2022-backdoors-in-neural-models-of-source-code,2023-Enhancing-Robustness-of-AI-Offensive-Code-Generators,2024-Vulnerabilities-in-AI-Code-Generators} \\ 
            \midrule

            DL-CAIS & 2018 & - & \cite{2022-RoPGen,2023-SHIELD} \\ 
            \midrule
            
            ASTNN & 2019 & \url{https://github.com/zhangj111/astnn} & \cite{2020-MHM,2021-Generating-adversarial-examples-of-source-code-classification-models-via-q-learning-based-markov-decision-process,2022-Towards-Robustness-of-Deep-Program-Processing-Models} \\ 
            \midrule
            
            Code2Seq & 2019 & \url{https://github.com/tech-srl/code2seq} & \cite{2020-STRATA,2021-A-Search-Based-Testing-Framework,2022-backdoors-in-neural-models-of-source-code,2022-Semantic-Robustness-of-Models-of-Source-Code} \\ 
            \midrule
            
            Code2Vec & 2019 & \url{https://github.com/tech-srl/code2vec} & \cite{2020-Adversarial-examples-for-models-of-code,2021-A-Search-Based-Testing-Framework,2023-Discrete-Adversarial-Attack} \\ 
            
            \midrule

            GPT-2 & 2019 & \url{https://huggingface.co/openai-community/gpt2} & \cite{2023-How-Robust-Is-a-Large-Pre-trained-Language-Model-for-Code-Generation,2021-you-autocomplete-me} \\ 
            \midrule
            
            CodeBERT & 2020 & \url{https://huggingface.co/microsoft/codebert-base} & \makecell[l]{\cite{2021-A-Search-Based-Testing-Framework, 2022-TABS,2022-Semantic-Preserving-Adversarial-Code-Comprehension,2022-Towards-Robustness-of-Deep-Program-Processing-Models, 2022-Natural-Attack-for-Pre-trained-Models-of-Code, 2023-CodeAttack, 2023-Discrete-Adversarial-Attack, 2023-Adversarial-Attacks-on-Code-Models, 2023-DIP} \\ \cite{2023-Enhancing-Robustness-of-AI-Offensive-Code-Generators,2023-RNNS, 2023-CodeBERT-Attack,2023-Code-Difference-Guided-Adversarial-Example-Generation-for-Deep-Code-Models,2024-poison-attack-and-poison-detection, 2022-you-see-what-I-want-you-to-see, 2023-BADCODE, 2024-Vulnerabilities-in-AI-Code-Generators} \\\cite{2024-stealthy-backdoor-attack, 2023-BadCS,2024-CodeTAE,2024-STRUCK,2024-MOAA,2024-dece,2023-OSEQL}} \\ 
            \midrule

            GraphCodeBERT & 2020 & \url{https://huggingface.co/microsoft/graphcodebert-base} & \cite{2022-Natural-Attack-for-Pre-trained-Models-of-Code,2022-Semantic-Preserving-Adversarial-Code-Comprehension,2023-CodeBERT-Attack,2023-CodeAttack,2023-Adversarial-Attacks-on-Code-Models,2023-DIP,2023-RNNS,2023-RNNS,2023-BadCS,2023-Code-Difference-Guided-Adversarial-Example-Generation-for-Deep-Code-Models,2024-CodeTAE,2024-STRUCK,2024-MOAA} \\ 
            \midrule
            
            CodeT5 & 2021 & \url{https://huggingface.co/Salesforce/codet5-base} & \cite{2023-CodeAttack,2023-DIP,2023-RNNS,2024-radar,2023-BADCODE,2023-Code-Difference-Guided-Adversarial-Example-Generation-for-Deep-Code-Models,2024-stealthy-backdoor-attack,2023-multi-target-backdoor-attacks,2024-CodeTAE,2024-MOAA,2024-dece,2023-OSEQL} \\ 
            \midrule
            
            CodeT5+ & 2021 & \url{https://huggingface.co/Salesforce/codet5-base} & \cite{2023-Enhancing-Robustness-of-AI-Offensive-Code-Generators,2024-Vulnerabilities-in-AI-Code-Generators,2024-radar} \\ 
            \midrule
            
            PLBART & 2021 & \url{https://github.com/wasiahmad/PLBART} & \cite{2024-stealthy-backdoor-attack,2023-multi-target-backdoor-attacks,2024-radar,2023-OSEQL} \\ 
            \midrule

            CodeGen & 2022 & \url{https://huggingface.co/Salesforce/codegen-350M-multi} & \cite{2023-Large-Language-Models-for-Code-Security-Hardening-and-Adversarial-Testing, 2024-radar, 2024-dece, 2024-TrojanPuzzle} \\
            \midrule

            GPT-3.5 & 2022 & \url{https://openai.com/} & \cite{2023-Transfer-Attacks-and-Defenses-for-Large-Language-Models-on-Coding-Tasks} \\ 
            \midrule

            GPT-4 & 2023 & \url{https://openai.com/} & \cite{2023-Transfer-Attacks-and-Defenses-for-Large-Language-Models-on-Coding-Tasks} \\ 
            \bottomrule
            
        \end{tabular}
    }
\end{table}

\textbf{\circled{1} CNN, RNN, LSTM, GRU, BiLSTM, BiRNN.}
CNN (Convolutional Neural Network) is a class of neural networks that excels at processing structured data, typically used in image recognition tasks, but CNN can also capture local features in code analysis tasks. By applying convolution operations to the abstract syntax tree (AST) of code, CNN can extract useful features from the code structure, making it applicable to tasks such as code classification and vulnerability detection. RNN (Recurrent Neural Network) and its variants, LSTM (Long Short-Term Memory), GRU (Gated Recurrent Unit), and BiLSTM (Bidirectional LSTM), are particularly suited for handling sequential data, especially for code generation and code understanding tasks. They pass information along the time dimension, capturing long-term dependencies in the code. BiRNN and BiLSTM further enhance context understanding by processing code sequences in both forward and backward directions, making them widely used in tasks such as code summarization and function classification.

\textbf{\circled{2} GNN, GCN, GGNN.} 
Graph Neural Networks (GNN) are designed to process graph-structured data, making them ideal for tasks like code analysis, such as Abstract Syntax Trees (ASTs) and Control Flow Graphs (CFGs). GNNs capture relationships between nodes, enabling a deeper understanding of code structure. Building on GNNs, Graph Convolutional Networks (GCN) use convolution operations to focus on local structural features, making them effective for code classification and similarity detection. Gated Graph Neural Networks (GGNN) enhance GNNs by incorporating LSTM-like gating mechanisms to better manage long-range dependencies, improving performance on complex code structures in tasks like vulnerability detection and pattern recognition.

\textbf{\circled{3} Random Forest.}
Random Forest is an ensemble learning model based on decision trees, improving accuracy and robustness by building multiple decision trees and combining their predictions. In code analysis, Random Forest is commonly used for static code analysis tasks such as code defect detection and classification. By randomly sampling code features, it reduces the risk of overfitting a single decision tree, making it particularly suited for handling high-dimensional and structurally complex code data.

\textbf{\circled{4} Transformer.}
The Transformer is a deep learning model based on the attention mechanism, widely applied in natural language processing and code generation tasks. Unlike traditional RNNs, the Transformer can process entire input sequences in parallel using a global attention mechanism, making it more efficient and performant when handling long sequences. The self-attention mechanism of the Transformer effectively captures long-range dependencies and complex syntax structures in code, making it excel in tasks such as code generation, code translation, and code repair. Variants of the Transformer, such as CodeBERT, GraphCodeBERT, and CodeT5, have demonstrated strong capabilities in code understanding and generation.

\textbf{\circled{5} DL-CAIS.}
DL-CAIS (Deep Learning Context-Aware Intelligent System) is an intelligent system that integrates deep learning and context-aware technology, used for program analysis and code detection tasks. DL-CAIS combines static and dynamic code analysis, leveraging deep learning models (such as CNNs and RNNs) to extract high-level features from code. It is applied in scenarios such as malware detection, code vulnerability identification, and automatic code generation. By deeply understanding the code context, the system significantly improves the precision and robustness of code analysis tasks.

\textbf{\circled{6} ASTNN.}
ASTNN (Abstract Syntax Tree Neural Networks) is a model designed specifically for code-related tasks, leveraging the abstract syntax tree (AST) to represent code structures. It is often applied in tasks such as code classification and function classification.

\textbf{\circled{7} Code2Vec.}
Code2Vec is the neural network model designed to represent code snippets as fixed-length vectors. It works by extracting paths from the abstract syntax tree (AST) of a code snippet and embedding them into a continuous vector space. Code2Vec captures both syntactic and semantic information, making it useful for tasks like method name prediction, variable renaming, and code similarity detection. The model's ability to create vector representations from code allows it to generalize across different programming tasks.

\textbf{\circled{8} Code2Seq.}
Similar to Code2Vec, Code2Seq is a model that converts code into a sequence representation by extracting paths from the AST. However, instead of generating a single vector, Code2Seq outputs a sequence of tokens, making it more suitable for tasks like code summarization, where generating a description of code functionality is required. Code2Seq is particularly effective in tasks that require sequence generation, such as method name generation and documentation generation, by understanding the structure and meaning of the code.

\textbf{\circled{9} CodeBERT and GraphCodeBERT.}
Pre-trained code models based on the BERT architecture, designed for program understanding, code summarization, vulnerability detection, and code search. GraphCodeBERT enhances the model's understanding of code by incorporating graph neural networks (GNNs) to capture both syntax and semantic structures in code.

\textbf{\circled{10} CodeT5 and CodeT5+.}
CodeT5 is a transformer-based model specifically designed for code-related tasks, built on the T5 architecture. 
CodeT5 follows a text-to-text format, allowing it to handle tasks such as code summarization, code generation, code translation, and code repair. It is pre-trained on a large dataset across multiple programming languages, which helps it capture both syntax and semantics. 
CodeT5 also utilizes masked span prediction and other pre-training techniques to improve its understanding of code structure. Its versatility and strong performance make it effective in various neural code model applications.

\textbf{\circled{11} PLBART.}
PLBART is a pre-trained sequence-to-sequence model specifically designed for programming languages based on the BART (Bidirectional and Auto-Regressive Transformers) architecture. PLBART is pre-trained using denoising autoencoding tasks on both programming languages and natural language. It excels in various code-related tasks such as code translation, generation, and repair by learning both the syntactic and semantic aspects of programming languages. Its bidirectional and autoregressive properties make it a powerful model for code understanding and manipulation tasks.

\textbf{\circled{12} CodeGen.}
CodeGen is a Transformer-based model designed specifically for code generation tasks. It is pre-trained on a large-scale dataset of mixed programming and natural languages, learning to generate corresponding code from natural language descriptions. CodeGen is suited for tasks such as automatic code completion, code generation, and code repair. Its autoregressive architecture allows it to predict subsequent code segments step by step during generation, improving the accuracy and syntactical consistency of the output.

\textbf{\circled{13} GPT-2, GPT-3.5, GPT-4.}
GPT-2 is an autoregressive language model based on the Transformer, primarily designed for natural language processing tasks. However, its powerful generation capabilities have also been widely applied to code generation tasks. GPT-3.5 and GPT-4 are enhanced versions of GPT-2, with larger model parameters and a more extensive training dataset, allowing them to handle more complex code generation tasks. The GPT series not only excels at automatic code completion but also learns the semantics and structure of code through large-scale pre-training, being widely applied to code generation, code repair, and code search tasks. As the model size increases, GPT-4 shows stronger cross-domain capabilities, especially in the understanding and generation of complex code.

\subsection{RQ4.3: Evaluation Metrics}
\label{subsec:performance_metrics}
To evaluate the performance of attack and defense techniques against \abbr{}s, existing studies have proposed many evaluation metrics. In this section, we summarize and present these metrics.
These metrics can be divided into two categories: those used for evaluating attack and defense techniques, and those used for assessing the performance of \abbr{}s themselves.

\subsubsection{Metrics used in attack and defense research on \abbr{}s.}
\

This section introduces the evaluation metrics used in attacks and defenses targeting \abbr{}s. These metrics are typically employed to measure attack success rates, model robustness, and the effectiveness of defense methods. \R{It is worth noting that these evaluation metrics are of interest to both attackers and defenders, as the same metrics can be used to assess the effectiveness of attack methods as well as the resilience of defense methods.}

\indent
\textbf{\circled{1} Backdoor False Positive Rate ($FPR$).}  
$FPR$ means the proportion of positive samples with wrong predictions to all positive samples. It is also called the false recognition rate and false alarm rate. This metric is used to evaluate the performance of a classification model and is often employed by backdoor defense methods. The calculation formula of $FPR$ is as follows:
\begin{equation}
    \mathrm{FPR}=\frac{FP}{TN+FP}
\end{equation}

\textbf{\D{\circled{6}}\R{\circled{2}} Attack Success Rate ($ASR$).} $ASR$ refers to the proportion of poisoned samples that are successfully predicted by the model to target label, which is frequently used to measure the effect of backdoor attacks \R{or defenses}. In general, the higher the $ASR$, the better the performance of the attack method\R{, and vice versa}. $M$ refers to clean model, $M'$ refers to backdoored model. $C$ refers to clean data, $C'$ refers to poisoned data. The calculation formula of $ASR$ is as follows:
\begin{equation}
    ASR =
    \frac{|\{C|M'(C')=y_{target}\wedge M(C)\neq y_{target}\}|}{|\{C\}|}
\end{equation}

\textbf{\D{\circled{2}}\R{\circled{3}} Average Normalized Rank ($ANR$).} This metric is used to measure the effectiveness of backdoor attacks \R{or defenses}\D{against} \R{on} code retrieval models. $ANR$ indicates how much the attack can improve the retrieval ranking of poisoned samples. The smaller $ANR$ value represents the more effective attack method\R{, and vice versa}. Specifically, $s'$ stands for the length of the injected trigger code snippet. $|S|$ stands for the length of the full ranking list. The calculation formula of $ANR$ is as follows:
\begin{equation}
    \mathrm{ANR} = \frac1{|Q|}\sum_{i=1}^{|Q|}\frac{Rank(Q_i, s^{\prime})}{|S|}
\end{equation}

Therefore, the number of queries is not only an important indicator for evaluating the efficiency of black-box adversarial attack methods but also a crucial metric for assessing the effectiveness of corresponding adversarial defense methods.

\textbf{\D{\circled{3}}\R{\circled{4}} Number of Queries.} This represents the average number of successful queries against the target model in an adversarial attack. For black-box adversarial attack methods, queries are the only way to access the target model. Therefore, the number of queries is \R{not only an}\D{ one of the} important indicators \D{to}\R{for} evaluat\D{e}\R{ing} the efficiency of black-box adversarial attack methods \R{, but also an important metric for assessing the effectiveness of corresponding adversarial defense methods}. Specifically, $q_i$ represents the number of queries for the $i_{th}$ successful attack. In detail, $i \in \{j|f(j) = 1\}$. The calculation formula of Number of Queries is as follows:
\begin{equation}
    Query=\frac{\sum q_i}{\sum f(i)}
\end{equation}

\begin{equation}
    f(j) = \begin{cases} 1, & \text{if  }M(C_j^{adv})\neq y_j\wedge M(C_j)=y_j\\
    0, & \text{otherwise}
    \end{cases}
\end{equation}

\textbf{\D{\circled{4}}\R{\circled{5}} Ratio of Perturbation ($Pert$).} This metric is used to represent the proportion of perturbation injected into the original source code during an adversarial attack. A lower perturbation ratio indicates that the generated adversarial samples have fewer perturbations. Specifically, $C_i^{adv}$ denotes for the adversarial example of $C_i$. $t(\cdot)$ denotes the number of tokens. The calculation formula of $Pert$ is as follows:
\begin{equation}
    Pert = \frac{\sum t(C_i^{adv})-t(C_i^{adv}\cap C_i)}{\sum t(C_i)}
\end{equation}

\textbf{\D{\circled{5}}\R{\circled{6}} Relative Degradation ($R_d$).} $R_d$ is used to evaluate the performance degradation of the model in the attack state. Specifically, $refs$ denotes for reference comment. $y$ denotes for original output. $y'$ denotes the output of the perturbed program. The calculation formula of $R_d$ is as follows:
\begin{equation}
    R_d = \frac{\mathrm{BLEU}(y, \mathrm{refs})-\mathrm{BLEU}(y^{\prime}, \mathrm{refs})}{\mathrm{BLEU}(y, \mathrm{refs})}
\end{equation}

\textbf{\D{\circled{6}}\R{\circled{7}} Valid Rate ($V_r$).} $V_r$ is defined as the percentage of adversarial samples that can be compiled~\cite{2016-Stealing-ML-Models-via-Prediction-APIs}. This metric is used to evaluate the quality of the adversarial samples generated \R{during adversarial attacks or adversarial defenses,}\D{quantity} as well as the efficiency of the generation process. Specifically, $Count_{valid}$ is the number of adversarial samples that can pass the compile check of the program. $Count_{all}$ is the number of all samples. The calculation formula of $V_r$ is as follows:
\begin{equation}
    V_r=\frac{Count_{valid}}{Count_{all}}
\end{equation}

\textbf{\D{\circled{7}}\R{\circled{8}} Success Rate ($S_r$).} $S_r$ is defined as the product of relative degradation ($R_d$) and valid rate ($V_r$), providing a comprehensive index of attack efficiency and sample quality generated. Essentially, a higher success rate indicates that the corresponding method can generate effective adversarial samples with better attack capabilities. The calculation formula of $S_r$ is as follows:
\begin{equation}
    S_r = R_d * V_r
\end{equation}

\textbf{\D{\circled{8}}\R{\circled{9}} Variable Change Rate ($VCR$).} $VCR$ is the proportion of variables that an attacker needs to modify for a successful attack. The lower the VCR, the better the attack method, and the more efficient the adversarial sample can be found only by modifying a small number of variables. 
\R{Conversely, the higher the $VCR$, the more effective the applied defense method.}
In the formula for \D{Variable Change Rate (VCR)}\R{$VCR$}, $n_i$ refers to the number of variables that an attacker has successfully modified in the $i$-th adversarial sample, while $m_i$ represents the total number of variables present in the $i$-th sample.
The calculation formula of $VCR$ is as follows:
\begin{equation}
    VCR=
    \frac{\sum_in_i}{\sum_im_i}
\end{equation}

\subsubsection{Metrics used in performance on \abbr{}s.}
\

This section focuses on the metrics used to evaluate the performance of \abbr{}s themselves, such as the accuracy of generated code, efficiency, and how closely the generated code aligns with human-written code.

\textbf{\circled{1} Clean Accuracy ($CA$).} Clean Accuracy means the accuracy of the model without any interference or attack. This metric is used to evaluate the performance of a model on clean data and is commonly used in backdoor attacks and adversarial attacks. For the backdoor attack, if the accuracy of the poisoned model is almost the same as that of the normal model, it can show that the attack method has good concealment. $M'$ refers to the backdoor model which is trained with the poisoned dataset. $C$ refers to clean data, $y$ refers to ground-true label. The calculation formula of $CA$ is as follows:
\begin{equation}
    CA= \frac{|\{C|M'(C)=y\}|}
        {|\{C\}|}
\end{equation}

\textbf{\circled{2} F1-score}. F1-score is the harmonic mean of precision and recall, and is commonly used to evaluate the performance difference of neural code classification models before and after attacks. The calculation formula of $F1-score$ is as follows:
\begin{equation}
    F1-score= \frac{2*\mathrm{precision*recall}}          {\mathrm{precision+recall}}
\end{equation}

\textbf{\circled{3} Mean Reciprocal Rank ($MRR$).} $MRR$ is the average of the reciprocal ranks of a set of search results, this metric provides an overall perspective to assess the validity of a search result. In general, the higher the $MRR$ value, the better the effect of model retrieval. Specifically, ${|Q|}$ is the size of the query set. $Rank_q$ is the rank of the code snippet that corresponds to the query $q$. The calculation formula of $MRR$ is as follows:
\begin{equation}
    MRR=\frac{1}{|Q|}\sum_{q\in Q}\frac{1}{Rank_q}
\end{equation}

\textbf{\circled{4} Bilingual Evaluation Understudy ($BLEU$).} BLEU~\cite{papineni-etal-2002-bleu} is used to evaluate the quality of machine-generated text by comparing the similarity between a machine-generated text and a human-generated reference text. It is a commonly used performance evaluation metric in generation tasks such as code summarization, code generation, and code completion. Specifically, $r$ is the number of words in a reference translation. $c$ is the number of words in a candidate translation. The calculation formula of $BLEU$ is as follows:
\begin{equation}
    BLEU=BP\cdot exp(\sum_{n=1}^N\omega_n\log(P_n))
\end{equation}
$BP$ stands for brevity penalty. When ${\displaystyle r\leq c}$, the brevity penalty ${\displaystyle BP=1}$, meaning that we do not punish long candidates and only punish short candidates. When ${\displaystyle r>c}$, the brevity penalty ${\displaystyle BP=e^{1-r/c}}$.
\begin{equation}
    BP = \begin{cases}1, c>r \\
    e^{1-\frac rc}, c\leq r&
    \end{cases}
\end{equation}
${P_n}$ is weighted n-gram precision, this quantity measures how many n-grams in the reference sentence are reproduced by the candidate sentence. where $n$ means the length of the n-gram, $C(i, i + n)$ is the n-gram from the position $i$ to the position $i + n$, and $Countclip(C(i, i + n))$ is the maximum number of n-grams co-occurring in a candidate code and a set of reference codes. $\mu_{n}^{i}$ denotes the weights of different keywords or n-gram.
\begin{equation}
    p_n=\frac{\sum\limits_{C\in\text{Candidates}} \sum\limits_ { i = 1 }^l\mu_n^i\cdot\text{Count}_{\text{clip}}(C(i,i+n))}{\sum\limits_{C'\in\text{Candidates}} ^ l \mu_n^i\cdot\text{Count}(C'(i,i+n))}
\end{equation}

\textbf{\circled{5} CodeBLEU.} $CodeBLEU$ is used to evaluate the quality of code generated by the model. This metric considers the functional and structural information of the code and weights the abstract syntax tree matching and data-flow matching scores. 
In the field of code synthesis, common evaluation metrics like $BLEU$ and perfect accuracy have limitations. $BLEU$, originally designed for natural language, overlooks the syntactic and semantic features of code, while perfect accuracy is too strict, underestimating outputs that share the same semantic logic.
To solve this problem, a new automatic evaluation metric called $CodeBLEU$ is created by Ren et al.~\cite{2020-CodeBLUE}. It absorbs the strengths of $BLEU$ in n-gram matching and further injects code syntax through an abstract syntax tree (AST) and code semantics through the data stream. The calculation formula of $CodeBLEU$ is as follows:
\begin{equation}
    \begin{aligned}
    CodeBLEU & = \alpha\cdot BLEU + \beta\cdot BLEU_{weight} + \gamma\cdot Match_{ast} + \delta\cdot Match_{df}
    \end{aligned}
\end{equation}
where $BLEU$ is calculated by standard $BLEU$~\cite{papineni-etal-2002-bleu}, $BLEU_{weight}$ is the weighted n-gram match, obtained by comparing the hypothesis code and the reference code tokens with different weights, $Match_{ast}$ is the syntactic AST match, exploring the syntactic information of code, and $Match_{df}$ is the semantic data-flow match, considering the semantic similarity between the hypothesis and the reference. The weighted n-gram match and the syntactic AST match are used to measure grammatical correctness, and the semantic data-flow match is used to calculate logic correctness.

\textbf{\D{\circled{7}}\R{\circled{6}} Top k Success Rate ($SuccessRate@k$).} This metric measures the average percentage of the retrieved model outputs that ranked in the top $k$ for the corresponding code snippet. In general, the higher the $SuccessRate@k$, the better the code retrieval model. Specifically, $|Q|$ is the size of the query set. $Rank_q$ refers to the ranking of the code snippet corresponding to the query $q$. $\delta(\cdot)$ is a function that returns 1 if the input is true, otherwise 0. The calculation formula of $SuccessRate@k$ is as follows:
\begin{equation}
    SuccessRate@k =\frac{1}{|Q|}\sum_{q\in Q}\delta({Rank}_q\leq k)
\end{equation}

\subsection{RQ4.4: Artifact Accessibility}
\label{subsec:replication_packages}

\begin{table*}[!t]
    \centering
    \caption{Replication open-source code repository links provided in \abbr{} studies.\R{check}}
    \resizebox{0.95\linewidth}{!}{
        \begin{tabular}{llll}
            \toprule
            \textbf{Technique} & \textbf{Year} & \textbf{Type} & \textbf{URL}  \\ 
            
            \midrule

            Ramakrishnan  et al.~\cite{2022-backdoors-in-neural-models-of-source-code} & 2022 & \makecell[l]{Data poisoning\\Pre-training defense} & \url{https://github.com/goutham7r/backdoors-for-code} \\ 
            
            \midrule

            Schuster et al.~\cite{2021-you-autocomplete-me} & 2021 & \makecell[l]{Data poisoning\\Model poisoning} & \url{https://github.com/ClonedOne/MalwareBackdoors} \\

            \midrule

            Severi et al.\cite{2021-Explanation-Guided-Backdoor-Poisoning-Attacks} & 2021 & Data poisoning & \url{https://github.com/ClonedOne/MalwareBackdoors} \\ 
            
            \midrule
            
            Wan et al.~\cite{2022-you-see-what-I-want-you-to-see} & 2022 & Data poisoning & \url{https://github.com/CGCL-codes/naturalcc} \\ 

            \midrule

            BadCode~\cite{2023-BADCODE} & 2023 & Data poisoning & \url{https://github.com/wssun/BADCODE} \\ 
            
            \midrule

            TrojanSQL~\cite{2023-TrojanSQL} & 2023 & Data poisoning & \url{https://github.com/ jc-ryan/trojan-sql} \\

            \midrule

            Cotroneo et al.~\cite{2024-Vulnerabilities-in-AI-Code-Generators} & 2023 & Data poisoning & \url{https://github.com/dessertlab/Targeted-Data-Poisoning-Attacks} \\ 
            
            \midrule

            Li et al.~\cite{2024-poison-attack-and-poison-detection} & 2024 & \makecell[l]{Data poisoning\\Pre-training defense} & \url{https://github.com/LJ2lijia/CodeDetector}   \\ 
            
            \midrule

            AFRAIDOOR~\cite{2024-stealthy-backdoor-attack} & 2024 & Data poisoning & \url{https://github.com/yangzhou6666/adversarial-backdoor-for-code-models}  \\ 
            
            \midrule

            TrojanPuzzle & 2024 & Data poisoning & \url{https://github.com/microsoft/CodeGenerationPoisoning} \\

            \midrule

            FDI~\cite{2024-FDI} & 2024 & Data poisoning & \url{https://github.com/v587su/FDI} \\

            \midrule

            LateBA~\cite{2024-LateBA} & 2024 & Data poisoning & \url{https://github.com/yxy-whu/LateBA} \\

            \midrule

            Li et al.~\cite{2023-multi-target-backdoor-attacks} & 2023 & Model poisoning & \url{https://github.com/Lyz1213/Backdoored_PPLM} \\ 
            
            \midrule

            PELICAN~\cite{2023-PELICAN} & 2023 & Model poisoning & \url{https://github.com/ZhangZhuoSJTU/Pelican} \\ 
            
            \midrule

            DeCE~\cite{2024-dece} & 2024 & In-training defense & \url{https://github.com/NTDXYG/DeCE}  \\ 
            
            \midrule
            
            DAMP~\cite{2020-Adversarial-examples-for-models-of-code} & 2020 & White-box attack & \url{https://github.com/tech-srl/adversarial-examples} \\ 
            
            \midrule
            
            Bielik et al.~\cite{2020-Adversarial-Robustness-for-code} & 2020 & White-box attack & \url{https://github.com/eth-sri/robust-code} \\ 
            
            \midrule

            Srikant et al.~\cite{2021-Generating-Adversarial-Computer-Programs} & 2021 & White-box attack & \url{https://github.com/ALFA-group/adversarial-code-generation} \\ 
            
            \midrule

            Quiring et al.~\cite{2019-Misleading-Authorship-Attribution} & 2019 & Black-box attack & \url{https://github.com/EQuiw/code-imitator} \\

            \midrule
            
            MHM~\cite{2020-MHM} & 2020 & Black-box attack & \url{https://github.com/Metropolis-Hastings-Modifier/MHM} \\ 
            
            \midrule
            
            Pour et al.~\cite{2021-A-Search-Based-Testing-Framework} & 2021 & Black-box attack & \url{https://github.com/MaryamVP/Guided-Mutation-ICST-2021} \\ 
            
            \midrule
            
            Ferretti et al.~\cite{2021-Deceiving-neural-source-code-classifiers-finding-adversarial-examples-with-grammatical-evolution} & 2021 & Black-box attack & \url{https://github.com/Martisal/adversarialGE} \\ 
            
            \midrule
            
            Anand et al.~\cite{2021-synthetic-code-generation} & 2021 & Black-box attack & \url{https://github.com/mtensor/neural_sketch} \\ 
            
            \midrule
            
            ACCENT~\cite{2022-Adversarial-Robustness-of-Deep-Code-Comment-Generation} & 2021 & Black-box attack & \url{https://github.com/zhangxq-1/ACCENT-repository} \\ 
            
            \midrule

            CARROT~\cite{2022-Towards-Robustness-of-Deep-Program-Processing-Models} & 2022 & \makecell[l]{Black-box attack\\Input modification} & \url{https://github.com/SEKE-Adversary/CARROT} \\ 
            
            \midrule
            
            ALERT~\cite{2022-Natural-Attack-for-Pre-trained-Models-of-Code} & 2022 & Black-box attack & \url{https://github.com/soarsmu/attack-pretrain-models-of-code} \\ 
            
            \midrule

            Li et al.~\cite{2022-RoPGen} & 2022 & \makecell[l]{Black-box attack\\Input modification} & \url{https://github.com/RoPGen/RoPGen} \\ 
            
            \midrule

            CodeAttack~\cite{2023-CodeAttack} & 2023 & Black-box attack & \url{https://github.com/reddy-lab-code-research/CodeAttack} \\ 
            
            \midrule

            RNNS~\cite{2023-RNNS} & 2023 & Black-box attack & \url{https://github.com/18682922316/RNNS-for-code-attack} \\

            \midrule

            CODA~\cite{2023-Code-Difference-Guided-Adversarial-Example-Generation-for-Deep-Code-Models} & 2023 & Black-box attack & \url{https://github.com/tianzhaotju/CODA} \\ 
            
            \midrule

            CodeBERT-Attack~\cite{2023-CodeBERT-Attack} & 2023 & Black-box attack & \url{https://gitfront.io/r/DrLC5417/5pBsYXgKinB3/CodeBERT-Attack/tree/codebert/} \\

            \midrule

            RADAR~\cite{2024-radar} & 2024 & \makecell[l]{Black-box attack\\Additional Models} & \url{https://github.com/NTDXYG/RADAR}  \\ 
            
            \midrule

            STRUCK~\cite{2024-STRUCK} & 2024 & Black-box attack & \url{https://github.com/zhanghaha1707/STRUCK} \\

            \midrule

            CodeTAE~\cite{2024-CodeTAE} & 2024 & Black-box attack & \url{https://github.com/yyl-github-1896/CodeTAE} \\

            \midrule

            MOAA~\cite{2024-MOAA} & 2024 & Black-box attack & \url{https://github.com/COLA-Laboratory/MOAA} \\ 

            \midrule
            
            MalwareTotal~\cite{2024-MalwareTotal} & 2024 & Black-box attack & \url{https://github.com/Optimus-He/MalwareTotal} \\

            \midrule
            
            Henkel et al.~\cite{2022-Semantic-Robustness-of-Models-of-Source-Code} & 2022 & Input modification & \url{https://github.com/jjhenkel/averloc} \\ 

            \midrule

            Cristina et al.~\cite{2023-Enhancing-Robustness-of-AI-Offensive-Code-Generators} & 2023 & Input modification & \url{https://github.com/dessertlab/Robustness-of-AI-Offensive-Code-Generators} \\
            
            \midrule

            SVEN~\cite{2023-Large-Language-Models-for-Code-Security-Hardening-and-Adversarial-Testing} & 2023 & Model modification & \url{https://github.com/eth-sri/sven}  \\ 
            
            \midrule
            
            SPACE~\cite{2022-Semantic-Preserving-Adversarial-Code-Comprehension} & 2022 & Additional models & \url{https://github.com/EricLee8/SPACE} \\ 
            
            \bottomrule
            
        \end{tabular}
        \label{tab:summary_studies_repositories}
    }
\end{table*}

In this section, we explore the open-source code repositories provided by the reviewed \abbr{} studies in their papers. 
The open-source code repositories offered by the authors will assist other researchers in reproducing the study and facilitate further utilization and research. The quality of these open-source code repositories depends on reproducibility, which refers to obtaining the same results using generally the same methods. 
Reproducibility is crucial for assessing the quality of research and validating its credibility, as it enhances our confidence in the results and helps us distinguish between reliable and unreliable findings. 
This is particularly important for technical methods that rely on datasets, data processing methods, and hyperparameter settings.

To verify the reproducibility of these \abbr{} studies, we systematically review each paper to ensure that the provided code links are publicly accessible and valid. 
We compile information from all the papers and include those with open-source code repositories along with their links in Table~\ref{tab:summary_studies_repositories} to facilitate future \abbr{} research.

\summary{In this section, we first introduce the experimental datasets used in \abbr{} security research, summarizing the commonly adopted datasets in both backdoor and adversarial attack and defense scenarios.
Next, we summarize the \abbr{}s evaluated in existing studies.
We then introduce the evaluation metrics used in attacks and defenses involving \abbr{}s, as well as the metrics used to assess the overall performance of the models themselves. Finally, we discuss artifact accessibility, highlighting the importance of open-source code repositories to ensure transparency and reproducibility in \abbr{} research, thereby facilitating the broader advancement of the field.}

\section{Challenges and Opportunities}
\label{sec:challenges_and_opportunities}

After analyzing and summarizing the existing research on \abbr{}s, we can suggest that further research will have both practical and academic significance. From a practical application perspective, these excellent techniques will undoubtedly help us improve the security of \abbr{}s. On the other hand, the security of \abbr{}s still faces many challenges that need to be addressed. We pose some of the challenges and opportunities identified from the relevant papers below.

\subsection{Challenges and Opportunities in Attacks against \abbr{}s}
\label{subsec:challenges_and_opportunities_in_attacks}

\noindent\textbf{Stealthiness of Backdoor Triggers.}
In backdoor attacks on \abbr{}s, the stealthiness of triggers is a key focus for attackers. They continually explore more covert trigger designs, evolving from early methods like dead code~\cite{2022-backdoors-in-neural-models-of-source-code, 2022-you-see-what-I-want-you-to-see} to variable/function names~\cite{2023-BADCODE, 2024-poison-attack-and-poison-detection}, and even adaptive triggers~\cite{2024-stealthy-backdoor-attack}, aiming to inject increasingly stealthy triggers into code. However, comprehensively evaluating the stealthiness of triggers remains a challenging task. Current research methods typically focus on specific aspects, such as syntactic or semantic visibility, or rely on human experiments. For instance, BadCode~\cite{2023-BADCODE} outperforms Wan et al.~\cite{2022-you-see-what-I-want-you-to-see} in human experiment evaluations, while AFRAIDOOR~\cite{2024-stealthy-backdoor-attack} uses automated evaluations to demonstrate that adaptive triggers are more stealthy. Nonetheless, these methods do not cover all possible detection dimensions, leaving room for improvement in both evaluation metrics and techniques.

\noindent\textbf{Backdoor Injection Methods for Large Language Models.}
The effectiveness of backdoor injection methods also faces several challenges. Existing methods often fail to work effectively for large code models. Current injection approaches rely on two scenarios: first, attackers cannot control the training process of the model, but the model is trained using poisoned data; second, attackers can control the model’s training process. However, large code language models, such as Codex and GPT-4, are typically closed-source, meaning attackers cannot control the training process or trace the training data. For open-source large \abbr{}s, the cost of injecting backdoors through training or fine-tuning significantly increases. Additionally, as large \abbr{}s become more complex and robust, it becomes increasingly difficult for attackers to insert backdoors without detection, and the effectiveness of these backdoors tends to diminish as the model size grows.

\noindent\textbf{Syntactic Correctness and Semantic Preservation of Adversarial Samples.}
In adversarial attacks on \abbr{}s, the generated adversarial samples need to maintain the syntactic correctness and preserve the semantics of the code. Current techniques typically achieve this by modifying/replacing variable names or applying transformations that do not alter the code's semantics. However, existing evaluation methods do not fully account for the syntactic correctness and semantic preservation of these adversarial samples after perturbation. Even if some adversarial samples appear to maintain the code's semantics on the surface, they may introduce syntactic or logical errors during execution. Therefore, comprehensively evaluating the syntactic correctness and semantic consistency of adversarial samples remains a significant challenge.

\noindent\textbf{Stealthiness of Adversarial Perturbations.}
The stealthiness of adversarial perturbations is a key metric for assessing the quality of adversarial samples. In both white-box and black-box attacks, current techniques often use similarity-based metrics (e.g., CodeBLEU) to evaluate the stealthiness or naturalness of adversarial samples. However, these metrics are not always ideal. Some perturbations may be imperceptible to humans but show significant differences in similarity metrics, and vice versa. Moreover, the current metrics do not cover all aspects that impact the stealthiness of adversarial samples, particularly when evaluating the real-world effects of the perturbations. Thus, designing more comprehensive metrics for evaluating stealthiness remains a challenge in this field.

\noindent\textbf{In-depth Understanding of the Principles Behind Attacks on \abbr{}s.}
The development of explainability may help us better understand the underlying principles of backdoor and adversarial attacks. Since the inception of neural networks, they have faced the issue of low interpretability. Small changes in model parameters can significantly impact prediction results, and people cannot directly comprehend how neural networks operate. In recent years, explainability has become a pressing area in deep learning. How to deeply understand neural networks (including \abbr{}s) themselves and explain how inputs affect outputs remain urgent problems to be addressed.
Currently, some work is being done to provide security and robustness proofs for adversarial attacks. However, more in-depth research is needed to explain the reasons behind prediction results, making the training and prediction processes no longer black boxes. Explainability can not only enhance the security of \abbr{}s but also unveil the mysteries of the model, making it easier for us to understand its mechanisms. However, this could also benefit attackers.
For example, for backdoor attacks, attackers may analyze the model for explainability to select triggers that are more likely to cause harm. For adversarial attacks, attackers could exclude input ranges that have been proven to be secure, thereby narrowing the search space and more effectively finding adversarial samples. They could also construct targeted attacks by gaining a deeper understanding of the model. Nevertheless, this field should not stagnate, as black-box models do not guarantee safety. Therefore, as explainability improves, the security of \abbr{}s may be enhanced in a convoluted manner.

\noindent\textbf{\R{Exploring the Security of CodeLLMs from Other Attack Perspectives.}}
\R{In addition to backdoor and adversarial attack threats, recent research has also explored other types of attack threats against CodeLLMs, such as privacy leakage~\cite{2023-CodexLeaks, 2024-Unveiling-Memorization-in-Code-Models, 2024-Traces-of-Memorisation-in-Large-Language-Models-for-Code, 2023-Uncovering-the-Privacy-Issue-of-Neural-Code-Completion-Tools}, membership inference attacks~\cite{2024-This-Model-Uses-My-Code, 2024-Traces-of-Memorisation-in-Large-Language-Models-for-Code, 2024-Code-Membership-Inference-for-Detecting-Unauthorized-Data} and watermark attack~\cite{2025-DeCoMa}.
Exploring new attack threats is crucial for a deeper understanding of the security of CodeLLMs. By studying these attack methods, researchers can identify potential vulnerabilities that the model may face in real-world applications, reveal its weaknesses in various scenarios, and better understand the model’s behavior. Therefore, conducting more research on different types of attacks will help both academia and industry more comprehensively assess the security of CodeLLMs and provide a foundation for future security research.}

\subsection{Challenges and Opportunities in Defenses against Attacks on \abbr{}s}
\label{subsec:challenges_and_opportunities_in_defenses}

\noindent\textbf{Balancing the Effectiveness of Backdoor Defenses with Maintaining Model Performance Stability.}
In Section~\ref{subsec:backdoor_defenses_on_LM4Code}, we present the existing backdoor defense techniques for \abbr{}s.  
These defense techniques are typically designed to protect different phases of \abbr{}s from backdoor attacks.  
However, accurately and efficiently detecting poisoned samples or removing backdoors from the \abbr{}s, while ensuring the model's regular performance, still presents many challenges. 
Firstly, for pre-training defense techniques~\cite{2022-backdoors-in-neural-models-of-source-code, 2024-poison-attack-and-poison-detection}, they primarily detect poisoned samples by identifying ``outlier'' characteristics in the training data. 
However, this approach often results in a high false positive rate and requires significant computational resources, making it difficult to achieve both high accuracy and efficiency in detecting poisoned samples. 
For complex triggers, such as variable/function name extension or adaptive triggers, current defense techniques face even greater difficulty in detecting and removing them.
Secondly, for post-training defense techniques, they primarily remove backdoors from the model or inputs through unlearning or input filtering. As the model size increases, these techniques require an unacceptable amount of time and computational resources. Additionally, these techniques can cause a certain level of damage to the model's regular performance. 
Applying these methods in real-world scenarios presents significant challenges.

\noindent\textbf{Balancing Model Performance and Robustness in Adversarial Defense Techniques.}
In Section~\ref{subsec:adversarial_defenses_on_LM4Code}, we introduced the existing adversarial defense techniques. The adversarial defense methods for \abbr{}s primarily use adversarial training or data augmentation techniques to improve the robustness of the model. However, maintaining model performance while enhancing robustness and security remains a significant challenge.
Current research employs gradient-based perturbations to transform programs in the worst-case scenarios. Compared to random perturbations, this approach is more likely to produce robust models. However, these methods often reduce the model’s normal performance while improving robustness. Although some studies attempt to enhance both model robustness and performance by combining gradient-based adversarial training with programming language data features or by designing specific loss functions, these methods tend to require more computational resources.

\noindent\textbf{Multi-Scenario Defense against Attacks on \abbr{}s.}
Compared to backdoor attacks and adversarial attacks on \abbr{}s, research progress on backdoor and adversarial defenses has lagged behind. Therefore, there are still numerous research opportunities in addressing attacks across different scenarios. Beyond single defense scenarios, multi-scenario defense techniques hold greater potential. From the perspective of the \abbr{} lifecycle, implementing defense strategies that address both data protection and model protection before, during, and after model training in a mixed-scenario approach can further enhance the security of \abbr{}s.

\noindent\textbf{Explainability in Defending against Attacks \abbr{}s.}
Advancements in explainability can help address the lag in defense methods. Since current research has yet to fully understand \abbr{}s (e.g., why inputs with triggers are predicted as target results, and how different data affect model weights), discovering vulnerabilities is often easier than preventing attacks in advance. This leads to a certain lag in the security of \abbr{}s. If we can thoroughly understand code models, it is believed that defenses will surpass or at least keep pace with advancements in attack techniques.

\noindent\textbf{\R{Defending Against Emerging Security Threats in CodeLLMs.}}
\R{With the emergence of new security threats such as privacy leakage and membership inference attacks, it has become crucial to extend existing defense techniques to address these risks. Although previous studies have proposed defense techniques to enhance the privacy protection of CodeLLMs~\cite{2023-InCoder, 2023-SantaCoder, 2023-StarCoder}, these defenses still face challenges in the face of the continuously emerging attack threats. Therefore, developing defense strategies that effectively address the evolving threats is essential to maintaining the security of CodeLLMs in real-world applications.}

Overall, the security threats to \abbr{}s can be seen as an ongoing and evolving struggle between attackers and defenders, with neither side able to secure absolute dominance. Nevertheless, both parties can leverage new techniques and applications to gain a strategic advantage. On the attacker’s side, an effective strategy involves probing for new attack vectors, uncovering novel attack scenarios, pursuing diverse objectives, and expanding the scope and impact of attacks. On the defender’s side, employing a combination of defense mechanisms presents a promising approach to mitigating attack risks. However, such integration may introduce additional computational or system overhead, which must be carefully addressed during the design phase.

\section{Threats to Validity}
\label{sec:threats_to_validity}

\R{\textbf{Potential Impact of Database Selection on the Inclusion of Relevant Papers.}
We conduct automated searches across six widely used databases, including Google Scholar, ACM Digital Library, IEEE Xplore, Springer, DBLP, and ArXiv, all of which are standard and commonly used online engines or databases.
One potential issue is that relevant papers may be omitted if the search relies solely on these databases.
To mitigate this impact, we also employ both forward and backward snowballing search methods to further enhance the search results. Through snowballing, we identify papers with transitive dependencies and expand our paper collection. Specifically, backward snowballing helps us uncover relevant research in the reference lists of the collected papers, while forward snowballing tracks subsequent studies that cite the papers we have collected, further extending the scope of our search. This approach ensures comprehensive coverage of relevant literature, reducing the risk of omissions.}

\R{\noindent\textbf{Potential Impact of Non-peer-reviewed Papers on the Reliability of the Conclusions.}
We include ArXiv in our search databases, considering that some researchers are inclined to disclose the latest techniques on ArXiv in advance. As a result, the final selection of papers may include non-peer-reviewed papers. One potential impact is that these non-peer-reviewed papers may affect the reliability of our research conclusions. 
To mitigate this impact, we ensure the quality of the selected papers.
We apply five exclusion criteria to rigorously filter all the collected papers.
For the final set of selected papers, the first two authors review them thoroughly and multiple times to ensure that each paper has a clear research motivation, provides detailed descriptions of the techniques, presents comprehensive experimental setups, and clearly confirms the experimental findings, thus ensuring that low-quality papers are excluded from our study.
Additionally, our research conclusions primarily encompass the development trends in the field (Section~\ref{sec:survey_methodology}) and summaries of various research directions (Sections~\ref{sec:Answer_to_RQ1} to \ref{sec:Answer_to_RQ4}). These conclusions do not include specific technique comparisons between the papers. Therefore, the inclusion of non-peer-reviewed papers does not significantly impact the reliability of our conclusion.}

\R{\noindent\textbf{Potential Bias in Study Selection.} In the study selection process, we establish five inclusion and exclusion criteria to filter the papers, combining automated and manual procedures. In the first and second phases of the study selection, we use automated filtering methods, as the inclusion and exclusion criteria in these stages are easily aligned with relevant fields in the BibTeX records. However, in the third to fifth phases, due to the difficulty in directly correlating the inclusion and exclusion criteria with specific fields in the BibTeX records, the risk of mislabeling papers is higher. Therefore, we use manual filtering. During this process, we carefully review the papers multiple times to mitigate potential biases introduced by manual filtering. However, manual filtering may still be influenced by the researchers’ subjective judgment biases, which could affect the accuracy of the quality assessment of papers. To address these concerns, the first two authors carefully review the selected papers multiple times, focusing on the abstract, introduction, related work, experiments, and conclusion sections, to fully understand the research objectives and categorization of each paper. This step aims to improve the accuracy of paper selection, minimize omissions, and reduce the potential bias in the study selection process.}

\section{Conclusion}
\label{sec:conclusion}
This paper systematically reviews the literature on the security of \abbr{}s and provides a detailed analysis of recent publications. The focus of this paper is on the security threats posed by \abbr{}s in the context of backdoor attacks and adversarial attacks. Backdoor attacks are categorized into data poisoning attacks and model poisoning attacks, while adversarial attacks are divided into white-box adversarial attacks and black-box adversarial attacks. A comprehensive summary and analysis of research on attacks targeting different categories of \abbr{}s are presented. Subsequently, the paper reveals the defense gaps corresponding to backdoor attacks and adversarial attacks. Additionally, a thorough analysis of the challenges faced by current research in this field is provided, offering valuable guidance for further investigation. Finally, the paper organizes and summarizes commonly used datasets, language models, evaluation metrics, and the accessibility of tools in the field of security for \abbr{}s, facilitating the convenience of readers.

\section*{Acknowledgments}
This research is supported by the National Key Research and Development Program of China (2024YFF0908001) and the National Natural Science Foundation of China (U24A20337 and 62372228), the National Research Foundation, Singapore, and DSO National Laboratories under the AI Singapore Programme (AISG Award No: AISG2-GC-2023-008), as well as the National Research Foundation, Singapore, and the Cyber Security Agency of Singapore under the National Cybersecurity R\&D Programme (NCRP25-P04-TAICeN). Any opinions, findings, conclusions, or recommendations expressed in this paper are those of the author(s) and do not reflect the views of the National Research Foundation, Singapore, or the Cyber Security Agency of Singapore.

\bibliographystyle{ACM-Reference-Format}
\bibliography{reference}





\end{document}